 \documentclass{article}
 \usepackage{amsfonts}

\def\nqq{\hspace{-2em}}

\newcommand{\beq}[1]{\begin{equation}\label{#1}}
\newcommand{\eeq}{\end{equation}}
\newcommand{\bear}[1]{\begin{eqnarray}\label{#1}}
\newcommand{\ear}{\end{eqnarray}}
\newcommand{\nn}{\nonumber}
\newcommand{\rf}[1]{(\ref{#1})}

\catcode`\@=11 \@addtoreset{equation}{section}\catcode`\@=12

\def\barr{\left(\begin{array}}
\def\earr{\end{array}\right)}
\def\beq#1{\begin{equation}\label{#1}}
\def\eeq{\end{equation}}
\def\ber#1{\begin{eqnarray}\label{#1} \nqq}
\def\eer{\end{eqnarray}}
\def\eern{\nonumber \end{eqnarray}}
\def\mm{\\ \nqq}

\newcommand{\N}{ {\mathbb N} }
\newcommand{\R}{ {\mathbb R} }
\newcommand{\C}{ {\mathbb C} }

\newcommand{\SL}{\mathop{\rm sl}\nolimits}
\newcommand{\SO}{\mathop{\rm so}\nolimits}
\newcommand{\SP}{\mathop{\rm sp}\nolimits}
\newcommand{\ds}{\displaystyle}

\newcommand{\sign}{ \mbox{\rm sign} }
\newcommand{\e}{ \mbox{\rm e} }
\newcommand{\eps}{ \varepsilon }
\newcommand{\rank}{ \mbox{\rm rank} }

\newcommand{\p}{\partial}
\newcommand{\btd}{\bigtriangledown}
\newcommand{\btu}{\bigtriangleup}
\newcommand{\tri}{\Delta}
\newcommand{\sq}[1]{\sqrt{|#1|}}

\newcommand{\sh}{\mathop{\rm sh}\nolimits}
\newcommand{\ch}{\mathop{\rm ch}\nolimits}
\newcommand{\im}{{\rm i}}
\newcommand{\fnm}{\footnotemark}
\newcommand{\fnt}{\footnotetext}

\begin{document}

\begin{center}
\large \bf
Exact solutions in multidimensional gravity \\
with  antisymmetric forms

\end{center}

\vspace{0.3truecm}

\begin{center}

\normalsize\bf
V. D. Ivashchuk\fnm[1]\fnt[1]{e-mail: ivas@rgs.phys.msu.su}
and V. N. Melnikov\fnm[2]\fnt[2]{e-mail: melnikov@rgs.phys.msu.su},

\it
Center for Gravitation and Fundamental Metrology,
VNIIMS, 3-1 M. Ulyanovoy Str., Moscow, 117313, Russia and

\it Institute of Gravitation and Cosmology,
Peoples' Friendship University of Russia,
6 Miklukho-Maklaya St., Moscow 117198, Russia

\end{center}

\begin{abstract}

This topical review deals with a multidimensional gravitational model
containing  dilatonic scalar fields
and antisymmetric forms. The manifold
is chosen in the form $M = M_0 \times M_1 \times \ldots
\times M_n$, where  $M_i$ are Einstein spaces ($i \geq 1$).
The sigma-model approach and
exact solutions in the model are reviewed
and the solutions with p-branes
(e.g. Majumdar-Papapetrou-type, cosmological, spherically symmetric,
black-brane and Freund-Rubin-type ones) are considered.

\end{abstract}
\hspace*{0.950cm} PACS number(s): \ 0450, \ 0465, \ 9880H, \ 0460K

\newpage

{\advance\baselineskip -0.7pt

\tableofcontents

\contentsname

}

\newpage

\section{\bf Introduction}
\setcounter{equation}{0}

\subsection{Multidimensional models}

The motivation for studying multidimensional models of gravitation and
cosmology \cite{Mel2,Mel} is quite apparent for several reasons.
Indeed, the
main trend of modern physics is the unification of all known fundamental
physical interactions: electromagnetic, weak, strong and gravitational
ones. During the recent decades there has been significant progress in
unifying weak and electromagnetic interactions, and some more modest
achievements in GUT, supersymmetric, string and superstring theories.

Now, theories of membranes, $p$-branes and (more vague) M- and F-theories
are being created and studied (see subsect 1.2 below).
Since no self-consistent successful theory of
unification is currently available,
it is desirable to study the common features of these
theories and their applications to solving basic problems of modern
gravity and cosmology.

Multidimensional gravitational models, as well as scalar-tensor
theories of gravity, are theoretical frameworks for describing
possible temporal and range variations of fundamental physical constants
\cite{StM,Mel3,deSMP,Mel4}. These ideas
originated from the earlier papers of
 Milne (1935) and  Dirac (1937) \cite{Dirac}
on relations between the phenomena of
micro- and macro-worlds, and up until now they
have been thoroughly studied both theoretically and experimentally.

On applying multidimensional gravitational models to
the basic problems
of modern cosmology and black-hole physics,
we hope to find answers to such long-standing problems as the
cosmological constant, acceleration, isotropization and graceful
exit problems, stability and nature of
fundamental constants \cite{Mel3},
the possible number of extra dimensions, their stable compactification,
etc.

Bearing in mind that multidimensional gravitational models are certain
generalizations of general relativity which is tested reliably for weak
fields up to 0.001 and partially in strong fields (binary pulsars), it is
quite natural to inquire about their possible observational or
experimental windows.  From what we already know, among these windows are:

-- possible deviations from the Newton and Coulomb laws, or new
interactions,

-- possible variations of the effective gravitational constant with
a time rate smaller than the Hubble one,

-- possible existence of monopole modes in gravitational waves,

-- different behaviour of strong field objects, such as multidimensional
black holes, wormholes and $p$-branes,

-- standard cosmological tests etc.

As no accepted unified model exists, in our approach we adopt
a simple,  but general (from the point of view of number of
dimensions) models based on multidimensional Einstein equations with or
without sources of a different nature:
cosmological constant, perfect and
viscous fluids, scalar and electromagnetic fields, fields of
antisymmetric forms (related to $p$-branes), etc.
Our programme's main objective was and is to obtain exact
self-consistent solutions (integrable models) for these models and then to
analyze them in cosmological, spherically and axially symmetric cases. In
our view this is natural and most reliable way to study highly nonlinear
systems. Here (in the bulk of the paper) we review only exact solutions
with scalar fields and antisymmetric forms.

The history of the multidimensional approach begins with the well known
papers of Kaluza and Klein on five-dimensional theories which
initiated interest in investigations in
multidimensional gravity. These ideas were
continued by Jordan who suggested considering the more general case
$g_{55}\ne{\rm const}$, leading to a theory with an additional scalar
field. They were in some sense a source of inspiration for Brans
and  Dicke in their well known work on a scalar-tensor gravitational
theory. After their work many investigations were performed
in models with
material or fundamental scalar fields, both conformal and non-conformal
(see details given in \cite{StM}).

A revival of the ideas of many dimensions
started in the 1970s and  continues
now, mainly due to the development of unified theories. In the 1970s
interest in multidimensional gravitational models was stimulated mainly
by
(i) the ideas of gauge theories leading to a non-Abelian
generalization of the Kaluza-Klein approach
and (ii) by supergravitational
theories. In the 1980s the supergravitational theories were ``replaced" by
superstring models. Now it is driven by expectations connected with
the overall M-theory. In all these theories,
four-dimensional gravitational models with extra fields were obtained from
some multidimensional model by a dimensional reduction based on the
decomposition of the manifold
$$ M=M^4\times M_{\rm int}, $$
where $M^4$ is a four-dimensional manifold and $M_{\rm int}$ is some
internal manifold (widely considered to be compact).

The earlier papers on multidimensional gravity and cosmology dealt with
multidimensional Einstein equations and with a block-diagonal cosmological
or spherically symmetric metric, defined on the manifold $M= \R \times
M_0\times \dots \times M_n$ of the form
$$     g=-dt\otimes dt+\sum_{r=0}^n a_r^2(t) g^r $$
where $(M_r,g^r)$ are Einstein spaces, $r=0,\dots,n$.
In some of them a
cosmological constant and simple scalar fields were also used \cite{BIMZ}.

Such models are usually reduced to pseudo-Euclidean Toda-like systems with
the Lagrangian \cite{IMi}
$$ L=\frac12G_{ij}\dot x^i\dot x^j-\sum_{k=1}^mA_k{\rm e}^{u_i^kx^i} $$
and the zero-energy constraint $E=0$.

It should be noted that pseudo-Euclidean Toda-like systems are not
well studied yet. There exists a special class of equations of state that
gives rise to Euclidean Toda models \cite{GIM}.

It is well known that cosmological solutions are closely
related to the solutions exhibiting spherical
symmetry, and relevant schemes to obtain these
solutions are quite similar to those
applied in the cosmological approach \cite{Mel2}. The first
multidimensional generalization of such a type was considered by
Kramer \cite{Kr}
and rediscovered by Legkii,  Gross and  Perry \cite{Le,GP}
(and also by
Davidson and Owen). In \cite{BrI} the Schwarzschild solution was
generalized to the case of $n$ internal Ricci-flat spaces, showing
that a black hole configuration takes place when the scale factors of
internal spaces are constants.
Another important feature is that a minimally
coupled scalar field is incompatible with the existence of black holes.
Additionally, in
\cite{FIM2} an analogous generalization of the Tangherlini solution
was obtained, and an investigation of singularities was performed in
\cite{IMB}.  These solutions were also generalized to the electrovacuum
case with and without a scalar field \cite{FIM3,IM8,BM}. Here, it was
also proved that BHs exist only when a scalar field is switched
off. Deviations from the Newton and Coulomb laws were obtained depending on
mass, charge and number of dimensions.
A theorem was proved in \cite{BM} that ``cuts'' all non-black-hole
configurations as being unstable under even monopole perturbations. In
\cite{IM13} the extremely charged dilatonic black hole solution was
generalized to a multicenter (Majumdar-Papapetrou) case when the
cosmological constant is non-zero.

We note that for $D =4$ the pioneering Majumdar-Papapetrou solutions with
a conformal scalar field and an electromagnetic field were considered in
\cite{Br}.

{\bf ``Brane-world'' approach.}
Recently, interest in the so-called  ``brane world''  models
(see \cite{RubS}-\cite{IMbrw} and  therein)
was greatly increased  after the papers of \cite{ArDDK,RS1,RS2}.
It is supposed that we are living on a $(1+3)$-dimensional
thin (or thick) layer (``3-brane'') in multidimensional space and
there exists a potential preventing us from escaping from
this layer, i.e. gauge and matter fields are localized on branes
whereas gravity ``lives'' in a multidimensional bulk.
Randall and Sundrum \cite{RS1} suggested a rather elegant
construction for the confining  potential
(see also \cite{MVV} )
using two symmetric copies of a part of 5-dimensional
bulk anti-deSitter space-time.
In framework of ``brane-world'' approach the  modifications of
Friedmann equations and Newton's law were obtained.
Nevertheless, at present the status quo of
this approach is unclear, since in this rapidly developing field
there exist a huge stream of publications, where any new paper may change
drastically the ``state of arts''.

\subsection{Solutions with $p$-branes}

In this review we consider  several classes of the
exact solutions for the multidimensional gravitational
model governed (up to some details)
by the Lagrangian
\beq{1.1}
{\cal L}  = R[g]- 2\Lambda - h_{\alpha\beta}
g^{MN}\partial_{M}\varphi^\alpha\partial_{N}\varphi^\beta
-\sum_{a}\frac{1}{n_a!}\exp(2\lambda_{a \alpha}\varphi^\alpha)
(F^a)^2,
\eeq
where $g$ is metric, $F^a = d A^a$ are forms of ranks $n_a$
and $\varphi^\alpha$ are scalar fields and $\Lambda$ is a cosmological
constant  (the matrix $h_{\alpha\beta}$ is invertible).

The simplest $D$-dimensional
theory with scalar field, 2-form and dilatonic coupling
$\lambda^2 = (D -1)/(D-2)$ may be obtained
by dimensionally reducing the $(D+1)$-dimensional
Kaluza-Klein theory  (in this case the scalar field $\varphi$ is
associated with the size of $(D+1)$ dimension).
We note that the cosmological constant term can be
imitated also by the form of ${\rm rank F} = D$.

{\bf Supergravities.} For certain field  contents with
distinguished values of total dimension $D$, ranks $n_a$,
dilatonic couplings $\lambda_{a}$  and $\Lambda = 0$
such Lagrangians appear as ``truncated'' bosonic sectors
(i.e. without Chern-Simons terms) of certain
supergravitational theories or low-energy limit of superstring models
\cite{CJS,SaSe,GrSW}. For $D=11$ supergravity
\cite{CJS} (that is considered now as a low-energy limit
of some conjectured $M$-theory
\cite{Tow1,Wit,HTW,Tow2,Sc,Du,Gibb}) we have
a metric and $4$-form in the bosonic sector
(there are no scalar fields). For $D = 10$ one may consider
type I supergravity with metric,  scalar field and 3-form,
type IIA supergravity with bosonic fields of type I supergravity
called as Neveu-Schwarz-Neveu-Schwarz (NS-NS) sector
and additionally 2-form and 4-form
Ramond-Ramond (R-R) sector,
type IIB supergravity with bosonic fields of type I supergravity
(NS-NS sector) and additionally 1-form,  3-form  and (self-dual) 5-form
(R-R sector). It is now believed that all five
string theories (I, IIA, IIB and two heterotic ones with
gauge groups $G = E_8 \times
E_8$ and ${\rm Spin(32)}/Z_2)$ \cite{GrSW} as well as 11-dimensional
supergravity \cite{CJS} are limiting case of $M$-theory. All these theories
are conjectured to be related by a set of duality transformations: $S-$,
$T-$ (and more general $U-$) dualities \cite{HTW,Gibb}.

The list of supergravitional theories is not restricted only
by dimensions $D =10,11$ and signature $(-,+, \ldots,+)$.
One may consider also supergravities in dimensions
$D < 11$ (e.g. those obtained by dimensional reduction
from $D=11$ supergravity \cite{St}), or
$D =12$ supergravity with two time dimensions \cite{Nish},
or even Euclidean supergravity model \cite{CLLPST}.

It was proposed earlier that $IIB$ string
may have its origin in a 12-dimensional theory, known as $F$-theory
\cite{Hull,Vafa}. In \cite{KKLP} a low energy effective
(bosonic) Lagrangian for $F$-theory was suggested.
The field content of this 12-dimensional field model
is the following one: metric, one scalar field
(with negative kinetic term), 4-form and 5-form.
In \cite{IMJ} a chain of so-called  $B_D$-models
in dimensions $D = 11, 12, \ldots$ was suggested.
$B_D$-model contains $l =D-11$ scalar fields with  negative
kinetic terms (i.e. $h_{\alpha\beta}$ in (\ref{1.1}) is negative
definite)
coupled to $(l+1)$ different forms of ranks $4, \ldots, 4 + l$.
These models were constructed using $p$-brane intersection
rules that will be discussed below. For $D=11$ ($l=0$) $B_D$-model
coincides with the truncated  bosonic sector
of $D=11$ supergravity. For $D=12$ $(l=1)$ it coincides with
truncated $D=12$ model from \cite{KKLP}. It was conjectured
in \cite{IMJ} that these $B_D$-models for $D > 12$ may correspond
to low energy limits of some unknown $F_D$-theories (analogues of
$M-$ and $F$-theories).

{\bf Setup for fields.}
Here we review certain classes of (non-localized)
$p$-brane  solutions to field equations corresponding to
the Lagrangian (\ref{1.1}).
These solutions have a block-diagonal metrics
defined on $D$-dimensional product manifold, i.e.
\beq{1.2}
g= e^{2\gamma} g^0  + \sum_{i=1}^{n} e^{2\phi^i} g^i, \qquad
M_0  \times M_{1} \times \ldots \times M_{n},
\eeq
where $g^0$  is  a metric on $M_0$  and $g^i$ are fixed
Ricci-flat (or Einstein) metrics on $M_i$ ($i >0$).  The moduli
$\gamma, \phi^i$ and scalar fields  $\varphi^{\alpha}$
are functions on $M_0$ and fields of forms are also governed by
several scalar functions on $M_0$. Any $F^a$ is supposed to be a
sum  of  (linear independent) monoms, corresponding to electric or
magnetic $p$-branes ($p$-dimensional analogues of membranes),
i.e. the so-called composite $p$-brane ansatz is considered.
(In non-composite case we have no more than one monom for each $F^a$.)
$p=0$ corresponds to a particle, $p=1$ to a string, $p=2$
to a membrane etc.
The $p$-brane worldvolume (worldline for $p=0$,
worldsurface for $p=1$ etc) is isomorphic to some product of
internal manifolds:
$M_I = M_{i_1} \times \ldots \times M_{i_k}$
where $1 \leq i_1 < \ldots < i_k \leq n$ and has dimension
$p + 1 = d_{i_1} + \ldots + d_{i_k} = d(I)$, where
$I = \{i_1, \ldots, i_k \}$ is a multiindex describing
the location of $p$-brane and $d_i = {\rm dim} M_i$.
Any $p$-brane is described   by the triplet ($p$-brane index)
$s = (a,v,I)$, where $a$ is the color index
labeling the form $F^a$, $v = e(lectric), m(agnetic)$
and $I$ is the multiindex defined above. For the electric
and magnetic branes corresponding to form $F^a$  the worldvolume
dimensions are $d(I)=n_a-1$ and $d(I)=D-n_a-1$, respectively.
The sum of this dimensions is $D - 2$. For $D =11$ supergravity
we get $d(I) =3$ and $d(I) = 6$, corresponding to electric $M2$-brane
\cite{DS} and magnetic $M5$-brane \cite{Guv}, respectively (see also
\cite{HS}).

{\bf Sigma model representation. Constraints.}
In  \cite{IMC}  the model  under consideration
was reduced to gravitating self-interacting sigma-model
with certain constraints imposed. These constraints
coincide with non-block-diagonal part of Hilbert-Einstein
equations. For $d_0 = {\rm dim} M_0 \neq 2$ there
are two groups of constraints:
$(ee + mm)$ electric-electric plus magnetic-magnetic
and $(em)$ electro-magnetic. In the first case
$(ee + mm)$ the number of constraints is $n_1(n_1 - 1)/2$,
where $n_1$ is number of 1-dimensional manifolds among $M_i$.
``Electro-magnetic'' $(em)$ constraints appear for ${\rm dim} M_0 = 1, 3$
and the number of these constraints is $n_1$.
We note that in  $d_0 \neq 2$ case
a generalized harmonic gauge is used
$\gamma= \sum_{j=1}^n d_j\phi^j/(2-d_0)$.
For $d_0 = {\rm dim} M_0 = 2$ the sigma model representation
with $(ee + mm)$ constraints is also valid when additional
restrictions on brane intersections are imposed.
Here (see Section 2)  we consider
the sigma-model representation in a simplified form:
all constraints are satisfied identically due to additional
restrictions on brane intersections. There
are three groups of restrictions:
$(ee)$ electric-electric, $(mm)$ magnetic-magnetic
and $(em)$ electric-magnetic.
The restrictions of $(ee)$- and $(mm)$-types forbid the following
intersections of two electric or two magnetic branes with the same color
index:  $d(I \cap J) =d-1$, where $d=d(I) =d(J)$.
Electro-magnetic restrictions forbid the
following intersections:
$d(I \cap J) = 0$  for $d_0 = 1$, and $d(I \cap J) = 1$ for $d_0 = 3$.
All restrictions are satisfied identically in non-composite
case when there are no two branes with the same color index.
The restrictions are satisfied also when $n_1$ is small enough:
$n_1 \leq 1$ in $(ee)$- and $(mm)$-cases and $n_1 =0$ for $(em)$-case.
(Notice, that
$(ee)$- and $(mm)$-restrictions were considered first in \cite{AR}).
The derivation of all constraints and restrictions on intersections
was considered in detail in \cite{IMC}.
(The sigma-model representation for
non-composite  electric case was obtained earlier in
\cite{IM11,IM12}, for  electric composite case see also
\cite{IMR}). We note that, recently, sigma model representation
for  non-block-diagonal  metrics  and  two (intersecting)
branes was obtained in \cite{GR}.

The $\sigma$-model Lagrangian has the form  \cite{IMC}
(see Section 2)
\bear{1.3}
{\cal L}_{\sigma } =
R[g^0]- \hat G_{AB} g^{0\mu\nu}\p_\mu\sigma^A\p_\nu\sigma^B
-\sum_{s}\eps_s \exp(-2U^s)g^{0\mu\nu} \p_\mu\Phi^s\p_\nu\Phi^s -2V,
\ear
where $(\sigma^A)=(\phi^i,\varphi^\alpha)$, $V$ is a potential,
$(\hat G_{AB})$ are components of (truncated) target space metric,
$\eps_s = \pm 1$,
$$
U^s =   U_A^s \sigma^A =
\sum_{i \in I_s} d_i \phi^i - \chi_s \lambda_{a_s \alpha} \varphi^{\alpha}
$$
are linear functions,
$\Phi^s$ are scalar functions on $M_0$ (corresponding to forms),
and $s=(a_s,v_s,I_s)$. Here parameter
$\chi_s = +1$ for the
electric brane ($v_s =e$) and $\chi_s = -1$ for the magnetic one ($v_s =m$).

A pure gravitational sector of the sigma-model
was considered earlier in \cite{Ber,RZ,IM0}
(notice, that ref. \cite{Ber} contains a typo in the potential).
For $p$-brane applications $g^0$ is Euclidean,
$(\hat G_{AB})$ is positive definite (for $d_0 > 2$) and
$\eps_s= -1$, if pseudo-Euclidean (electric and magnetic) $p$-branes
in a pseudo-Euclidean space-time are considered.
The sigma-model (\ref{1.3}) may be also considered
for the pseudo-Euclidean metric $g^0$ of signature
$(-,+, \ldots, +)$ (e.g. in investigations of gravitational
waves). In this case for a positive definite matrix $(\hat G_{AB})$ and
$\eps_s= 1$ we get a non-negative kinetic energy terms.

{\bf Brane $U$-vectors.} The co-vectors $U^s$ play a key role in
studying the integrability of the field equations \cite{IMC,IMBl,IK}
and possible existence of stochastic behaviour
near the singularity \cite{IMb1}. An important
mathematical characteristic here
is the matrix of scalar products
$(U^s,U^{s'}) =\hat G^{AB} U_A^s U_B^{s'}$,
where $(\hat G^{AB}) = (\hat G_{AB})^{-1}$.
The scalar products for co-vectors
$U^s$  were calculated in \cite{IMC}
(for electric  case see \cite{IM11,IM12,IMR} )
$$
(U^s,U^{s'})=d(I_s\cap I_{s'})+\frac{d(I_s)d(I_{s'})}{2-D}+
\chi_s\chi_{s'}\lambda_{a_s \alpha} \lambda_{a_{s'} \beta} h^{\alpha \beta},
$$
where $(h^{\alpha\beta})=(h_{\alpha\beta})^{-1}$; $s=(a_s,v_s,I_s)$,
$s'=(a_{s'},v_{s'},I_{s'})$. They  depend upon
brane intersections (first term), dimensions of brane worldvolumes
and total dimension $D$ (second term),
scalar products of dilatonic coupling vectors
and electro-magnetic types of branes (third term).
As will be shown below the so-called
``intersections rules''(i.e. relations for
$d(I_s\cap I_{s'})$)  are defined by scalar
products  of $U^s$-vectors.

{\bf Solutions with harmonic functions.}
In Section 2 we consider a family of Majumdar-Papapetrou
(MP) type  solutions \cite{MP} for gravitating sigma-model
(\ref{1.3}).  An important class of solutions appear when all "internal
spaces" $(M_i,g^i)$ are Ricci-flat, $\Lambda = 0$ (in this case the
potential is trivial $V =0$) and  brane vectors are orthogonal
\beq{1.6}
(U^s,U^{s'})=0
\eeq
for $s \neq s'$ \cite{IMC} and all $\eps_s(U^s,U^s)<0$.
These solutions are
governed by a set of harmonic functions $H_s$ defined
on $M_0$. In non-composite electric case these solutions
were obtained in \cite{IM11,IM12} and in the composite electric
case in \cite{IMR}. In Section 2 we present exact solutions for  $D =
11$ supergravity and  related 12-dimensional theory ($F$-theory)
\cite{IMC}.

In \cite{IMC} the solutions with Ricci-flat spaces $M_i$ and $M_0$
were generalized to the case of
non-Ricci-flat $M_0$, when  some additional
"internal" Einstein spaces of non-zero curvature were
added to $M$ (the $p$-branes "live" only on products of
Ricci-flat spaces). In this case the solution
is a superposition of pure gravitational solution and $p$-brane
one.

{\bf Special solutions.}
When the space $(M,g)$ is the pseudo-Euclidean one,
and spaces $(M_i,g^i)$ are flat, the special solutions
with intersection rules (\ref{1.6}) were considered earlier
in numerous  papers, e.g. in  \cite{V} (for one form $F$ and two branes:
one electric and one magnetic), \cite{AVV} (in non-composite case,
for two forms and two branes) and \cite{AV}  (in electric case
for one form  $F$) (see also \cite{AVVV}), \cite{AR} (for one and two
forms), \cite{AEH,AIR} ($d_1 = \ldots = d_n =1)$.  In these solutions all
``branes'' contain a common manifold, say $M_1$ and time-submanifold
belongs to $M_1$.

We note that there exists a lot of $p$-brane solutions
in supergravitational theories,
governed by harmonic functions with intersection rules
(\ref{1.6}) with flat $M_i$,
see \cite{DKL,St,Gaun} for reviews and also
\cite{Dab}-\cite{OhZ} and references therein.
Some of them were obtained using supersymmetry arguments,
duality transformations, dimensional reductions etc.
For example, the brane solutions of type IIA supergravity
have an eleven dimensional interpretation \cite{Tow1}:
the fundamental string ($FS1$) \cite{Dab} and  five-brane ($NS5$)
\cite{CHS} are double dimensional reduction of
$M2$-brane \cite{DS} and the direct dimensional reduction
of $M5$-brane \cite{Guv}, respectively;
the Dirichlet $D2$- and $D4$-branes  may be obtained
from $M2$- and $M5$-branes using direct and double dimensional
reduction, respectively.
$M$-branes (for $D=11$) and  $D$-branes (corresponding to
Ramond-Ramond sectors of $D=10$ supergravities) play a rather important
role in non-perturbative analysis of superstring theories \cite{Polch}.
We note, that
the  $p$-brane solutions of IIA supergravity (e.g. $D4$- and $NS5$-branes)
play a rather important role in
the so-called MQCD \cite{EGK}--\cite{BIKSY}.

General solutions may be also considered for theories in
dimensions $D \geq 12$.
In \cite{IMC} a general composite solution for truncated $12$-dimensional
model (with $4$-form and $5$-form and one scalar field) from \cite{KKLP}
(corresponding to low-energy limit of $F$-theory) was obtained. This
solution contains electrically charged $2$- and $3$-branes and magnetically
charged $5$- and $6$-branes.  In \cite{IMJ} a chain of  $B_D$-models
in dimensions $D=11,12, \ldots $ was
constructed using intersection rules (\ref{1.6}): the scalar products
of dilatonic couplings vectors were obtained from the requirement
of the existence of binary configurations for any two $p$-branes.
(It looks rather promising that
the intersection rules may be used for constructing new
Lagrangians with fields of forms and scalar fields.)

{\bf Symmetries of target space metric.}
The target space of the model is $(\R^K,{\cal G})$ ($K$ is integer),
where
$${\cal G}=\hat G_{AB}d\sigma^A\otimes d\sigma^B+
\sum_{s}\eps_s \exp(-2U_A^s\sigma^A) d\Phi^s\otimes d\Phi^s.$$
It was proved in \cite{Iv3,Ivas} that the target space
${\cal T}=(\R^K,{\cal G})$ is a homogeneous (coset) space $G/H$
($G$ is the isometry group of ${\cal T}$,
$H$ is the isotropy subgroup of $G$). ${\cal T}$ is (locally)
symmetric (i.e. the Riemann tensor is covariantly constant:
$\nabla_MR_{M_1M_2M_3M_4}[{\cal G}]=0$) if and only if
$$(U^{s_1}-U^{s_2})(U^{s_1},U^{s_2})=0$$
for all $s_1,s_2\in S$, i.e. when any two vectors $U^{s_1}$ and $U^{s_2}$,
$s_1\ne s_2$, are either coinciding $U^{s_1}=U^{s_2}$ or orthogonal
$(U^{s_1},U^{s_2})=0$.
For nonzero
noncoinciding $U$-vectors the Killing equations were solved.
Using a block-orthogonal decomposition of the set of the $U$-vectors
it was shown that under rather general assumptions
the algebra of Killing vectors is a direct sum of several copies of
$sl(2,\R)$ algebras (corresponding to 1-vector blocks), several
solvable Lie algebras ${\bf g_i}$
(with ${\bf g_i}^{(2)} = [{\bf g_i}^{(1)}, {\bf g_i}^{(1)}] = {\bf 0}$,
where ${\bf g_i}^{(1)} = [{\bf g_i},{\bf g_i}]$)
corresponding to multivector blocks
and  the Killing algebra of a flat space.
The target space manifold was decomposed in
a product of a flat space, several 2-dimensional spaces
of constant curvature (e.g. Lobachevsky space, part of
anti-de Sitter space) and several solvable Lie group manifolds.
We note that recently solvable Lie algebras were
studied for supergravity models in numerous papers
(see \cite{torino,torin} and references therein).

{\bf Block-orthogonal solutions.}
In \cite{IMBl} the ``orthogonal'' (in target-space sense)
solutions from \cite{IMC} were generalized to
a more general ``block-orthogonal''  case (see subsect. 3.1)
$$
(U^s,U^{s'})=0, \quad s\in S_i, \quad s'\in S_j, \quad i\ne j;
$$
$i,j=1,\dots,k$, where the index  set $S$ is a union of $k$
non-intersecting (non-empty) subsets $S_i$: $S=S_1 \cup\dots\cup S_k$.
This means that
the set of  brane vectors $(U^s,s\in S)$ has a
block-orthogonal structure with respect to
the scalar product:  it splits into $k$ mutually
orthogonal blocks $(U^s,s\in S_i)$, $i=1,\dots,k$.
The number of independent harmonic functions is $k$.
The solutions do exist, when the matrix of scalar products
$((U^{s_1},U^{s_2}))$ and parameters $\eps_s$ satisfy the relations
$$
\sum_{s'}(U^s,U^{s'})\eps_{s'}\nu_{s'}^2=-1
$$
for some set of real $\nu_s$ (meanwhile, the elementary monoms in
composite forms $F^a$ are proportional to $\nu_s$).
We note that first block-orthogonal solutions
appeared for black and wormhole branes in \cite{Br1}.
When all $(U^s,U^s) \neq 0$ one can introduce
a quasi-Cartan matrix
$$
A_{ss'} \equiv \frac{2(U^s,U^{s'})}{(U^{s'},U^{s'})},
$$
$s,s'\in S$,
which coincides with the Cartan matrix,
when $U^s$ are  simple roots  of some Lie algebra
and $(.,.)$ is standard  bilinear form on the root space.
It should be stressed here, that
the quasi-Cartan matrix is a rather
convenient tool for classification
of $p$-brane solutions. For example,
in \cite{IMBl}  (see
subsection 3.1.2 below) we analyzed
three possibilities, when quasi-Cartan matrix  coincides with
the Cartan matrix of a (simple):
a) finite-dimensional Lie algebra, ${\rm det}(A_{ss'}) > 0)$;
b) hyperbolic  Kac-Moody (KM) algebra, ${\rm det}(A_{ss'}) < 0)$;
c) affine KM algebra, ${\rm det}(A_{ss'}) = 0)$ \cite{Kac,FS}.
It was shown that all $\eps_s=-1$ in the case a) and all
$\eps_s=+1$ in the case b). The last possibility  c) does
not appear in the solutions under consideration.
For fixed $A$-matrix  the intersection rules" read
\beq{1.12}
d(I_{s}\cap I_{s'})=
\frac{d(I_s)d(I_{s'})}{D - 2} - \chi_s\chi_{s'}\lambda_{a_s \alpha}
\lambda_{a_{s'} \beta} h^{\alpha \beta}
+\frac12 (U^{s'},U^{s'}) A_{s s'},  \quad s \neq s'.
\eeq
In supergravity examples  all $(U^s,U^s) =2$.
Intersecting $p$-brane solutions
with ``orthogonal'' intersection rules correspond to the
Lie algebras ${\bf A_1} \oplus \ldots \oplus{\bf A_1} $,
where  ${\bf A_1} =sl(2,\C)$.
In supergravitational  models these solutions correspond to the
so-called BPS saturated states preserving fractional supersymmetries.
The MP solution \cite{MP} in this
classification corresponds to the algebra ${\bf A_1}$.
For Lie algebra case $A_{s s'} = 0, -1, \dots,  \quad s \neq
s'$, and hence, dimensions of $p$-brane intersections in this case are
not greater than in the ``orthogonal''  case (if all $(U^{s},U^{s})$
are positive).
In \cite{IMBl,IKM,IMKor} some examples of solutions
corresponding to hyperbolic and finite dimensional Lie algebras
were considered (see subsect. 3.1.2):  among them ${\bf A_2}$-dyon
solutions  in $D=11$ supergravity and in $B_D$-models.
The dyon solution for
$D=11$ supergravity contains electric 2-brane and magnetic 5-brane.
This solution has ${\bf A_2} = sl(3,\C)$ intersection rule:
$3 \cap 6 = 1$ instead of the ``orthogonal'' one $3 \cap 6 = 2$
(here $3$ and $6$ are worldvolume dimensions). An analogous dyon
solution for $D=10$ $IIA$ supergravity was considered in \cite{GrI}.
For all simple  finite-dimensional Lie algebras the parameters
$\nu_s$ were calculated in \cite{GrI}.  These parameters
are related to integers $n_s$ coinciding with the components
of twice the  dual Weyl vector in the basis of simple coroots
\cite{FS} (see Appendix 3).

The quasi-Cartan matrix should not obviously coincide
with the Cartan one. For instance, one may consider the dyon
solution for truncated bosonic sector (i.e. without Chern-Simons term)
of $D=11$ supergravity with intersection $3\cap6=3$.

In  \cite{IMBl} (see subsect. 3.1.3)  the behavior of the Riemann tensor
squared (Kretschmann scalar) for multicenter  solutions
$$
H_s(x)=1+\sum_{b}\frac{q_{sb}}{|x-b|^{d_0-2}},
$$
was investigated   and criteria for the existence of horizon
and finiteness of the
Kretschmann scalar were established.  For $d_0>2$ and
"multicenter" harmonic functions the so-called
$\eta$- and $\xi$- indicators
were introduced. These indicators describe,
under certain assumptions, the existence of a
curvature singularity and a horizon, respectively,
for $x\to b$.

{\bf One-block solutions from null-geodesics.}
In  \cite{IK} (see subsec. 3.2) the null-geodesic method
(for $D=4$ see \cite{NK,KSMH}) was applied to the
sigma-model with zero potential
(i.e. when all spaces $M_i$ are Ricci-flat)
and  a general class of solutions was obtained.
These solutions are governed  by one harmonic function
$H$ on $M_0$ and functions $f_s(H)=
\exp(-q^s(H))$, where $q^s$ are solutions to Toda-type equations
\beq{1.14}
\ddot{q^s} = - B_s \exp(\sum_{s'} A_{s s'} q^{s'}),
\eeq
$B_s \neq 0$.  When $(A_{s s'})$ is a
Cartan matrix of some semisimple finite-dimensional Lie algebra we get
Toda lattice equations that are integrable \cite{B,K,OP}.
Moreover, according to
Adler-van Moerbeke criterion \cite{AM} the condition of
the integrability in quadratures for Toda-type systems singles out
Cartan matrices $(A_{ss'})$ among quasi-Cartan ones.
Thus intersection rules (\ref{1.12}) are restricted
by condition of integrability of Toda-like system.
In \cite{IK}  special solutions  corresponding to ${\bf A_m}$
Toda lattices (in parametrization of \cite{And}) were considered.
These solutions contain a class of
"cosmological" solutions  with Ricci-flat internal spaces
and the so-called "matryoshka" solution from \cite{GR}.

{\bf Cosmological and spherically symmetric solutions.}
A family of general cosmological type $p$-brane solutions with
$n$ Ricci-flat internal spaces was considered in \cite{IK},
where also a generalization to the case of $n-1$ Ricci-flat spaces and one
Einstein space of non-zero curvature (say $M_1$)
was obtained (see subsect.
4.2). These solutions are defined up to solutions to
Toda-type equations (\ref{1.14}) and may be obtained using
the Lagrange dynamics following from the sigma-model approach
\cite{IMJ} (see subsect. 4.1). The solutions from \cite{IK}
contain a subclass a spherically symmetric solutions
(for $M_1 = S^{d_1}$). Special solutions with
orthogonal and block-orthogonal sets of $U$-vectors were
considered earlier in \cite{IMJ} and \cite{IMJ2,IMJ1}, respectively.
(For non-composite case, see \cite{LPX,BGIM,GrIM,BKR}) and references
therein.)

{\bf Toda solutions.}
In \cite{IMJ} the reduction of $p$-brane cosmological
type solutions to Toda-like systems was performed (see also \cite{IK}).
General classes of $p$-brane solutions
(cosmological and spherically symmetric ones)
related to Euclidean Toda lattices associated with Lie algebras
(mainly ${\bf A_m}$, ${\bf C_m}$ ones)  were obtained in
\cite{GM1,GM2,IK,IMp2,IMp3,GM3}. Special $p$-brane
configurations were considered earlier in \cite{LPX,LP1,LMMP}
(see also refs. therein).

{\bf Static solutions.}
In \cite{Cosm,I2} (see subsect. 4.3) general exact solutions
defined on product of Einstein internal spaces
with intersecting composite p-branes and constant scale factors
were obtained and
the effective cosmological constant was generated via p-branes.
These solutions generalize
well-known Freund-Rubin-type solutions \cite{FR,DNP} of $D=11$
supergravity (for special supergravity solutions see
also  \cite{DLP,torin} and refs. therein).
It is remarkable, that special
solutions of such kind with anti-de-Sitter
factor-spaces appear in the ``near-horizon''
limit of extremely charged $p$-brane configurations
(certain $p$-branes interpolate between flat and AdS spaces \cite{GT}).
The interest to static configurations
appeared due to papers on $AdS/CFT$ correspondence,
i.e. the duality between certain limit of some
superconformal theories in $d$-dimensional space and
string or M-theory compactified on $AdS_{d+1} \times
W$, where $W$ is a compact manifold \cite{M,GKP,Wi}.

{\bf Quantum cosmology.}
In \cite{IMJ,IMAJ} (see subsect. 4.4) the Wheeler-DeWitt (WDW) equation
for the quantum cosmology with composite electro-magnetic $p$-branes
defined on product of Einstein spaces
was obtained (for non-composite  electric case see also \cite{GrIM}).
As in the pure gravitational case \cite{IMZ} this equation
has a covariant  and conformally covariant form (see also \cite{Mis,Hal}).
Moreover, in \cite{IMJ,IMAJ} the WDW equation
was integrated for intersecting $p$-branes with orthogonal $U$-vectors,
when $n-1$ internal spaces are Ricci-flat and one is
the Einstein space of a
non-zero curvature (for non-composite electric case see \cite{GrIM}).
It should be mentioned also, that
a slightly different approach with classical field of forms (and rather
special brane setup) was suggested in \cite{LMMP}.
In \cite{IKenM}  the solutions from \cite{IMJ} were used
for constructing quantum analogues of black brane
solutions (e.g. for $M2$ and $M5$ extremal branes).

{\bf Black brane solutions.}
In \cite{IMp2,IMp3} (see Section 5 below) a family of spherically-symmetric
solutions from \cite{IK} was investigated and a subclass of black-hole
configurations related to Toda-type equations with certain asymptotical
conditions  imposed  was singled out.  These black hole solutions are
governed by functions $H_s(z) > 0$, defined
on the interval $(0, (2 \mu)^{-1})$,
where $\mu > 0$ is the extremality parameter,
and obey a set of differential equations
(equivalent to Toda-type ones)
$$
\frac{d}{dz} \left( \frac{(1 - 2\mu z)}{H_s} \frac{d}{dz} H_s \right) =
 \bar B_s \prod_{s' }  H_{s'}^{- A_{s s'}},
$$
with  the following boundary conditions imposed:
${\bf (i)} \quad H_{s}((2\mu)^{-1} -0) = H_{s0} \in (0, + \infty)$;
${\bf (ii)}  \quad H_{s}(+ 0) = 1$,
$s \in S$. Here  $\bar B_s \neq 0$ and
$(A_{s s'})$ is a quasi-Cartan matrix.
It was shown, that
for the positive definite scalar field metric $(h_{\alpha \beta})$ all
$p$-branes in this solution should contain a time manifold
(see Proposition 1 in \cite{IMp2,IMp3} and Theorem  3 from \cite{Br2};
for ``orthogonal'' case see also \cite{BIM}).
In refs. \cite{IMp1,IMp2,IMp3} the following hypothesis
was suggested: the functions $H_s$
are polynomials when intersection rules (\ref{1.12}) correspond to
semisimple Lie algebras, i.e.  when $(A_{s s'})$ is a Cartan matrix.
This hypothesis
was verified for Lie algebras: ${\bf A_m }$, ${\bf C_{m+1}}$,
$m = 1,2, \ldots$, in \cite{IMp2,IMp3}.
It was also confirmed by special  black-hole "block orthogonal"
solutions considered earlier in \cite{Br1,IMJ2,CIM}.
An analogue of this conjecture for extremal black holes
was considered earlier in \cite{LMMP}.  In Sect. 5  explicit formulas for
the solution corresponding to the algebra ${\bf A_2}$ are presented.
These
formulas are illustrated by two examples of ${\bf A_2}$-dyon solutions:
a dyon
in $D = 11$ supergravity (with $M2$ and $M5$ branes intersecting at a point)
and Kaluza-Klein dyon. In subsect. 5.4 we deal with extremal configurations
(e.g. with multi-black-hole extension.)
We note, that special black hole solutions with orthogonal $U$-vectors
were considered in \cite{AIV,Oh}
($d_1 = \ldots = d_n =1)$, \cite{BIM}
(for non-composite case) and \cite{IMJ}
(for earlier supergravity solutions  see \cite{CT,OS} and refs. therein).
In \cite{BIM,Br1} some propositions related to i)
interconnection between the Hawking temperature and the singularity
behaviour, and ii) multitemporal configurations were proved.

Now we briefly overview several important topics that are not
considered in this  short review.

{\bf PPN parameters.}
In \cite{IMp1,IMp2} the (parametrized) post-Newtonian
(Eddington) parameters $\beta$ and $\gamma$ for $4$-dimensional section of
the metric were calculated. It was shown that $\beta$
does not depend upon the $p$-brane intersections, while
$\gamma$ does depend. These results agree with the earlier
calculations for block-orthogonal case \cite{IMJ2,CIM}
(see also \cite{IMM}).

{\bf Stability of spherically symmetric solutions.} It was shown in
\cite{BrM} that single-brane black hole solutions are stable under
spherically symmetric perturbations, whereas  similar
solutions possessing naked singularities turn out to be catastrophically
unstable (this conclusion may be also extended to some configurations
with intersecting branes). For possible other kinds of instabilities
in multidimensional models (e.g. caused by waves in extra dimensions) see
\cite{GrL}.

{\bf Billiard representation near the singularity.}
It is well-known, that
the cosmological models with $p$-branes may have
a ``never ending'' oscillating behaviour near the cosmological
singularity as it takes place in Bianchi-IX model \cite{BLK}.
Remarkably, this oscillating behaviour may
be described using the so-called billiard representation
near the singularity (for multidimensional case see \cite{IKM1,IKM2,IM}
and refs. therein). In  \cite{IMb1,IMb2} the billiard representation for a
cosmological model with a set of electro-magnetic composite $p$-branes in
a theory with Lagrangian (\ref{1.1}) and metric (\ref{1.2})  was
obtained. Some examples with billiards of a finite volume  in
multidimensional Lobachevsky space (e.g. triangle billiard imitating
Bianchi-IX model) and hence oscillating behaviour near the singularity
were considered.  In terms of the Kasner parameters $\alpha = (\alpha^{A}) =
(\alpha^{i}, \alpha^{\gamma})$, satisfying relations
$$\sum_{i=1}^{n} d_i \alpha^i =
\sum_{i=1}^{n} d_i (\alpha^i)^2 +
\alpha^{\beta} \alpha^{\gamma} h_{\beta \gamma}= 1,$$
the existence of never ending oscillating  behaviour near the
singularity takes place if for any $\alpha$ there exists
$s \in S$ such that
\beq{1.17}
U^s(\alpha) =  U_A^{s} \alpha^A = \sum_{i\in I_s}d_i\alpha^i
 -\chi_s\lambda_{a_s \gamma}\alpha^{\gamma} \leq 0
\eeq
\cite{IMb1}.
Thus, $U$-vectors play a key role in determination
of possible oscillating behaviour near the singularity.
In \cite{IMb1} the relations (\ref{1.17})
were also interpreted in terms of
illumination of a (Kasner) sphere by point-like sources.
We note that recently in   \cite{DamH1,DamH2,DamH3} the
relations (\ref{1.17}) were applied, in fact,
to $D=10,11$ supergravities and the never ending
oscillatory  behaviour of the generic solution near the cosmological
singularity was announced to be established.

{\bf Supersymmetries.}
For flat internal spaces $M_i$ and $M_0$ the
supersymmetric (SUSY) solutions in supergravitational models
were considered  in numerous
publications (see  \cite{DKL,St,Gaun,Ts1,GKT,BREJS,BREJSa}
and refs. therein).  As is well-known,
the major part of these solutions
preserve a fractional number of SUSY of the form
$N = 2^{-k}$, where
$k$ is the number of intersecting $p$-branes.
The "$N = 2^{-k}$-rule" was explained in \cite{Iv2}
using the so-called $2^{-k}$-splitting theorem for a
family of $k$ commuting linear operators.
(We note that this rule is not a general one: there
exists certain counterexamples.)
Recently, certain  SUSY solutions in $D= 10, 11$ supergravities
with several internal Ricci-flat internal spaces were considered
in \cite{DLPS}-\cite{Iv2}.
Hovewer, some of them may be obtained by a simple
replacing of flat metrics by Ricci-flat ones.
The major part of these solutions are not new ones but are special
cases of those obtained before (see Section. 2).
For example,
the magnetic $5$-brane solution from \cite{K1}
with $N =1/4$ SUSY is a special case of solutions
from \cite{IM11,IM12,IMC} etc.
Notice, that in \cite{Iv2}
the fractional number of SUSY were obtained
for (non-marginal) $M2$- and $M5$-brane
solutions, defined on product  of two Ricci-flat spaces.

{\bf Reviews.}
We note, that there exist  several good reviews devoted to
certain aspects of solutions with $p$-branes (see, for example,
\cite{DKL,St,Y}). However these reviews  deal mainly with more or less
special classes of $p$-brane solutions and their applications in
superstring, M-theories, etc. Moreover, these reviews do
not consider later results dealing with more general
classes of $p$-brane solutions
(see for references below). It should be stressed that
one of the aims of our review is to improve
the situations with citations in this area.  Here we try to overview
more general families of solutions with composite non-localized
electro-magnetic
$p$-branes, when the block-diagonal metrics on product manifolds
are considered. The main part of the solutions under consideration
deals with Ricci-flat internal spaces of arbitrary dimensions and
signatures (though certain solutions with Einstein internal
spaces are also considered).

From a mathematical point of view, here we  concentrate mainly on the
following key items:
$(a)$ sigma-model representation;
$(b)$ Toda-like Lagrangians and Toda chains;
$(c)$ general intersection rules related
to Cartan matrices in integrable cases.

Our approach cover more or less uniformly such topics as
solutions with harmonic functions, classical and quantum cosmological
solutions, spherically symmetric and black hole configurations, etc.

Meanwhile, some topics are out of consideration. Among them,
there are localized branes, branes at angles,
stability,  "brane-world" models based on $p$-branes, global
properties of solutions (see \cite{St} and refs.  therein),
black holes and branes in string theory \cite{Y} and microscopic origin
of the Bekenstein-Hawking entropy \cite{SVa,CaMa,Sken}
etc. The inclusion
of these topics may be a subject of a future more extensive review.


\newpage

\section{\bf The model}
\setcounter{equation}{0}

\subsection{The action and equations of motion}

We consider the model governed by an action
\bear{2.1}
S =&& \frac{1}{2\kappa^{2}}
\int_{M} d^{D}z \sqrt{|g|} \{ {R}[g] - 2 \Lambda - h_{\alpha\beta}
g^{MN} \partial_{M} \varphi^\alpha \partial_{N} \varphi^\beta
\\ \nn
&& - \sum_{a \in \Delta}
\frac{\theta_a}{n_a!} \exp[ 2 \lambda_{a} (\varphi) ] (F^a)^2_g \}
+ S_{GH},
\ear
where $g = g_{MN} dz^{M} \otimes dz^{N}$ is the metric
on the manifold $M$, ${\dim M} = D$, $\varphi=(\varphi^\alpha)\in \R^l$
is a vector from dilatonic scalar fields,
$(h_{\alpha\beta})$ is a non-degenerate symmetric
$l\times l$ matrix ($l\in \N$),
$\theta_a  \neq 0$,
$$
F^a =  dA^a
=\frac{1}{n_a!} F^a_{M_1 \ldots M_{n_a}}
dz^{M_1} \wedge \ldots \wedge dz^{M_{n_a}}
$$
is a $n_a$-form ($n_a \geq 2$) on a $D$-dimensional manifold $M$,
$\Lambda$ is a cosmological constant
and $\lambda_{a}$ is a $1$-form on $\R^l$ :
$\lambda_{a} (\varphi) =\lambda_{a \alpha}\varphi^\alpha$,
$a \in \Delta$, $\alpha=1,\ldots,l$.
In (\ref{2.1})
we denote $|g| = |\det (g_{MN})|$,
$(F^a)^2_g =
F^a_{M_1 \ldots M_{n_a}} F^a_{N_1 \ldots N_{n_a}}
g^{M_1 N_1} \ldots g^{M_{n_a} N_{n_a}},
$
$a \in \Delta$, where $\Delta$ is some finite set, and $S_{\rm GH}$ is the
standard Gibbons-Hawking boundary term \cite{GH}. In the models
with one time all $\theta_a =  1$  when the signature of the metric
is $(-1,+1, \ldots, +1)$.

The equations of motion
corresponding to  (\ref{2.1}) have the following
form
\bear{2.4}
R_{MN} - \frac{1}{2} g_{MN} R  =   T_{MN} - \Lambda g_{MN},
\\
\label{2.5}
{\btu}[g] \varphi^\alpha -
\sum_{a \in \Delta} \theta_a  \frac{\lambda^{\alpha}_a}{n_a!}
e^{2 \lambda_{a}(\varphi)} (F^a)^2_g = 0,
\\
\label{2.6}
\nabla_{M_1}[g] (e^{2 \lambda_{a}(\varphi)}
F^{a, M_1 \ldots M_{n_a}})  =  0,
\ear
$a \in \Delta$; $\alpha=1,\ldots,l$.
In (\ref{2.5}) $\lambda^{\alpha}_{a} = h^{\alpha \beta}
\lambda_{\beta a}$, where $(h^{\alpha \beta})$
is matrix inverse to $(h_{\alpha \beta})$.
In (\ref{2.4})
\bear{2.7}
T_{MN} =   T_{MN}[\varphi,g]
+ \sum_{a\in\Delta} \theta_a  e^{2 \lambda_{a}(\varphi)} T_{MN}[F^a,g],
\ear
is the stress-energy tensor where
\bear{2.8}
T_{MN}[\varphi,g] =
h_{\alpha\beta}\left(\p_{M} \varphi^\alpha \p_{N} \varphi^\beta -
\frac{1}{2} g_{MN} \p_{P} \varphi^\alpha \p^{P} \varphi^\beta\right),
\\
T_{MN}[F^a,g] = \frac{1}{n_{a}!}
\left[ - \frac{1}{2} g_{MN} (F^{a})^{2}_{g}
+ n_{a}  F^{a}_{M M_2 \ldots M_{n_a}} F_{N}^{a, M_2 \ldots M_{n_a}}
 \right].
\label{2.9}
\ear
In (\ref{2.5}) and  (\ref{2.6}) operators ${\btu}[g]$ and ${\btd}[g]$
are Laplace-Beltrami and covariant derivative operators, respectively,
corresponding to  $g$.

\subsection{Ansatz for composite  p-branes }

Let us consider the manifold
\beq{2.10}
M = M_{0}  \times M_{1} \times \ldots \times M_{n},
\eeq
with the metric
\beq{2.11}
g= e^{2{\gamma}(x)} \hat{g}^0  +
\sum_{i=1}^{n} e^{2\phi^i(x)} \hat{g}^i ,
\eeq
where $g^0  = g^0 _{\mu \nu}(x) dx^{\mu} \otimes dx^{\nu}$
is an arbitrary metric with any signature on the manifold $M_{0}$
and $g^i  = g^{i}_{m_{i} n_{i}}(y_i) dy_i^{m_{i}} \otimes dy_i^{n_{i}}$
is a metric on $M_{i}$  satisfying the equation
\beq{2.13}
R_{m_{i}n_{i}}[g^i ] = \xi_{i} g^i_{m_{i}n_{i}},
\eeq
$m_{i},n_{i}=1, \ldots, d_{i}$; $\xi_{i}= {\rm const}$,
$i=1,\ldots,n$. Here $\hat{g}^{i} = p_{i}^{*} g^{i}$ is the
pullback of the metric $g^{i}$  to the manifold  $M$ by the
canonical projection: $p_{i} : M \rightarrow  M_{i}$,
$i = 0,\ldots, n$. Thus, $(M_i, g^i )$  are Einstein spaces,
$i = 1,\ldots, n$.
The functions $\gamma, \phi^{i} : M_0 \rightarrow \R $ are smooth.
We denote $d_{\nu} = {\rm dim} M_{\nu}$; $\nu = 0, \ldots, n$;
$D = \sum_{\nu = 0}^{n} d_{\nu}$.
We put any manifold $M_{\nu}$, $\nu = 0,\ldots, n$,
to be oriented and connected.
Then the volume $d_i$-form
\beq{2.14}
\tau_i  \equiv \sqrt{|g^i(y_i)|}
\ dy_i^{1} \wedge \ldots \wedge dy_i^{d_i},
\eeq
and signature parameter
\beq{2.15}
\varepsilon(i)  \equiv {\rm sign}( \det (g^i_{m_i n_i})) = \pm 1
\eeq
are correctly defined for all $i=1,\ldots,n$.

Let $\Omega = \Omega(n)$  be a set of all non-empty
subsets of $\{ 1, \ldots,n \}$.
The number of elements in $\Omega$ is $|\Omega| = 2^n - 1$.
For any $I = \{ i_1, \ldots, i_k \} \in \Omega$, $i_1 < \ldots < i_k$,
we denote
\bear{2.16}
\tau(I) \equiv \hat{\tau}_{i_1}  \wedge \ldots \wedge \hat{\tau}_{i_k},  \\
 \label{2.17}
\eps(I) \equiv \eps(i_1) \ldots \eps(i_k),  \\
\label{2.18}
M_{I} \equiv M_{i_1}  \times  \ldots \times M_{i_k}, \\
\label{2.19}
d(I) \equiv  \sum_{i \in I} d_i.
\ear
Here $\hat{\tau}_{i} = p_{i}^{*} \hat{\tau}_{i}$ is the
pullback of the form $\tau_i$  to the manifold  $M$ by the
canonical projection: $p_{i} : M \rightarrow  M_{i}$,
$i = 1,\ldots, n$. We also put $\tau(\emptyset)= \eps(\emptyset)=
1$ and $d(\emptyset)=0$.

In the Appendix 1 we outline for completeness all relations
for Riemann and  Ricci tensors corresponding to the metric (\ref{2.11}).

For fields of forms we consider the following composite electromagnetic
ansatz
\ber{2.1.1}
F^a=\sum_{I\in\Omega_{a,e}}{\cal F}^{(a,e,I)}+
\sum_{J\in\Omega_{a,m}}{\cal F}^{(a,m,J)}
\eer
where
\bear{2.1.2}
{\cal F}^{(a,e,I)}=d\Phi^{(a,e,I)}\wedge\tau(I), \\
\label{2.1.3}
{\cal F}^{(a,m,J)}= e^{-2\lambda_a(\varphi)}*(d\Phi^{(a,m,J)}
\wedge\tau(J))
\ear
are elementary forms of electric and magnetic types respectively,
$a\in\tri$, $I\in\Omega_{a,e}$, $J\in\Omega_{a,m}$ and
$\Omega_{a,v} \subset \Omega$, $v = e,m$. In (\ref{2.1.3})
$*=*[g]$ is the Hodge operator on $(M,g)$

$$
(* \omega)_{M_1...M_{D-k}} = \frac{|g|^{1/2}}{k!}
\eps_{M_1...M_{D-k} N_{1}...N_{k}} \omega^{N_{1}...N_{k}},
$$
where ${\rm rank} \omega = k$.
For scalar functions we put
\ber{2.1.5}
\varphi^\alpha=\varphi^\alpha(x), \quad
\Phi^s=\Phi^s(x),
\eer
$s\in S$. Thus $\varphi^{\alpha}$ and $\Phi^s$ are functions on $M_0$.

Here and below
\ber{2.1.6}
S=S_e \sqcup S_m, \quad
S_v=\sqcup_{a\in\tri}\{a\}\times\{v\}\times\Omega_{a,v},
\eer
$v=e,m$. Here and in what follows $\sqcup$ means the union
of non-intersecting sets.
The set $S$ consists of elements $s=(a_s,v_s,I_s)$,
where $a_s \in \tri$ is colour index, $v_s = e, m$ is electro-magnetic
index and set $I_s \in \Omega_{a_s,v_s}$ describes the location
of brane.

Due to (\ref{2.1.2}) and (\ref{2.1.3})
\ber{2.1.7}
d(I)=n_a-1, \quad d(J)=D-n_a-1,
\eer
for $I\in\Omega_{a,e}$ and $J\in\Omega_{a,m}$ (i.e. in electric
and magnetic case, respectively). The sum of worldvolume dimensions
for electric and magnetic branes corresponding to the same form is equal
to $D-2$, it does not depend upon the rank of the form.

\subsection{The sigma model}

Let $d_0 \neq 2$ and
\ber{2.2.1}
\gamma=\gamma_0(\phi) \equiv
\frac1{2-d_0}\sum_{j=1}^nd_j\phi^j,
\eer
i.e. the generalized harmonic gauge (frame) is used.
As we shall see below the equations of motions have a
rather simple form in this gauge. Moreover harmonic gauge (\ref{2.2.1})
preserve the harmonicity of coordinates: harmonic
coordinates on $(M_0,g^0)$ (i.e. obeying $\tri[g^0] x^{\mu}=0$)
have  harmonic pullbacks $\hat{x}^{\mu}$
on $(M,g)$ (i.e. $\tri[g] \hat{x}^{\mu} =0$) and
anologous statement is valid for harmonic coordinates
$y^{m_i}$ on $(M_i,g^i)$, $i = 1, \ldots, n$. It was shown
recently in \cite{Gibbion} that the choice of harmonic coordinates is
the most convenient coordinate choice for studying $p$-branes.

\subsubsection{Restrictions on $p$-brane configurations.}

Here we present two restrictions on the sets of $p$-branes
that guarantee  the block-diagonal form of the  energy-momentum tensor
and the existence of the sigma-model representation (without additional
constraints).

We denote
$w_1\equiv\{i|i\in\{1,\dots,n\},\quad d_i=1\}$, and
$n_1=|w_1|$ (i.e. $n_1$ is the number of 1-dimensional spaces among
$M_i$, $i=1,\dots,n$).

{\bf Restriction 1.} {\em For any $a\in\tri$ and $v= e,m$ there are no
$I,J \in\Omega_{a,v}$ such that
$ I= \{i\} \sqcup (I \cap J)$, and $J= (I \cap J) \sqcup \{ j \}$
for some $i,j \in w_1$, $i \neq j$.}

Let us define $\bar I$ as follows
\ber{2.2.5}
\bar I \equiv  I_0 \setminus I,  \qquad I_0 = \{1,\ldots,n\}.
\eer

{\bf Restriction 2} (only for $d_0=1,3$).
{\em For any $a\in\tri$ there are no
$I\in\Omega_{a,e}$ and $J\in\Omega_{a,m}$ such that
$\bar J=\{i\}\sqcup I$ for $d_0 = 1$,  and
$I=\{i\}\sqcup \bar J$  for $d_0 = 3$,
where $i \in w_1$.}

Restriction 1  is satisfied for $n_1 \leq 1$ and also in
the non-composite case: $|\Omega_{a,e}|+ |\Omega_{a,m}| = 1$ for all
$a\in\tri$.  For $n_1\ge2$ it forbids the following
pairs of two electric or two magnetic $p$-branes,
corresponding to the same form $F^a, a \in \tri$:

\begin{center}
\begin{tabular}{cccc}
\cline{1-2}
\multicolumn{1}{|c|}{$i$} &
\multicolumn{1}{|c|}{\hspace*{1cm}} & & $\quad I$ \\
\cline{1-2}
 & & & \\
\cline{2-3}
 & \multicolumn{1}{|c|}{\hspace*{1cm}} &
\multicolumn{1}{|c|}{$j$} & $\quad J$ \\
\cline{2-3}
\end{tabular}
\end{center}

\begin{center}

{\bf \small Figure 1.
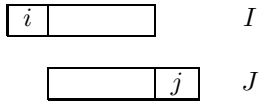
A forbidden by Restriction 1 pair of two electric or two
magnetic p-branes.
}

\end{center}

Here $d_i = d_j =1$, $i \neq j$, $i,j =1,\dots,n$. Restriction 1
may be also rewritten in terms of intersections
\beq{2.2.2a}
{\bf (R1)} \quad d(I \cap J) \leq d(I)  - 2,
\eeq
for any $I,J \in\Omega_{a,v}$, $a\in\tri$, $v= e,m$ (here $d(I) = d(J)$).

Restriction 2 is satisfied for $n_1=0$  or when $d_0 \neq 1,3$. For
$n_1\ge1$ it forbids the following electro-magnetic pairs,
corresponding to the same form $F^a, a \in \tri$:

\begin{center}
\begin{tabular}{ccc}
\cline{1-2}
\multicolumn{1}{|c|}{\hspace*{1cm}} &
\multicolumn{1}{|c|}{$i$}  & $\quad\bar J$ \\
\cline{1-2}
 & & \\
\cline{1-1}
\multicolumn{1}{|c|}{\hspace*{1cm}} & & $\quad I$ \\
\cline{1-1}
\end{tabular} \hspace{2cm}
\begin{tabular}{ccc}
\cline{1-2}
\multicolumn{1}{|c|}{\hspace*{1cm}} &
\multicolumn{1}{|c|}{$i$}  & $\quad I$ \\
\cline{1-2}
 & & \\
\cline{1-1}
\multicolumn{1}{|c|}{\hspace*{1cm}} & & $\quad\bar J$ \\
\cline{1-1}
\end{tabular}
\end{center}

\begin{center}

{\bf \small Figure 2.
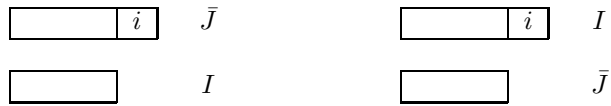
Forbidden by Restriction 2 electromagnetic pairs of
p-branes for $d_0 =1$ and $d_0 =3$, respectively.
}

\end{center}

for $d_0=1$ and $d_0=3$ respectively.
Here $d_i =1$, $i =1,\dots,n$. In terms of intersections
Restriction 2 reads
\beq{2.2.3a}
{\bf (R2)} \quad d(I \cap J) \neq 0 \ for \ d_0 = 1, \qquad
d(I \cap J) \neq 1  \quad for \ d_0 = 3
\eeq
(see (\ref{2.1.7})).

It should be noted that possible non-diagonality
of stress-energy tensor $T^M_N$ for composite $p$-branes usually is
not discussed in publications devoted to $p$-brane solutions.
To our knowledge this topic was considered first in \cite{AR}
and (in detailed) in \cite{IMC}.

\subsubsection{Sigma-model action for harmonic gauge}

It was proved in \cite{IMC} that equations of motion for the model
(\ref{2.1}) and the Bianchi identities:
\ber{2.2.6}
d{\cal F}^s=0,
\eer
$s \in S_m$, for fields from (\ref{2.11}),
(\ref{2.1.1})--(\ref{2.1.5}), when Restrictions 1 and 2
are  imposed, are equivalent to equations of motion for the $\sigma$-model
governed by the action
\bear{2.2.7}
S_{\sigma 0} = && \frac{1}{2 \kappa_0^2}
\int d^{d_0}x\sqrt{|g^0|}\biggl\{R[g^0]-\hat G_{AB}
g^{0\mu\nu}\p_\mu\sigma^A\p_\nu\sigma^B  \\ \nn
&&-\sum_{s\in S}\eps_s \exp{(-2U_A^s\sigma^A)}
g^{0\mu\nu} \p_\mu\Phi^s\p_\nu\Phi^s - 2V \biggr\},
\ear
where $(\sigma^A)=(\phi^i,\varphi^\alpha)$, $k_0 \neq 0$, the index set
$S$ is defined in (\ref{2.1.6}),
\beq{2.2.8}
 V = {V}(\phi)
= \Lambda e^{2 {\gamma_0}(\phi)}
-\frac{1}{2}   \sum_{i =1}^{n} \xi_i d_i e^{-2 \phi^i
+ 2 {\gamma_0}(\phi)}
\eeq
is the potential,
\ber{2.2.9}
(\hat G_{AB})=\barr{cc}
G_{ij}& 0\\
0& h_{\alpha\beta}
\earr,
\eer
is the target space metric
with
\ber{2.2.10}
G_{ij}= d_i \delta_{ij}+\frac{d_i d_j}{d_0-2},
\eer
and co-vectors
\ber{2.2.11}
U_A^s =
U_A^s \sigma^A = \sum_{i \in I_s} d_i \phi^i - \chi_s \lambda_{a_s}(\varphi), \quad
(U_A^s) =  (d_i \delta_{iI_s}, -\chi_s \lambda_{a_s \alpha}),
\eer
$s=(a_s,v_s,I_s)$.
Here $\chi_e=+1$ and $\chi_m=-1$;
\ber{2.2.12}
\delta_{iI}=\sum_{j\in I}\delta_{ij}
\eer
is an indicator of $i$ belonging
to $I$: $\delta_{iI}=1$ for $i\in I$ and $\delta_{iI}=0$ otherwise; and
\bear{2.2.13}
\eps_s=(-\eps[g])^{(1-\chi_s)/2}\eps(I_s) \theta_{a_s},
\ear
$s\in S$, $\eps[g]\equiv\sign\det(g_{MN})$. More explicitly
(\ref{2.2.13}) reads
\beq{2.2.13a}
\eps_s=\eps(I_s) \theta_{a_s} \ {\rm for} \ v_s = e; \qquad
\eps_s = -\eps[g] \eps(I_s) \theta_{a_s}, \ {\rm for} \ v_s = m.
\eeq

For finite internal space volumes $V_i$ (e.g. compact $M_i$)
and electric $p$-branes  (i.e. all $\Omega_{a,m} = \emptyset$)
the action (\ref{2.2.39}) coincides with the action (\ref{2.1}) when
$\kappa^{2} = \kappa^{2}_0 \prod_{i=1}^{n} V_i$.
This may be readily verified using  the relations
from Appendices 1 and 2. In general electro-magnetic
case  relation (\ref{2.2.39}) can not be obtained from the action
(\ref{2.1}) by a straightforward
integration over internal compact spaces $M_i$,
since such integration will give wrong signs for kynetic terms
corresponding to magnetic scalars $\Phi^s$. A simple
explanation of this fact was given in \cite{IMC} (see Sect. 5
therein).  Thus in general case one should be careful in writing the
sigma-model representation (especially when dealing with magnetic and
composite branes). To our knowledge this (subtle) point is not widely
discussed in literature.

Equations of motion corresponding to the action (\ref{2.2.7})
with the potential (\ref{2.2.8}) have the following form
\bear{2.2.14}
R_{\mu\nu}[g^0]=\hat G_{AB}\p_\mu\sigma^A\p_\nu\sigma^B+
\sum_{s\in S}\eps_s
\exp{(-2U_A^s\sigma^A)}\p_\mu\Phi^s\p_\nu\Phi^s
+ \frac{2V}{d_0 -2} g^0_{\mu \nu}, \\
\label{2.2.15}
\hat G_{AB}\tri[g^0]\sigma^B +
\sum_{s\in S}\eps_sU_A^s \exp{(-2U_C^s\sigma^C)}
g^{0\mu\nu} \p_\mu\Phi^s\p_\nu\Phi^s = \frac{\p V}{\p \sigma^A}, \\
\label{2.2.16}
\p_\mu\left(\sqrt{|g^0|}g^{0\mu\nu}\exp{(-2U_A^s\sigma^A)}
\p_\nu\Phi^s\right)=0,
\ear
$s\in S$. Here  $\tri[g^0]$ is the Laplace-Beltrami operator
corresponding to $g^0$.

{\bf Sigma-model  with constraints.}
In \cite{IMC} a general proposition concerning the sigma-model
representation when the {\bf Restrictions 1} and {\bf 2} are removed
is presented.  In this case the stress-energy tensor $T_{MN}$
is not identically block-diagonal  as it takes place for $R_{MN}$ and due
to equations of motion the off-block-diagonal components of $T_{MN}$
should be zero, hence, several additional constraints (or
restrictions) on the field configurations appear \cite{IMC} .

\subsubsection{General conformal gauges and $d_0 = 2$ case.}

We may also fix the gauge
$\gamma = {\gamma}(\phi)$  (where ${\gamma}(\phi)$ is a smooth function)
by arbitrary manner or do not fix it. In this case the
 action (\ref{2.2.7}) is simply replaced by the action
\bear{2.2.39}
S_{\sigma} =  \frac{1}{2 \kappa^{2}_0}
     \int_{M_0} d^{d_0}x   \sq {g^{0}}
     e^{f(\gamma, \phi)} \biggl\{ {R}[g^0 ]
- \sum_{i =1}^{n} d_i (\p \phi^i)^2 -  (d_0 - 2) (\p \gamma)^2
  \\ \nn
+ (\p f) \p (f + 2\gamma) - 2 \Lambda e^{2 \gamma}
+ \sum_{i =1}^{n} \xi_i d_i e^{-2 \phi^i +2\gamma}
-\sum_{s\in S}\eps_s \exp{(-2U_A^s\sigma^A)}
(\p \Phi^s)^2  \biggr\},
\ear
where
\beq{2.2.40}
f = {f}(\gamma, \phi)  = (d_0 - 2) \gamma +\sum_{j=1}^{n} d_j  \phi^j .
\eeq
Here
$\p f_1 \p f_2 = g^{0 \mu\nu} \p_\mu f_1 \p_\nu f_2$ and
$\p f \p f = (\p f)^2$.

Now let us consider the case $d_0=2$. In this case the sigma-model
representation   holds if the {\bf Restriction $2$} is replaced by the
following restriction \cite{IMC}.

{\bf Restriction $2^{*}$} ($d_0= 2$).
{\em For any $a\in\tri$ there are no  $I \in \Omega_{a,e}$,
$J\in\Omega_{a,m}$ such that
$ \bar I= J $ or $\bar J= \{i\} \sqcup (\bar J \cap
I)$, and $I =( \bar J \cap I) \sqcup \{ j \}$
for some $i,j \in w_1$, $i \neq j$.}

\newpage

\section{Solutions governed by harmonic functions}

\subsection{Solutions with orthogonal and block-orthogonal
$U^s$ and Ricci-flat $(M_{\nu}, g^{\nu})$.}

Here we consider a special class of solutions to equations of motion
governed by several harmonic functions when all factor spaces are
Ricci-flat and cosmological constant is zero, i.e.
$\xi_i = \Lambda = 0$,
$i = 1,\ldots,n$.
In certain situations these solutions describe extremal $p$-brane black holes
charged by fields of forms.

The solutions crucially depend upon  scalar products
of $U^s$-vectors $(U^s,U^{s'})$; $s,s' \in S$, where
\ber{3.1.1}
(U,U')=\hat G^{AB} U_A U'_B,
\eer
for $U = (U_A), U' = (U'_A) \in \R^N$, $N = n + l$ and
\beq{3.1.2}
(\hat G^{AB})=\left(\begin{array}{cc}
G^{ij}&0\\
0&h^{\alpha\beta}
\end{array}\right)
\eeq
is matrix inverse to  the matrix
(\ref{2.2.9}). Here (as in \cite{IMZ})
\beq{3.1.3}
G^{ij}=\frac{\delta^{ij}}{d_i}+\frac1{2-D},
\eeq
$i,j=1,\dots,n$.

The scalar products (\ref{3.1.1}) for vectors
$U^s$  were calculated in
\cite{IMC}
\ber{3.1.4}
(U^s,U^{s'})=d(I_s\cap I_{s'})+\frac{d(I_s)d(I_{s'})}{2-D}+
\chi_s\chi_{s'}\lambda_{a_s \alpha} \lambda_{a_{s'} \beta} h^{\alpha \beta},
\eer
where $(h^{\alpha\beta})=(h_{\alpha\beta})^{-1}$; and  $s=(a_s,v_s,I_s)$,
$s'=(a_{s'},v_{s'},I_{s'})$ belong to $S$. This relatio`n is a
very important one since it encodes  $p$-brane data
(e.g. intersections) in
scalar products of $U$-vectors. This relation simplify calculations,
clarify the algebraic structure of the equations of motion
and make transparent a reduction to Toda-like systems.

Let
\beq{3.1.5}
S=S_1\sqcup\dots\sqcup S_k,
\eeq
$S_i\ne\emptyset$, $i=1,\dots,k$, and
\beq{3.1.6}
(U^s,U^{s'})=0
\eeq
for all $s\in S_i$, $s'\in S_j$, $i\ne j$; $i,j=1,\dots,k$. Relation
(\ref{3.1.5}) means that the set $S$ is a union of $k$ non-intersecting
(non-empty) subsets $S_1,\dots,S_k$. According to (\ref{3.1.6}) the set of
vectors $(U^s, s \in S)$ has a block-orthogonal structure with respect to
the scalar product (\ref{3.1.1}), i.e. it  splits into $k$ mutually
orthogonal blocks $(U^s, s \in S_i)$, $i=1,\dots,k$.

Here we consider exact solutions in the
model (\ref{2.1}), when vectors $(U^s,s\in S)$ obey the block-orthogonal
decomposition (\ref{3.1.5}), (\ref{3.1.6}) with scalar products defined in
(\ref{3.1.4}) \cite{IMBl}. These solutions may be obtained from the
corresponding solutions to the $\sigma$-model equations
(\ref{2.2.14})-(\ref{2.2.16}).

{\bf Proposition  1 \cite{IMBl}.} {\em Let $(M_0,g^0)$ be Ricci-flat:
$R_{\mu\nu}[g^0]=0$.
Then the field configuration
\ber{3.1.7}
g^0, \qquad \sigma^A=\sum_{s\in S}\eps_sU^{sA}\nu_s^2\ln H_s, \qquad
\Phi^s=\frac{\nu_s}{H_s},
\eer
$s\in S$, satisfies to field equations (\ref{2.2.14})--(\ref{2.2.16}) with
$V=0$ if (real) numbers $\nu_s$ obey the relations
\ber{3.1.8}
\sum_{s'\in S}(U^s,U^{s'})\eps_{s'}\nu_{s'}^2=-1
\eer
$s\in S$, functions $H_s >0$ are harmonic, i.e.
$\tri[g^0]H_s=0$,
$s\in S$ and $H_s$ are coinciding inside blocks:
$H_s=H_{s'}$
for $s,s'\in S_i$, $i=1,\dots,k$.}

The Proposition 1 can be readily verified by a straightforward
substitution of (\ref{3.1.7})--(\ref{3.1.8}) into equations of motion
(\ref{2.2.14})--(\ref{2.2.16}). In the special (orthogonal) case, when any
block contains only one vector (i.e. all $|S_i|=1$) the Proposition 1
coincides with Proposition 1 of \cite{IMC}. In general case vectors
inside each
block $S_i$ are not orthogonal. The solution under consideration
depends on $k$ independent harmonic functions. For a given set of
vectors $(U^s,s\in S)$ the maximal number $k$ arises for the irreducible
block-orthogonal decomposition (\ref{3.1.5}), (\ref{3.1.6}), when any block
$(U^s,s\in S_i)$ does not split into two mutually-orthogonal
subblocks.

Using the sigma-model solution from Proposition 1
and relations for contravariant components \cite{IMC}:
\ber{2.2.38}
U^{si}=\delta_{iI_s}-\frac{d(I_s)}{D-2}, \quad
U^{s\alpha}=-\chi_s\lambda_{a_s}^\alpha,
\eer
$s=(a_s,v_s,I_s)$,  we get \cite{IMBl}:
\bear{3.1.11}
g= \left(\prod_{s\in S}H_s^{2d(I_s)\eps_s\nu_s^2}\right)^{1/(2-D)}
\left\{\hat{g}^0+ \sum_{i=1}^n
\left(\prod_{s\in S}H_s^{2\eps_s\nu_s^2\delta_{iI_s}}\right)
\hat{g}^i\right\}, \\
\label{3.1.14}
\varphi^\alpha=-\sum_{s\in S}\lambda_{a_s}^\alpha\chi_s
\eps_s\nu_s^2\ln H_s, \\
\label{3.1.15}
F^a=\sum_{s\in S}{\cal F}^s\delta_{a_s}^a,
\ear
where $i=1,\dots,n$, $\alpha=1,\dots,l$, $a \in \Delta$ and
\ber{3.1.16}
{\cal F}^s=\nu_sdH_s^{-1}\wedge\tau(I_s), \mbox{ for } v_s=e, \mm
\label{3.1.17}
{\cal F}^s=\nu_s(*_0dH_s)\wedge\tau(\bar I_s), \mbox{ for } v_s=m,
\eer
$H_s$ are harmonic functions on $(M_0,g^0)$ coinciding inside blocks
(i.e. $H_s=H_{s'}$ for $s,s'\in S_i$, $i=1,\dots,k$)
and relations  (\ref{3.1.8}) on parameters $\nu_s$ are imposed. Here
the matrix $((U^s,U^{s'}))$ and parameters $\eps_s$, $s\in S$,
are defined in (\ref{3.1.4}) and
(\ref{2.2.13}), respectively;
$\lambda_a^\alpha=  h^{\alpha\beta}\lambda_{\beta a}$,
 $*_0=*[g^0]$ is the Hodge operator  on $(M_0,g^0)$ and $\bar I$  is defined  (\ref{2.2.5}).
In (\ref{3.1.17}) we redefined the sign of $\nu_{s}$-parameter
(compared to (\ref{2.1.3}))  as following:
$\nu_{s} \mapsto  -\eps(I)\mu(I)\nu_s$.

Relation  (\ref{3.1.17})  may be obtained from (\ref{2.1.3})
by use  of the following identity
\beq{3.1.22}
\exp(-2 U^s_A \sigma^A)= H_s^2,
\eeq
$s\in S$, following from Proposition 1.
For certain models the latter
appears as so-called "no-force" condition of the vanishing of the static
fource on a $p$-brane probe in the background of another  $p$-brane
\cite{DKL,Dab,DS,CaKh,Ts1A}.

{\bf Remark 1}. The solution (\ref{3.1.11})-(\ref{3.1.17})
is also valid for $d_0=2$, if Restriction 2 from previous section
is replaced by Restriction $2^{*}$.  It may be verified using the
sigma-model representation (\ref{2.2.39}).

\subsubsection{Solutions with orthogonal $U^s$ }

Let us consider the orthogonal case \cite{IMC}
\beq{3.1.23}
(U^s,U^{s'})=0, \qquad  s \neq s',
\eeq
$s, s' \in S$. Then relation (\ref{3.1.8}) reads as follows
\beq{3.1.24}
(U^s,U^{s}) \eps_{s} \nu_{s}^2=-1,
\eeq
$s \in S$. This implies $(U^s,U^{s}) \neq 0$ and
\beq{3.1.25}
\eps_{s} (U^s,U^{s}) < 0,
\eeq
for all $s \in S$.
For $d(I_s) < D - 2$ and
$\lambda_{a_s \alpha} \lambda_{a_{s} \beta} h^{\alpha \beta} \geq 0$ we get
from (\ref{3.1.4})
$(U^s,U^{s}) > 0$,
and, hence, $\eps_{s} < 0$,
$s \in S$. If $\theta_a > 0$ for all $a \in \Delta$,
then
\beq{3.1.28}
\eps(I_s) = -1 \ {\rm for} \ v_s = e; \qquad
\eps(I_s) = \eps[g] \ {\rm for} \ v_s = m.
\eeq
For pseudo-Euclidean metric $g$ all $\eps(I_s) = -1$ and, hence,
all $p$-branes should contain time manifold. For the  metric $g$ with
the Euclidean signature only magnetic $p$-branes can exist in this case.

From  scalar products (\ref{3.1.4}) and the
orthogonality condition (\ref{3.1.23}) we get the "orthogonal"
intersection rules \cite{AR,AEH,IMC}
\ber{3.1.30}
d(I_s\cap I_{s'}) =
\frac{d(I_s)d(I_{s'})}{D - 2} - \chi_s\chi_{s'}\lambda_{a_s \alpha}
\lambda_{a_{s'} \beta} h^{\alpha \beta} \equiv
\Delta(s,s'),
\eer
for $s=(a_s,v_s,I_s) \neq s'=(a_{s'},v_{s'},I_{s'})$.
(For pure electric case see also \cite{IM11,IM12,IMR}.)

{\bf Example 1: D = 11 supergravity \cite{CJS}.}
The action for the bosonic sector of  D = 11 supergravity
with omitted  Chern-Simons term has the  following form
\beq{3.1.31}
\hat{S}_{11} =
\int_{M} d^{11}z \sqrt{|g|} \{ {R}[g] -  \frac{1}{4!}  F^2 \}.
\eeq
Here  ${\rm rank} F = 4$ and the signature of $g$ is $(-,+, \ldots, +)$.

The dimensions of p-brane worldvolumes are (see (\ref{2.1.7}))
\bear{3.1.32}
d(I_s) = &&3,  \ {\rm for} \    v_s = e,   \\ \nn
         &&6,  \ {\rm for} \   v_s = m.
\ear
The model describes  electrically charged $2$-branes and magnetically charged  $5$-branes.

From (\ref{3.1.30}) we obtain the intersection rules \cite{Ts1}
\beq{3.1.34}
\begin{array}{rll}\displaystyle
d(I_s \cap I_{s'}) = &1,  \ {\rm for} \    v_s = v_{s'} = e;   \\
                     &2,   \ {\rm for} \   v_s =e, \  v_{s'} = m; \\
                     &4,  \ {\rm for} \    v_s = v_{s'} = m.                ,
\end{array}
\eeq
The {\bf Restrictions 1} and {\bf 2} from Section 2 are
satisfied in this case
(see also (\ref{2.2.2a}) and (\ref{2.2.3a}). Here
 using the relation for the bosonic part of action
for $D = 11$ supergravity \cite{CJS}
\beq{3.1.40}
S_{11} =  \hat{S}_{11} + c_{11} \int_{M} A \wedge F  \wedge F
\eeq
where $c_{11} = {\rm const}$ and $\hat{S}_{11}$ is defined in \rf{3.1.31}
($F  =d A$). Indeed, the only modification
of equations of motion is  related to "Maxwell" equation
\ber{3.1.41}
d*F = {\rm const} \ F \wedge F,
\eer
with  $F \wedge F = 0$ in the solutions under consideration.
These solutions  \cite{IM11} coincide with those obtained in
\cite{BREJS,Ts1}) for
flat $(M_{\nu}, g^{\nu})$, $\nu = 0, \ldots, n$.

\subsubsection{Solutions related to Lie algebras}

Now we study the  solutions (\ref{3.1.11})-(\ref{3.1.17})
in more detail and show that some of them
may be related to different Lie algebras.
Here we put
\ber{3.1.2.1}
K_s \equiv (U^s,U^s)\neq 0,
\eer
for all $s\in S$ and introduce the quasi-Cartan matrix $A=(A_{ss'})$:
\ber{3.1.2.2}
A_{ss'} \equiv \frac{2(U^s,U^{s'})}{(U^{s'},U^{s'})},
\eer
$s,s'\in S$. Here  some ordering in $S$ is assumed.

Using this definition and (\ref{3.1.4}) we obtain the intersection rules
\ber{3.1.2.3}
d(I_{s}\cap I_{s'})=\Delta(s,s')+\frac12 K_{s'} A_{s s'}
\eer
$s \ne s'$, where  $\Delta(s,s')$ is defined in (\ref{3.1.30}).
These rules is a generalization of "orthogonal" intersection
rules to the case of

For $\det A \neq 0$ relation (\ref{3.1.8}) may be rewritten in the
equivalent form
\ber{3.1.2.5}
- \eps_s\nu_s^2(U^s,U^s)=2\sum_{s'\in S} A^{ss'} \equiv b_s,
\eer
$s\in S$, where $(A^{ss'})=A^{-1}$. Thus, eq. (\ref{3.1.8})
may be resolved in terms of $\nu_s$ for certain $\eps_s=\pm1$, $s\in S$.
We note that due to $(\ref{3.1.6})$ the matrix $A$ has a block-diagonal
structure  and, hence, for any $i$-th block the set of parameters $(\nu_s,
s \in S_i)$ depend upon the matrix inverse to the matrix  $(A_{s s'}; s,
s' \in S_i)$.

Now  we consider one-block case when
the $p$-brane intersections are related to some  Lie algebras.

{\bf 3.1.2.1. Finite dimensional Lie algebras \cite{GrI}.}

Let $A$ be a Cartan matrix of a simple
finite-dimensional Lie algebra. In this case $A_{ss'}\in\{0,-1,-2,-3\}$,
$s\ne s'$. The elements of inverse matrix $A^{-1}$ are positive
(see Ch. 7 in \cite{FS}) and hence we get from (\ref{3.1.2.5}) the same
signature relation (\ref{3.1.25}) as in orthogonal case.
Moreover, all $b_s$ are natural numbers:
\ber{3.1.2.6}
b_s = n_s \in \N,
\eer
$s\in S$.
Integers $n_s$ coincide with the components
of twice the so-called dual Weyl vector in the basis of simple coroots
(see Ch.3.1.7 in \cite{FS}).
Explicit formulas for $n_s$, corresponding to simple finite dimensional
Lie algebras  are outlined in Appendix 3.

Here we consider three examples of solutions in $D = 11$ and
$D = 10$ ($IIA$)
supergravities and so-called $B_D$-models  (see Example 4 below)
corresponding to ${\bf A_2}$-algebra with the Cartan matrix
\ber{3.1.2.7}
A=\barr{cc}
2& - 1\\
- 1& 2
\earr
\eer
and $n_1 =n_2 = 2$.

{\bf Example 2: ${\bf A_2}$-dyon in $D=11$ supergravity.}
For the $D=11$ supergravity with the bosonic part of the action
(\ref{3.1.40}) we get for ${\bf A_2}$-solutions with two branes
the intersections
\ber{3.1.2.19}
3\cap6=1, \quad 6 \cap 6=3.
\eer
Here and in what follows $(d_1 \cap d_2= d)\Leftrightarrow
(d(I)=d_1, d(J)= d_2, d(I\cap J) = d)$.
The electromagnetic dyon (i.e. the bound state of electric
and magnetic branes) with the intersection $3\cap6=1$ reads
\bear{3.1.2.16}
g=H^2 \hat{g}^0 - H^{-2} dt\otimes dt+ \hat{g}^1+\hat{g}^2, \\
\label{3.1.2.17}
F =\nu_1dH^{-1}\wedge dt\wedge\hat{\tau}_1+
\nu_2(*_0dH)\wedge\hat{\tau}_1,
\ear
where $H$ is a harmonic function on $(M_0,g^0)$, $d_0=3$,
$d_1=2$, $d_2= 5$, $\nu_1^2=\nu_2^2=1$,
and metrics $g^{\nu}$ have Euclidean signature.

For $g^0 = \sum_{\mu =1}^{3} dx^{\mu} \otimes dx^{\mu}$ and
\ber{3.1.2.17a}
H=1+\sum_{j=1}^{N} \frac{q_j}{|x- x_j|},
\eer
the  4-dimensional section of the metric (\ref{3.1.2.16}) coincides with the
metric of  Majumdar-Papapetrou solution \cite{MP} describing $N$ extremal
charged black holes with horizons at points $x_i$, and charges $q_i>0$,
$i=1,\ldots,N$. The solution (\ref{3.1.2.16})--(\ref{3.1.2.17}) with $H$ from
(\ref{3.1.2.17a}) describes $N$ extremal $p$-brane dyonic black holes.
Any dyon contains one electric "brane" and one magnetic "brane" with equal
charge densities.

{\bf Example 3: ${\bf A_2}$-dyon in $IIA$ supergravity.}  The bosonic part of action
for $D=10$ IIA supergravity reads
\ber{3.1.2.11}
S=\int d^{10}z\sqrt{|g|}\biggl\{R[g]-(\p\varphi)^2-\sum_{a=2}^4
e^{2\lambda_a\varphi}(F^a)^2\biggr\}-\frac12\int F^4 \wedge F^4\wedge A^2,
\eer
where $F^a=dA^{a-1}+\delta^a_4 A^1 \wedge F^3$ is an $a$-form,
$a=2,3,4$, and
$\lambda_3=-2\lambda_4$, $\lambda_2=3\lambda_4$,
$\lambda_4^2=  1/8$.
The dimensions of $p$-brane worldvolumes are
\ber{3.1.2.13}
d(I)=\left\{\begin{array}{ll}
1,2,3&\mbox{in electric case}, \\
7,6,5&\mbox{in magnetic case},
\end{array}\right.
\eer
for $a=2,3,4$, respectively.

We consider here the sector corresponding to $a=3,4$ describing electric
$p$-branes: fundamental string (FS), $D2$-brane and magnetic $p$-branes:
$NS5$- and $D4$-branes. We get $(U^s,U^s)=2$ for all $s$.
The solutions with ${\bf A_2}$ intersection
rules corresponding to relations
\beq{3.1.2.14}
2 \cap 6=3 \cap 6=3 \cap 5=1, \
6\cap5=6\cap6=3, \ 5\cap 5= 2
\eeq
are valid in the "truncated case" (without Chern-Simons term) and in
a general case (\ref{3.1.2.11}) as well.
Let us consider the solution describing the electromagnetic dyon with
one of intersections
$$2\cap6= 3\cap5 =1$$.
The solution is given by relations (\ref{3.1.2.16}) and (\ref{3.1.2.17})
with $F = F^a$ and
$H$ being harmonic function on $(M_0,g^0)$, $d_0=3$,
$\nu_1^2=\nu_2^2=1$, $d_1=a-2$, $d_2=8-a$, $a=3,4$, and metrics $g^{\nu}$
having Euclidean signatures.

{\bf Example 4: {\bf $A_2$-dyon} in $B_D$-models.}
Now we consider  examples of solutions  for  $B_D$-models with the action
\cite{IMJ}
\ber{3.1.2.09}
S_D=\int d^Dz\sqrt{|g|}\biggl\{R[g]+g^{MN}\p_M\vec\varphi\p_N\vec\varphi-
\sum_{a=4}^{D-7}\frac1{a!}\exp[2\vec\lambda_a\vec\varphi](F^a)^2\biggr\},
\eer
where $\vec\varphi=(\varphi^1,\dots,\varphi^l)\in\R^l$, $\vec\lambda_a=
(\lambda_{a1},\dots,\lambda_{al})\in\R^l$, $l=D-11$, $\rank F^a=a$,
$a=4,\dots,D-7$. Here vectors $\vec\lambda_a$ satisfy the relations
\ber{3.1.2.010}
\vec\lambda_a\vec\lambda_b=N(a,b)-\frac{(a-1)(b-1)}{D-2}, \\
\label{3.1.2.011}
N(a,b)=\min(a,b)-3,
\eer
$a,b=4,\dots,D-7$ and $\vec\lambda_{D-7}=-2\vec\lambda_4$.
For $D>11$ vectors $\vec\lambda_4,\dots,\vec\lambda_{D-8}$ are linearly
independent.

The model (\ref{3.1.2.09}) contains $l$ scalar fields with a negative kinetic
term (i.e. $h_{\alpha\beta}=-\delta_{\alpha\beta}$ in (\ref{2.1}))
coupled to $(l+1)$ forms. For $D=11$ ($l=0$) the model (\ref{3.1.2.09})
coincides with the truncated  bosonic sector
of $D=11$ supergravity. For $D=12$ $(l=1)$ (\ref{3.1.2.09}) coincides with
truncated $D=12$ model from \cite{KKLP} (see also \cite{IMC}).

For $p$-brane worldvolumes we have the following dimensions (see
(\ref{2.1.7}))
\ber{3.1.2.013}
d(I)=3,\dots,D-8, \quad I\in\Omega_{a,e}, \\
\label{3.1.2.014}
d(I)=D-5,\dots,6, \quad I\in\Omega_{a,m}.
\eer
Thus, there are $(l+1)$ electric and $(l+1)$ magnetic $p$-branes, $p=d(I)-1$.
In $B_D$-model all $K_s = 2$.

Let us consider $B_D$-model, $D\ge 11$. Let
$a\in\{4,\dots,D-7\}$, $g^3=-dt\otimes dt$, $d_1=a-2$, $d_2=D-2-a$,
$d_0=3$ and metrics $g^0,g^1,g^2$ are Ricci-flat. The
${\bf A_2}$-solution describing a dyon configuration with electric $d_1$-brane
and magnetic $d_2$-brane, corresponding to $F^a$-form and
intersecting in $1$-dimensional time manifold  reads
as given by relations  (\ref{3.1.2.16}), (\ref{3.1.2.17})
and $\vec\varphi=0$, where $H$ is the harmonic function on $(M_0,g^0)$
and $\nu_1^2 = \nu_2^2 =1$. The case $D=11$ was considered in Example 2.
For $D=12$ we have two possibilities:
a) $a=4$, $d_1=2$, $d_2=6$; b) $a=5$, $d_1=3$, $d_2=5$. The signature
restrictions on $g^1$ and $g^2$ are the following: $\eps_1 =+1$,
$\eps_2 = -\eps[g]$. They are satisfied when $g^0$ and $g^1$
are metrics of Euclidean signature.

{\bf Remark 3.} In Examples 2, 3 and 4 the  ${\bf A_2}$-dyon solutions
do not satisfy the {\bf Restriction 2}  (or, equivalently, (\ref{2.2.3a}))
that guarantees the vanishing of non-block-diagonal components
of stress-energy tensor, i.e. $T_{1_i \mu_0} = 0$, $(\mu =1,2,3$, $d_i = 1)$.
Nevertheless this vanishing  does take place (due to formulas for
non-diagonal components of stress-energy tensor in \cite{IMC}, see
also Appendix 2).

{\bf 3.1.2.2. Hyperbolic algebras}

Let $\det A <0$ and
\beq{3.1.2.20a}
 A_{s s'} = 0, -1, -2, \ldots
\eeq
$s \neq s'$. Among quasi-Cartan matrices there exists a large
subclass of Cartan matrices, corresponding to infinite-dimensional
simple hyperbolic generalized  Kac-Moody (KM) algebras of ranks
$r=2,\dots,10$ \cite{Kac,FS}.

For the  hyperbolic algebras the following relations are satisfied
\ber{3.1.2.20}
\eps_s(U^s,U^s) >0,
\eer
$s\in S_i$. This relation is valid, since
$A^{ss'} \leq 0$, $s,s' \in S$, for any hyperbolic algebra
\cite{NikP}.

For $(U^s,U^{s}) > 0$ we get
$\eps_{s} > 0$,
$s \in S$. If $\theta_{a_s} > 0$ for all $s \in S$,
then
\beq{3.1.28a}
\eps(I_s) = 1   \ {\rm for} \ v_s = e; \qquad
\eps(I_s) = - \eps[g] \ {\rm for} \ v_s = m.
\eeq
For pseudo-Euclidean metric $g$ all $\eps(I_s) = 1$ and, hence,
all $p$-branes are Euclidean or should contain
even number of time directions: $2,4, \ldots$. For $\eps[g] = 1$
only magnetic $p$-branes may be pseudo-Euclidean.

{\bf Example 5. ${\cal F}_3$ algebra \cite{IKM}.}
Now we consider an example of the
solution corresponding to the  hyperbolic KM algebra ${\cal F}_3$ with the
Cartan matrix
\bear{3.1.2.21}
A=\barr{ccc}
2 &-2 & 0 \\
-2 & 2  & -1 \\
0  & -1 & 2
\earr,
\ear
${\cal F}_3$ is an infinite dimensional Lie algebra generated by the
(Serre) relations \cite{Kac,FS}
\ber{3.1.2.22}
[h_i,h_j] =0, \quad [e_i,f_j] = \delta_{ij} h_j, \mm
\label{3.1.2.23}
[h_i,e_j] = A_{ij} e_j, \quad [h_i,f_j] = -A_{ij} f_j,
\mm
\label{3.1.2.24}
({\rm ad} e_i)^{1- A_{ij}}(e_j) = 0 \quad (i \neq j),
\mm
\label{3.1.2.25}
({\rm ad} f_i)^{1- A_{ij}}(f_j) = 0 \quad (i \neq j).
\eer
${\cal F}_3$ contains  ${\bf A_1^{(1)} }$ affine Kac-Moody subalgebra
(it corresponds to the Geroch group) and  ${\bf A_2}$  subalgebra.

There exists an example of the solution with the $A$-matrix
(\ref{3.1.2.21}) for $11$-dimensional model governed by the
action
\ber{3.1.2.28}
S= \int d^{11}z \sqrt{|g|} \biggl\{R[g] - \frac{1}{4!} (F^4)^2
- \frac{1}{4!} (F^{4*})^2 \biggr\},
\eer
where ${\rm rank } F^{4} = {\rm rank} F^{4*} = 4$.
Here $\Delta = \{ 4, 4* \}$.  We consider a configuration with
two  magnetic $5$-branes  corresponding to the form $F^4$ and
one  electric $2$-brane corresponding to the form  $F^{4*}$.
We denote $S = \{s_1,s_2,s_3 \}$,
$a_{s_1} = a_{s_3} = 4$, $a_{s_2} = 4*$ and
$v_{s_1} = v_{s_3} = m$, $v_{s_2} = e$, where
$d(I_{s_1}) = d(I_{s_3}) = 6$ and $d(I_{s_2}) = 3$.
The intersection rules (\ref{3.1.2.3}) read
\ber{3.1.2.29}
d(I_{s_1} \cap I_{s_2}) = 0, \quad
d(I_{s_2} \cap I_{s_3}) = 1, \quad
d(I_{s_1} \cap I_{s_3}) = 4.
\eer

For the manifold (\ref{2.10}) we put
$n= 5$ and $d_1 =2$, $d_2 =4$, $d_3 = d_4 =1$, $d_5 = 2$.
The corresponding sets for $p$-branes are the following:
$I_{s_1} = \{1,2 \}$, $I_{s_2} = \{4,5 \}$, $I_{s_3} = \{2,3,4 \}$.

The corresponding solution reads
\bear{3.1.2.30}
g=H^{-12} \left\{ - dt \otimes dt + H^9 \hat{g}^1 + H^{13} \hat{g}^2
+ H^4 \hat{g}^3  + H^{14} \hat{g}^4  + H^{10} \hat{g}^5  \right\}, \mm
\label{3.1.2.31a}
F^4= \frac{dH}{dt} \left\{
\nu_{s_1} \hat{\tau}_3 \wedge \tau_4 \wedge \hat{\tau}_5 +
\nu_{s_3} \hat{\tau}_1 \wedge \hat{\tau}_5 \right\}, \\  \label{3.1.2.31b}
F^{4*} = \frac{dH}{dt} \frac{\nu_{s_2}}{H^2} dt \wedge
\hat{\tau}_4 \wedge \hat{\tau}_5,
\ear
where  $\nu_{s_1}^2  = \frac{9}{2}$, $\nu_{s_2}^2  = 5$ and
$\nu_{s_3}^2 = 2$ (see (\ref{3.1.2.5})).

All metrics  $g^i$ are Ricci-flat ($i = 1, \ldots, 5$) with the Euclidean
signature (this agrees with relations (\ref{3.1.2.20}) and  (\ref{2.2.13})),
and  $H = ht + h_0 > 0$,
where $h, h_0$ are constants. The metric (\ref{3.1.2.30}) may be also rewritten using
the synchronous time variable $t_s$
\ber{3.1.2.30c}
g= - dt_s \otimes dt_s + f^{3/5} \hat{g}^1 + f^{-1/5} \hat{g}^2
+ f^{8/5} \hat{g}^3  + f^{-2/5} \hat{g}^4  + f^{2/5} \hat{g}^5,
\eer
where $f = 5h t_s = H^{-5} > 0$, $h > 0$ and $t_s > 0$.
The metric describes the power-law "inflation" in $D =11$. It is singular
for $t_s \to +0$. The powers  in scale-factors  $f^{2 \alpha_i}$
do not satisfy Kasner-like relations:
$\sum_{i=1}^{5} d_i \alpha_i  = \sum_{i=1}^{5} d_i (\alpha_i)^2 = 1$.
For flat $g^i$ the calculation of the Riemann tensor squared gives
us (see Appendix 1)
\ber{3.1.2.31d}
{\cal K}[g]  = 2,1428 t_s^{-4},
\eer
where
\ber{3.1.2.31e}
{\cal K}[g] \equiv R_{MNPQ}[g]R^{MNPQ}[g]
\eer
is also called  the Kretschmann scalar.

{\bf Example 6: $H_2(q,q)$ algebra.} Let
\ber{3.1.2.32}
A=\barr{cc}
2& -q_1\\
-q_2& 2
\earr, \quad q_1q_2>4,
\eer
$q_1,q_2\in \N$. This is the Cartan matrix for the hyperbolic KM algebra
$H_{2}(q_1,q_2)$ \cite{Kac}. Let us consider $B_D$-model
An example of the solution for $B_D$-model  with
two electric $p$-branes  ($p=d_1,d_2$),
corresponding to $F^a$ and $F^b$ fields and intersecting in time manifold,
is the following:
\ber{3.1.2.33}
g=H^{-2/(q-2)}\hat{g}^0-H^{2/(q-2)}dt\otimes dt+\hat{g}^1+\hat{g}^2, \mm
\label{3.1.2.34}
F=\nu_{1}dH^{-1}\wedge dt\wedge \hat{\tau}_1+
\nu_{2}dH^{-1}\wedge dt\wedge \hat{\tau}_2, \mm
\label{3.1.2.35}
\vec\varphi=-(\vec\lambda_a+\vec\lambda_b)(q -2)^{-1}\ln H
\eer
where $d_0=3$, $d_1=a - 2$, $a=q+4$, $b\ge a$, $d_2=b-2$, $d_0 = 3$, $D=a+b$. Here
$F=F^a+F^b$ for $a<b$ and $F=F^a$ for $a=b$. The signature
restrictions are : $\eps_1= \eps_2 = -1$. Thus, the space-time
$(M,g)$ should contain at least three time directions.
The minimal $D$ is 14. For $D=14$ we get $a=b=7$, $d_1=d_2=5$,
$q=3$. In this case $6\cap6=1$.

{\bf 3.1.2.3. Affine  Lie algebras.}

We note that  affine KM algebras
(with $\det A = 0$) do not appear in the solutions
(\ref{3.1.11})--(\ref{3.1.17}).
Indeed, any affine Cartan matrix satisfy the
relations
\beq{3.1.2.36}
\sum_{s' \in S} a_{s'} A_{s's}=0,
\eeq
with $a_s > 0$ called Coxeter labels \cite{FS}, $s \in S$.
This relation make impossible the existence of the solution to
eq. (\ref{3.1.8}).

Thus, affine Cartan matrices do not arise in our solutions and hence
some configurations are forbidden. Let us consider ${\bf A_1^{(1)}}$
affine KM algebra with the Cartan matrix
\ber{3.1.2.37}
A=\barr{cc}
2& -2\\
-2& 2
\earr.
\eer
For $D=11$ supergravity the intersections: $3 \cap 6=0$, $6 \cap 6=2$,
corresponding to the $A$-matrix (\ref{3.1.2.37}), are forbidden.

{\bf Remark 6.} In \cite{ETT} new solutions in the affine
case were obtained. These solutions contain as a
special case a solution in $D =11$ supergravity from \cite{GKT}
with the intersection $6 \cap 6=2$. The solutions from \cite{ETT}
use some modified  ansatz for fields of forms
(the ansatz for localized branes)
and  do not belong to scheme under consideration.
The solutions of this section in the special case of
$D= 10,11$ supergravities are also
different from the so-called non-marginal bound state
solutions, since the latter have non-trivial Chern-Simons
terms (see, for example, \cite{ILPT,RT}
and references therein), although the rules for binary intersections
may look similar.

\subsubsection{ Kretschmann scalar, horizon and generalized MP solutions}

Let $M_0=\R^{d_0}$, $d_0>2$ and $g^0=\delta_{\mu\nu}dx^\mu\otimes dx^\nu$.
For
\ber{3.1.3.28}
H_s=1+\sum_{b\in X_s}\frac{q_{sb}}{|x-b|^{d_0-2}},
\eer
where $X_s$ is finite non-empty subset $X_s\subset M_0$, $s\in S$, all
$q_{sb}>0$, and $X_s=X_{s'}$, $q_{sb} = q_{s'b}$ for $b \in X_s=X_{s'}$,
$s, s' \in S_j$, $j =1, \ldots, k$.  The harmonic functions (\ref{3.1.3.28}) are
defined in domain $M_0\setminus X$, $X=\bigcup_{s\in S}X_s$,
and  generate the solutions (\ref{3.1.11})--(\ref{3.1.17}).

Denote $S(b)\equiv\{s\in S|\quad b \in X_s\}$. We also put  $M_i=\R^{d_i}$,
$R[g^i]= {\cal K}[g^i]=0$, $i=1,\dots,n$ (see definition (\ref{3.1.2.31e})).
Then for the metric   (\ref{3.1.11}) we obtain
\beq{3.1.3.29}
{\cal K}[g] = \frac{C' + o(1)}{U^2|x-b|^4}
= [C + o(1)]|x-b|^{4(d_0-2)\eta(b)}
\eeq
for $x\to b \in X$, where
\ber{3.1.3.30}
\eta(b)\equiv\sum_{s\in S(b)}(-\eps_s)\nu_s^2 \frac{d(I_s)}{D-2}-
\frac{1}{d_0-2},
\eer
and $C = C(b) \geq 0$ is  given in Appendix 1 (see (\ref{A.15.1})).
In what follows we consider
non-exceptional $b \in X$ defined by relations $C = C(b) > 0$.

{\bf Remark 7.}
It follows from relation (\ref{A.15.1})
of Appendix 1 that an exceptional point $b \in X$,
defined by relation $C = C(b) = 0$, if and only if
\ber{3.1.3.30a}
U(x) \sim c |x-b|^{ -2 \alpha}, \quad   U(x) U_i(x)  \sim c_i,
\eer
for $x\to b$, where  $\alpha = 0,2$ and $c, c_i \neq 0$
are constants, $i=1,\dots, n$.

Due to (\ref{3.1.3.29}) the metric (\ref{3.1.11}) has no
curvature singularity when $x\to b \in X$, $C(b) > 0$, if and only if
\ber{3.1.3.31}
\eta(b)\ge 0.
\eer

From (\ref{3.1.3.30}) we see that the metric (\ref{3.1.11})
is regular at a ``point'' $b\in X$
for $\eps_s=-1$ and large enough values of $\nu_s^2$, $s\in S(b)$. For
$\eps_s=+1$, $s\in S(b)$, we have a curvature singularity at
non-exceptional point $b\in X$.

Now we consider a special case: $d_1=1$, $g^1=-dt\otimes dt$.
In this case we have a horizon w.r.t. time $t$,
when $x\to b\in X$, if and only if
\ber{3.1.3.32}
\xi_1(b)\equiv \sum_{s\in S(b)}(-\eps_s) \nu_s^2 \delta_{1I_s}
-\frac1{d_0-2}\ge0.
\eer
This relation follows from the requirement of infinite time
propagation of light to $b \in X$.
If $\eps_s= -1$, $1 \in I_s$ for all $s\in S(b)$, we get
\ber{3.1.3.33}
\eta(b)<\xi_1(b),
\eer
$b\in X$. This follows from the inequalities $d(I_s)<D-2$ ($d_0>2$).

We note that $g_{tt} \to 0$
for $x \to b \in X$, if (\ref{3.1.3.33}) is satisfied.
This follows from the relation
\ber{3.1.3.33a}
g_{tt} \sim {\rm const} |x-b|^{2(d_0-2)(\xi_1(b)- \eta(b))},
\eer
$x \to b$.

{\bf Remark 8.}
Due to relations (\ref{3.1.3.30a}) and (\ref{3.1.3.33a})
the point $b \in X$ is non-exceptional if
$g^1=-dt\otimes dt$ and $1\in I_s$, $\eps_s= -1$ for
all $s \in S(b)$.

Thus, for the metric (\ref{3.1.11}) with $H_s$ from (\ref{3.1.3.28})
there are two dimensionless ators at the
non-exceptional point $b\in X$:
a) horizon indicator $\xi_1(b)$ (corresponding to time $t$)
and b) curvature singularity indicator
$\eta(b)$.  These indicators  define (for our assumptions)
the existence of a horizon
and the singularity of the Kretschmann scalar (when $(M_i,g^i)$ are flat,
$i=1, \ldots,n$) at non-exceptional $b \in X$.

{\bf Generalized MP solutions.}
Here we consider special black hole solutions
for the model  (\ref{2.1})   with  all $\theta_a =  1$, $a \in \tri$,
when the signature of the   metric  $g$ is $(-1,+1, \ldots, +1)$.
We put $\eps_s = \eps(I_s)= -1$ and $1 \in I_s$
for all $s \in S$ and   $M_0 = \R^{d_0}$, $g^0 =
\sum_{\mu=1}^{d_0} dx^{\mu} \otimes dx^{\mu}$,
$M_1 = \R$, $g^1 = -dt \otimes dt$. Then, the metric
(\ref{3.1.11}) reads \cite{IMBl}
\ber{3.1.3.34}
g= \Bigl(\prod_{s \in S} H_s^{2d(I_s)\nu_s^2} \Bigr)^{1/(D-2)}
\Bigl\{ \sum_{\mu=1}^{d_0} dx^{\mu} \otimes dx^{\mu} \\ \nn
- \Bigl(\prod_{s \in S} H_s^{-2 \nu_s^2 }\Bigr) dt \otimes dt
+ \sum_{i=2}^{n}  \Bigl( \prod_{s\in S}
H_s^{-2\nu_s^2\delta_{iI_s}}\Bigr) \hat{g}^i \Bigr\},
\eer
where $(M_i,g^i)$  are flat Euclidean spaces, $i = 2, \ldots, n$,
Here all branes have a common time submanifold $M_1 = \R$,
for all $a \in \tri$, and
\beq{3.1.3.38}
\eta(b) = \sum_{s\in S(b)}\nu_s^2\frac{d(I_s)}{D-2}-
\frac1{d_0-2} \geq 0,
\eeq
$b \in X$. This solution describes a set of extreme
$p$-brane black holes with horizons at $b \in X$.
The Riemann tensor squared has a finite limit
at any $b \in X$.

Calculation of the Hawking "temperature"
corresponding to $b\in X$ using standard formula (see, for
example, \cite{W,BIM})
gives us
\ber{3.1.3.39}
T_H(b)=0,
\eer
for any $b\in X$ satisfying $\xi_1(b) > 0$.

{\bf Example 7: MP solution.} The standard 4-dimensional
Majumdar-Papapetrou solution \cite{MP} in our notations reads
\ber{3.1.3.40}
g=H^2\hat{g}^0-H^{-2}dt\otimes dt, \mm
\label{3.1.3.41}
F=\nu dH^{-1} \wedge dt,
\eer
where $\nu^2 = 2$, $g^0 = \sum_{i=1}^{3} dx^i \otimes dx^i$
and $H$ is a harmonic function. We have one electric $0$-brane (point)
``attached'' to the time manifold; $d(I_s) =1$, $\eps_s= -1$ and
$(U^s,U^s) = 1/2$. In this case
(e.g. for extremal Reissner-Nordstr\"om black hole) we get
\ber{3.1.3.41a}
\eta(b)=0, \qquad \xi_1(b)=1,
\eer
and $T_H(b)=0$, $b\in X$.

{\bf Example 8: $D = 11$ supergravity.}
In this case  there are a lot of  $p$-brane
MP-type solutions with orthogonal intersection rules
e.g.
(i) solution with one electric $2$-brane ($d(I_s)=3$) and
$d_0 =8$;
(ii) solution with one magnetic $5$-brane ($d(I_s)=6$) and
$d_0 =5$;
(iii) solution with  one electric $2$-brane and one
 magnetic $5$-brane ($d(I_{s_1}\cap I_{s_2})=2$) and
 $d_0 =4$;
(iv) solution with  two electric $2$-branes
    ($d(I_{s_1}\cap I_{s_2})=1$) and $d_0 =5$.
In the examples  (iii), (iv) the harmonic functions
$H_{s_1}$ and $H_{s_2}$ from (\ref{3.1.3.28}) should
have the coinciding sets of poles, i.e. $X_{s_1} = X_{s_2}$,
to maintain the relation (\ref{3.1.3.38}). The
Chern-Simons terms are zero for these solutions.
In all these examples $\eta(b)=0$, $b \in X_{s}$, and
$\nu_s^2 = 1/2$, $s \in S$. In examples (i) and (ii)
$M2$ and $M5$ solutions are non-marginal, i.e. they have no
internal spaces $M_i$ (margins) that are not occupied by branes.
For marginal analogs of these two solutions we get
$\eta(b) < 0$ for all $b \in X_{s}$. Thus, marginal $M$ branes
have singularities at "points" $b$
(that are horizons for $d_0 \geq 4$, since all $\xi_1(b)\geq 0$).
Let us compare the situation with $IIA$  supergravity.
Performing the calculations for
single $FS$-, $NS5$-, $D2$- and $D4$-branes
in $IIA$  supergravity one
get $\eta(b) < 0$ for all $b \in X_{s}$ in all (marginal and
non-marginal) cases, i.e. all $b$ are singular "points"
(horizons for $d_0 \geq 4$).

{\bf Fundamental matrix.} Let
\ber{3.1.3.3}
N(a,b)\equiv\frac{(n_a-1)(n_b-1)}{D-2}-\lambda_a\cdot\lambda_b,
\eer
where $\lambda_a \cdot \lambda_b = \lambda_{a \alpha}
\lambda_{b \beta} h^{\alpha \beta}$,
$a,b\in\tri$. The matrix (\ref{3.1.3.3}) is called the fundamental
matrix of the model (\ref{2.1}) \cite{IMJ}.
It depends only on basic parameters of the model (\ref{2.1}),
i.e. ranks of forms, total dimensions and dilatonic couplings.
For $s_1,s_2\in S$, $s_1\ne s_2$, the symbol (\ref{3.1.30})
of orthogonal intersection  may be expressed by means of the
fundamental matrix \cite{IMJ}
\ber{3.1.3.4}
\Delta(s_1,s_2)=\bar D\bar\chi_{s_1}\bar\chi_{s_2}+
\bar{n}_{a_{s_1}}\chi_{s_1}\bar\chi_{s_2}+
\bar{n}_{a_{s_2}}\chi_{s_2}\bar\chi_{s_1}+
N(a_{s_1},a_{s_2})\chi_{s_1}\chi_{s_2},
\eer
where $\bar D=D-2$, $\bar n_a=n_a-1$, $\bar\chi_s=\frac12(1-\chi_s)$.
More explicitly (\ref{3.1.3.4}) reads
\ber{3.1.3.4a}
\Delta(s_1,s_2)
=N(a_{s_1},a_{s_2}), \quad v_{s_1}=v_{s_2}=e; \mm
\label{3.1.3.4b}
\Delta(s_1,s_2) =
\bar{n}_{a_{s_1}}-N(a_{s_1},a_{s_2}), \quad v_{s_1}=e, \quad v_{s_2}=m; \mm
\label{3.1.3.4c}
\Delta(s_1,s_2) =
\bar{D}-\bar{n}_{a_{s_1}}-\bar{n}_{a_{s_2}}+N(a_{s_1},a_{s_2}),
\quad v_{s_1}=v_{s_2}=m.
\eer
This follows from the relations
$d(I_s)=\bar D\bar\chi_s+\bar n_{a_s}\chi_s$,
equivalent to (\ref{2.1.7}). Relation (\ref{3.1.3.4a}) means
that $N(a,b)$ defines the dimension of intersection
of two electric $p$-branes case corresponding to forms
$F^a$ and $F^b$.
Let
\ber{3.1.3.6}
K(a)\equiv n_a-1-N(a,a)=\frac{(n_a-1)(D-n_a-1)}{D-2}+
\lambda_a\cdot\lambda_a,
\eer
$a\in\tri$.  The parameters play a rather important
role in supergravitational theories, since they are preserved under
Kaluza-Klein reduction \cite{St} and define the norms of $U^s$-vectors:
$K_s = (U^s,U^s)=K(a_s)$,
$s\in S$.

{\bf Intersection rules in $B_D$-models and $F$-theory.}
Let us apply the general relations
(\ref{3.1.3.4a})-(\ref{3.1.3.4c}) to the $B_D$-model
from (\ref{3.1.2.09}), $D \geq 12$. The orthogonal intersection rules
(\ref{3.1.30}) read
\ber{3.1.2.e}
(a-1)_e \cap (b-1)_e = N(a,b),  \mm
\label{3.1.2.em}
(a-1)_e \cap (D-1 -b)_m = a -1 - N(a,b),  \mm
\label{3.1.2.mm}
(D-1 -a)_m \cap (D-1 -b)_m = D -a -b + N(a,b),
\eer
$a,b = 4, \ldots, D-7$,
where subscrits $e$ and $m$ mark electric and magnetic branes,
respectively ($N(a,b)$ is defined in (\ref{3.1.2.011})).
For $D = 12$ (in the $F$-theory case) we
get the following intersections \cite{IMC} \beq{3.1.2.F}
\begin{array}{rll}\displaystyle
d(I \cap J) = &1, \ \{ d(I), d(J)\} =& \{ 3, 3\}, \{ 3, 4\},  \\
              &2,                   & \{ 3, 6\},
                  \{ 3, 7\}, \{ 4, 4\},\{ 4, 6\},  \\
              &3,                   & \{ 4, 7\},   \\
              &4,                   & \{ 6, 6\}, \{ 6, 7\},   \\
              &5,                   & \{ 7, 7\},
              \end{array}
\eeq

In most models including $D=11$ supergravity, $B_{12}$ theory,
$D < 11$ supergravities \cite{St},  all $K(a)=2$
and (\ref{3.1.2.3}) has the following form
\ber{3.1.3.13a}
d(I_{s_1}\cap I_{s_2})= \tri(s_1,s_2) + A_{s_1s_2},
\eer
$s_1\ne s_2$, and get $A_{s_1s_2}=A_{s_2s_1}$, i.e. the Cartan matrix is
symmetric. In a finite dimensional case we
are led to the so-called simply laced or ${\bf A-D-E}$ Lie algebras. The
intersection rules are totally defined by the corresponding Dynkin diagram:
$d(I_{s_1}\cap I_{s_2})= \tri(s_1,s_2)-1$, when the vertices corresponding
to $s_1$ and $s_2$ are connected by a line and $d(I_{s_1}\cap
I_{s_2})=\tri(s_1,s_2)$ otherwise (since in ${\bf A-D-E}$ case
$A_{s_1s_2}=0, -1$, $s_1 \neq s_2$).

{\bf Generalization to non-Ricci-flat internal spaces.}
In \cite{IMC} we presented a generalization of the solutions
\rf{3.1.11}-\rf{3.1.17} to the
case of non-Ricci-flat space $(M_0,g^0)$, when some additional internal
Einstein spaces of non-zero curvature are
included.

\subsection{General Toda-type solutions with harmonic functions}

It is well known
that geodesics of the target space equipped with some
harmonic function on a three-dimensional space generate a solution
to the $\sigma$-model equations \cite{NK,KSMH}.
(It was  observed in
\cite{GC} that  null geodesics of the target space of stationary
five-dimensional Kaluza-Klein theory may be used to generate multisoliton
solutions similar to the Israel-Wilson-Perj\`es solutions
of Einstein-Maxwell theory.)
Here we apply this null-geodesic method  to our sigma-model and
obtain a  new class of solutions in
multidimensional gravity with  $p$-branes
governed by one harmonic function $H$. The solutions
from this class correspond to
null-geodesics of the target-space metric and are defined
by some functions $f_s(H)= \exp(-q^s(H))$
with $q^s(u)$  being  solutions to Toda-type equations.

\subsubsection{Toda-like Lagrangian}

Action (\ref{2.2.7}) may be also written in the form
\beq{3.2.1n}
S_{\sigma 0} =  \frac{1}{2 \kappa_0^2} \int d^{d_0}x\sqrt{|g^0|}
\{ R[g^0]-  {\cal G}_{\hat A\hat B}(X)
g^{0 \mu \nu} \p_\mu X^{\hat A} \p_\nu X^{\hat B}   - 2V \}
\eeq
where $X = (X^{\hat A})=(\phi^i,\varphi^\alpha,\Phi^s)\in{\bf
R}^{N}$, and minisupermetric
$
{\cal G}=
{\cal G}_{\hat A\hat B}(X)dX^{\hat A}\otimes dX^{\hat B}
$
on minisuperspace
${\cal M}={\bf R}^{N}$,  $N = n+l+|S|$
($|S|$ is the number of elements in $S$) is defined by the relation
\beq{3.2.3n}
({\cal G}_{\hat A\hat B}(X))=\left(\begin{array}{ccc}
G_{ij}&0&0\\[5pt]
0&h_{\alpha\beta}&0\\[5pt]
 0&0&\eps_s \exp(-2U^s(\sigma))\delta_{ss'}
\end{array}\right).
\eeq

Here we consider exact solutions to field equations corresponding
to the action  (\ref{3.2.1n})
\bear{3.2.4}
R_{\mu\nu}[g^0]=
{\cal G}_{\hat A \hat B}(X) \p_{\mu} X^{\hat A} \p_\nu  X^{\hat B}
+ \frac{2V}{d_0-2}g_{\mu\nu}^0,    \\
\label{3.2.5}
\frac{1}{\sqrt{|g^0|}} \p_\mu [\sqrt{|g^0|}
{\cal G}_{\hat C \hat B}(X)g^{0\mu\nu }  \p_\nu  X^{\hat B}]
- \frac{1}{2} {\cal G}_{\hat A \hat B, \hat C}(X)
g^{0,\mu \nu} \p_{\mu} X^{\hat A} \p_\nu  X^{\hat B} = V_{,\hat C},
\ear
$s\in S$. Here  $V_{,\hat C} = \p V / \p  X^{\hat C}$.

We put
\ber{3.2.6}
X^{\hat A}(x) =  F^{\hat A}(H(x)),
\eer
where $F: (u_{-}, u_{+}) \rightarrow \R^{N}$  is a smooth function,
$H: M_0 \rightarrow \R $ is a harmonic function on $M_0$ (i.e.
$\tri[g^0]H=0$), satisfying  $u_{-} < H(x)  < u_{+}$ for all $x \in M_0$.
Let all factor spaces are  Ricci-flat
and cosmological constant is zero, i.e. relation
$\xi_i = \Lambda = 0$
is satisfied. In this case the potential is zero : $V = 0$.
It may be verified that the field equations (\ref{3.2.4}) and (\ref{3.2.5})
are satisfied identically  if $F = F(u)$ obey the
Lagrange equations  for the Lagrangian
\beq{3.2.12}
L =  \frac{1}{2} {\cal G}_{\hat A\hat B}(F) \dot F^{\hat A}
\dot F^{\hat B}
\eeq
with the zero-energy constraint
\beq{3.2.13}
E =  \frac{1}{2} {\cal G}_{\hat A\hat B}(F) \dot F^{\hat A}
\dot F^{\hat B} = 0.
\eeq
This means that  $F: (u_{-}, u_{+}) \rightarrow  \R^N$
is a null-geodesic map for the minisupermetric ${\cal G}$.
Thus, we are led to the Lagrange system (\ref{3.2.12})
with the minisupermetric  ${\cal G}$ defined in (\ref{3.2.3n}).

The problem of integrability will be simplified if we integrate
the Lagrange equations corresponding to $\Phi^s$
(i.e. the Maxwell equations for $s\in S_e$ and
Bianchi identities for $s\in S_m$):
\bear{3.2.14}
\frac d{du}\left(\exp(-2U^s(\sigma))\dot\Phi^s\right)=0
\Longleftrightarrow
\dot\Phi^s=Q_s \exp(2U^s(\sigma)),
\ear
where $Q_s$ are constants, $s \in S$.
Here $(F^{\hat A})= (\sigma^A, \Phi^s)$.
We put $Q_s\ne0$ for all  $s \in S$.

For fixed $Q=(Q_s,s\in S)$ the Lagrange equations for the Lagrangian
(\ref{3.2.12})  corresponding to $(\sigma^A)=(\phi^i,\varphi^\alpha)$,
when equations (\ref{3.2.14}) are substituted, are equivalent to the
Lagrange equations for the Lagrangian
\beq{3.2.16}
L_Q=\frac12\hat G_{AB}\dot \sigma^A\dot \sigma^B-V_Q,
\eeq
where
\beq{3.2.17}
V_Q=\frac12  \sum_{s\in S}  \eps_s Q_s^2 \exp[2U^s(\sigma)],
\eeq
the matrix $(\hat G_{AB})$ is defined in (\ref{2.2.9}). The zero-energy
constraint (\ref{3.2.13}) reads
\beq{3.2.18}
E_Q= \frac12 \hat G_{AB}\dot \sigma^A \dot \sigma^B+ V_Q =0.
\eeq

\subsubsection{ Toda-type solutions}

Here we are interested in exact solutions for a special case
when the vectors $U^s $ have non-zero length, i.e.
$K_s =(U^s,U^s)\neq 0$,
for all $s\in S$, and the quasi-Cartan matrix (\ref{3.1.2.2})
is a non-degenerate one. Here  some ordering in $S$ is assumed.
It follows from the non-degeneracy of the matrix (\ref{3.1.2.2})
that the vectors $U^s, s \in S,$ are  linearly
independent. Hence, the number of the vectors
$U^s$ should not exceed the dimension of
$\R^{n+ l}$, i.e.
$|S| \leq n+ l$.

The  exact solutions  could be readily obtained
using the relations from Appendix 4 \cite{IK}
\bear{3.2.63}
g= \biggl(\prod_{s \in S} f_s^{2d(I_s)h_s/(D-2)}\biggr)
\biggl\{ \exp(2c^0 H+ 2\bar  c^0) \hat{g}^0  \\ \nn
+ \sum_{i =1}^{n} \Bigl(\prod_{s \in S }
f_s^{- 2h_s \delta_{i I_s} }\Bigr)
\exp(2c^i H+ 2\bar  c^i) \hat{g}^i \biggr\},
\\  \label{3.2.64}
\exp(\varphi^\alpha) =
\left( \prod_{s\in S} f_s^{h_s \chi_s\lambda_{a_s}^\alpha} \right)
\exp(c^\alpha H +\bar c^\alpha),
\ear
$\alpha=1,\dots,l$ and $F^a=\sum_{s\in S}{\cal F}^s\delta_{a_s}^a$
with
\bear{3.2.57}
{\cal F}^s= Q_s
\left( \prod_{s' \in S}  f_{s'}^{- A_{s s'}} \right) dH \wedge\tau(I_s),
\qquad s\in S_e, \\
\label{3.2.58}
 {\cal F}^s
=  Q_s (*_0 d H) \wedge \tau(\bar I_s),
\qquad s\in S_m,
\ear
where  $*_0 = *[g^0]$ is the Hodge operator on $(M_0,g^0)$.
Here
\beq{3.2.65}
f_s = f_s(H) = \exp(- q^s(H)),
\eeq
where $q^s(u)$ is a solution to  Toda-like equations
\beq{3.2.35}
\ddot{q^s} = -  B_s \exp( \sum_{s' \in S} A_{s s'} q^{s'} ),
\eeq
with $ B_s =  K_s \eps_s Q_s^2$,
$s \in S$, and $H = H(x)$ ($x \in M_0$) is a harmonic function on
$(M_0,g^0)$. Vectors $c=(c^A)$ and $\bar c=(\bar c^A)$ satisfy the linear
constraints (see   Appendix 4)
\bear{3.2.47}
U^s(c)=
\sum_{i \in I_s}d_ic^i-\chi_s\lambda_{a_s\alpha}c^\alpha=0,
\quad
U^s(\bar c)= 0,
\ear
$s\in S$, and
\beq{3.2.52a}
c^0 = \frac1{2-d_0}\sum_{j=1}^n d_j c^j,
\quad  \bar  c^0 = \frac1{2-d_0}\sum_{j=1}^n d_j \bar c^j.
\eeq
The zero-energy constraint reads
(see Appendix 4)
\bear{3.2.53}
2E_{TL} + h_{\alpha\beta}c^\alpha c^\beta+ \sum_{i=1}^n d_i(c^i)^2+
\frac1{d_0-2}\left(\sum_{i=1}^nd_ic^i\right)^2 = 0,
\ear
where
\beq{3.2.54a}
E_{TL} = \frac{1}{4}  \sum_{s,s' \in S} h_s
A_{s s'} \dot{q^s} \dot{q^{s'}}
  + \sum_{s \in S} A_s  \exp( \sum_{s' \in S} A_{s s'} q^{s'} ),
\eeq
is an integration constant (energy) for the solutions from
(\ref{3.2.35}) and $A_s =  \frac12  \eps_s Q_s^2$.

We note that the equations
(\ref{3.2.35}) correspond to the Lagrangian
\beq{3.2.37}
L_{TL} = \frac{1}{4}  \sum_{s,s' \in S} h_s  A_{s s'} \dot{q^s}
\dot{q^{s'}}
-  \sum_{s \in S} A_s  \exp( \sum_{s' \in S} A_{s s'} q^{s'} ),
\eeq
where $h_s = K_s^{-1}$.

Thus the solution is presented
by relations  (\ref{3.2.63})-(\ref{3.2.65})
with the functions  $q^s$  defined in
(\ref{3.2.35}) and the relations on the parameters of
solutions $c^A$, $\bar c^A$ $(A= i,\alpha,0)$,
imposed in  (\ref{3.2.47}),
(\ref{3.2.52a}), (\ref{3.2.53}).

\subsubsection{Solutions corresponding to $A_m$  Toda  chain}

Here we  consider exact solutions to Toda-chain equations
(\ref{3.2.35}) corresponding to the Lie algebra
${\bf A_{m}}= \SL(m+1, \C)$ \cite{T,And} , ($m \geq 1$)
where
\beq{3.2.B.1a}
\left(A_{ss'}\right)=
\left( \begin{array}{*{6}{c}}
2&-1&0&\ldots&0&0\\
-1&2&-1&\ldots&0&0\\
0&-1&2&\ldots&0&0\\
\multicolumn{6}{c}{\dotfill}\\
0&0&0&\ldots&2&-1\\
0&0&0&\ldots&-1&2
\end{array}
\right)\quad
\eeq
is the Cartan matrix  of the Lie algebra ${\bf A_{m}}$ and $B_s > 0$,
$s,s' = 1, \ldots, m$.  Here we put $S =  \{1, \ldots, m \}$.

The equations of motion   (\ref{3.2.35})
correspond
to the Lagrangian
\beq{3.2.B.2}
 L_T = \frac{1}{2} \sum_{s,s'=1}^{m} A_{ss'} \dot q^s  \dot q^{s'}  -
\sum_{s=1}^{m}  B_s \exp \left( \sum_{s'=1}^{m} A_{ss'} q^{s'}  \right).
\eeq
This Lagrangian may be obtained from the standard one \cite{T}
by separating a coordinate describing the  motion  of the center of mass.

Using the result of A. Anderson \cite{And}
we present the solution to eqs. (\ref{3.2.35}) in the following form
\beq{3.2.B.3}
C_s \exp(-q^s(u)) =
\sum_{r_1< \dots <r_s}^{m+1} v_{r_1}\cdots v_{r_s}
\Delta^2( w_{r_1}, \ldots, w_{r_s}) \exp[(w_{r_1}+\ldots +w_{r_s})u],
\eeq
$s = 1, \ldots, m$, where
$\Delta( w_{r_1}, \ldots, w_{r_s})  =
\prod_{i<j}^{s} \left(w_{r_i}-w_{r_j}\right)$;
$\Delta(w_{r_1}) \equiv 1$,
denotes the  Vandermonde determinant.
The real constants $v_r$ and $w_r$, $r = 1, \ldots, m + 1$, obey the
relations
\beq{3.2.B.5}
\prod_{r=1}^{m+1} v_r= \Delta^{-2}(w_1,\ldots, w_{m+1}), \qquad
\sum_{r=1}^{m+1}w_r=0.
\eeq
In (\ref{3.2.B.3})
$C_s = \prod_{s'=1}^{m} B_{s'}^{-A^{s s'}}$,
where $(A^{s s'})=(A_{ss'})^{-1}$ is presented in
(\ref{A3.12}) of Appendix 3.
Here
$v_r \neq 0, \qquad w_r \neq w_{r'}$, $r \neq r'$,
$r, r' = 1, \ldots, m +1$.
We note that the solution with $B_s > 0$ may be obtained
from the solution  with $B_s =1$ (see \cite{And}) by a certain shift $q^s
\mapsto q^s + \delta^s$.

The energy reads \cite{And}
\beq{3.2.B.9}
 E_T= \frac{1}{2} \sum_{s,s'=1}^{m} A_{s s'} \dot q^s  \dot q^{s'} +
\sum_{s=1}^{m}  B_s \exp \left( \sum_{s'=1}^{m} A_{ss'} q^{s'}  \right) =
\frac{1}{2}\sum_{r=1}^{m+1} w^2_r.
\eeq

If  $B_s > 0$, $s \in S$, then  all $w_r, v_r$ are real and, moreover, all
$v_r > 0$, $r =1, \ldots, m+1$.  In a general case $B_s \neq 0$, $s \in
S$, relations  (\ref{3.2.B.3}) also describe real
solutions to eqs. (\ref{3.2.35}) for suitably chosen complex parameters
$v_r$ and $w_r$. The parameters $w_i$ are either real or belong to pairs of
complex conjugate (non-equal) numbers, i.e., for example, $w_1 = \bar
w_2$. When some of $B_s$ are negative, there are also
some special (degenerate) solutions to eqs. (\ref{3.2.35}) that are not
described by relations (\ref{3.2.B.3},  but may be
obtained from the latter by certain limits of parameters $w_i$.

For the energy (\ref{3.2.54a}) we get $E_{TL} = \frac{1}{2K} E_T$.
Here $K_s = K$.
In the ${\bf A_m}$ Toda chain case
eqs.  (\ref{3.2.B.3}) should be substituted into
relations  (\ref{3.2.63})-(\ref{3.2.57}).

To our knowledge $p$-brane  solutions governed by open Toda lattices with
${\bf A_n}$ Lie algebras were studied first in \cite{LMPX,LPX}.
In  \cite{LMMP}
${\bf E_N}$ open Toda lattices  in maximal supergravities in $D$ dimensions
coming from $D =11$ supergravity  were considered.
The appearance of ${\bf A-D-E}$ algebras
is rather typical for supergravitational models
(since all $K_s = 2$ and hence the roots of the Lie algebra have equal
lengths).

\pagebreak

\section{Classical and quantum cosmological-type solutions}

Here we consider the case $d_0 =1$, $M_0 = \R$, i.e.
we are interesting in applications to
the sector with one-variable dependence.
We consider the manifold
\beq{4.1}
M = (u_{-},u_{+})  \times M_{1} \times \ldots \times M_{n}
\eeq
with the metric
\beq{4.2}
g= w \e^{2{\gamma}(u)} du \otimes du +
\sum_{i=1}^{n} \e^{2\phi^i(u)} {\hat g}^i ,
\eeq
where $w=\pm 1$, $u$ is a distinguished coordinate which, by
convention, will be called ``time'';
$(M_i,g^i)$ are oriented and connected Einstein spaces
(see (\ref{2.13})), $i=1,\dots,n$.
The functions $\gamma,\phi^i$: $(u_-,u_+)\to \R$ are smooth.

Here we adopt the p-brane ansatz from Sect. 2.
putting $g^0= w  du \otimes du$.

\subsection{ Lagrange dynamics}

It follows from Subsect. 2.3 that the
equations of motion  and the Bianchi
identities for the field configuration under consideration
(with the restrictions from Sect. 2.3.1 imposed)
are equivalent to equations of motion for
1-dimensional $\sigma$-model with the action
\beq{4.1.1}
S_{\sigma} = \frac{\mu}2
\int du {\cal N} \biggl\{G_{ij}\dot\phi^i\dot\phi^j
+h_{\alpha\beta}\dot\varphi^{\alpha}\dot\varphi^{\beta}
+\sum_{s\in S}\eps_s\exp[-2U^s(\phi,\varphi)](\dot\Phi^s)^2
-2{\cal N}^{-2}V_{w}(\phi)\biggr\},
\eeq
where $\dot x\equiv dx/du$,
\beq{4.1.2}
V_{w} = -w V = -w\Lambda\e^{2\gamma_0(\phi)}+
\frac w2\sum_{i =1}^{n} \xi_i d_i \e^{-2 \phi^i + 2 {\gamma_0}(\phi)}
\eeq
is the potential with
$\gamma_0(\phi)
\equiv\sum_{i=1}^nd_i\phi^i$,
and
${\cal N}=\exp(\gamma_0-\gamma)>0$
is the lapse function, $U^s = U^s(\phi,\varphi)$ are defined
in (\ref{2.2.11}), $\eps_s$ are defined in (\ref{2.2.13})
for $s=(a_s,v_s,I_s)\in S$, and
$G_{ij}=d_i\delta_{ij}-d_id_j$
are components of the "pure cosmological" minisupermetric, $i,j=1,\dots,n$
\cite{IMZ}.

In the electric case $({\cal F}^{(a,m,I)}=0)$ for finite internal space
volumes $V_i$ the action (\ref{4.1.1}) coincides with the
action (\ref{2.1}) if
$\mu=-w/\kappa_0^2$, $\kappa^{2} = \kappa^{2}_0 V_1 \ldots V_n$.

Action (\ref{4.1.1}) may be also written in the form
\beq{4.1.6}
S_\sigma=\frac\mu2\int du{\cal N}\left\{
{\cal G}_{\hat A\hat B}(X)\dot X^{\hat A}\dot X^{\hat B}-
2{\cal N}^{-2}V_w \right\},
\eeq
where $X = (X^{\hat A})=(\phi^i,\varphi^\alpha,\Phi^s)\in
{\R}^{N}$, $N = n +l + |S|$, and minisupermetric
${\cal G}$ is defined in (\ref{3.2.3n}).

{\bf Scalar products.}
The minisuperspace metric (\ref{3.2.3n}) may be also written in the form
${\cal G}=\hat G+\sum_{s\in S}\eps_s
\e^{-2U^s(\sigma)}d\Phi^s\otimes d\Phi^s$,
where $\sigma = (\sigma^A) = (\phi^i,\varphi^\alpha)$,
\bear{4.1.8}
\hat G=\hat G_{AB}d \sigma^A
\otimes d \sigma^B=G_{ij}d\phi^i\otimes d\phi^j+
h_{\alpha\beta}d\varphi^\alpha\otimes d\varphi^\beta,
\ear
is truncated minisupermetric and $U^s(\sigma)=U_A^s \sigma^A$ is defined in
(\ref{2.2.11}).
The potential (\ref{4.1.2})
reads
\beq{4.1.9}
V_w=(-w\Lambda)\e^{2U^\Lambda(\sigma)}+\sum_{j=1}^n\frac w2\xi_jd_j
\e^{2U^j(\sigma)},
\eeq
where
\bear{4.1.10}
U^j(\sigma)=U_A^j \sigma^A=-\phi^j+\gamma_0(\phi),
\qquad (U_A^j)=(-\delta_i^j+d_i,0),
\\ \label{4.1.11}
U^\Lambda(\sigma)=U_A^\Lambda \sigma^A=\gamma_0(\phi),
\qquad (U_A^\Lambda)=(d_i,0).
\ear

The integrability of the Lagrange system (\ref{4.1.6}) crucially depends
upon the scalar products of co-vectors $U^\Lambda$, $U^j$, $U^s$
(see (\ref{3.1.1})).
These products are defined by
(\ref{3.1.4}) and the following relations \cite{IMC}
\bear{4.1.14}
(U^i,U^j)=\frac{\delta_{ij}}{d_j}-1,
\\
\label{4.1.15}
(U^i,U^\Lambda)=-1,
\qquad
(U^\Lambda,U^\Lambda)=-\frac{D-1}{D-2},
\\ \label{4.1.17}
(U^s,U^i)=-\delta_{iI_s}, \qquad
(U^s,U^\Lambda)=\frac{d(I_s)}{2-D},
\ear
where $s=(a_s,v_s,I_s) \in S$;
$i,j= 1,\dots,n$.

{\bf Toda-like representation.}
We put $\gamma= \gamma_0(\phi)$, i.e. the harmonic
time gauge is considered.  Integrating
the Lagrange equations corresponding to $\Phi^s$
(see (\ref{3.2.14})) we are led to the Lagrangian
from (\ref{3.2.16}) and the zero-energy constraint (\ref{3.2.18})
with the modified potential
\beq{4.1.19}
V_Q=V_w +\frac12\sum_{s\in S} \eps_sQ_s^2\exp[2U^s(\sigma)],
\eeq
where $V_w$ is defined in  (\ref{4.1.2}).

\subsection{Classical solutions with $\Lambda = 0$}

Here we consider classical solutions with $\Lambda = 0$.

\subsubsection{Solutions with Ricci-flat spaces}

Let all spaces be Ricci-flat, i.e. $ \xi_1 =\dots=\xi_n=0$.

Since  $H(u)= u$ is a harmonic function on $(M_0,g^0)$
with $g^0 = w du\otimes du$  we get
for the metric and scalar fields from
(\ref{3.2.63}), (\ref{3.2.64}) \cite{IK}
\bear{4.3.3}
g= \biggl(\prod_{s \in S} f_s^{2d(I_s)h_s/(D-2)}\biggr)
\biggl\{ \exp(2c^0 u+ 2\bar  c^0) w du \otimes du   \\ \nonumber
+ \sum_{i =1}^{n} \Bigl(\prod_{s\in S }
f_s^{- 2h_s \delta_{i I_s} }\Bigr)
\exp(2c^i u + 2\bar  c^i) {\hat g}^i \biggr\},
\\  \label{4.3.4}
\exp(\varphi^\alpha) = \left( \prod_{s\in S}
f_s^{h_s \chi_s\lambda_{a_s}^\alpha} \right)
\exp(c^\alpha u +\bar c^\alpha),
\ear
$\alpha=1,\dots,l$, where   $f_s = f_s(u) = \exp(- q^s(u))$ and
$q^s(u)$ obey  Toda-like equations (\ref{3.2.35}).

Relations (\ref{3.2.52a}) and  (\ref{3.2.53}) take the form
\bear{4.3.5}
c^0 = \sum_{j=1}^n d_j c^j,
\qquad  \bar  c^0 = \sum_{j=1}^n d_j \bar c^j,
\\  \label{4.3.6c}
2E_{TL} + h_{\alpha\beta}c^\alpha c^\beta+ \sum_{i=1}^n d_i(c^i)^2
- \left(\sum_{i=1}^nd_ic^i\right)^2 = 0,
\ear
with $E_{TL}$  from (\ref{3.2.54a}) and all other relations
(e.g. constraints (\ref{3.2.47}) and
relations  for forms (\ref{3.2.57}) and (\ref{3.2.58}) with $H = u$)
are unchanged.  In a special  ${\bf A_m}$ Toda chain case this solution
was considered previously in \cite{GM2}.

\subsubsection{Solutions with one curved space}

The cosmological solution with Ricci-flat spaces
may be also  modified to the following case:
$ \xi_1 \ne0, \quad \xi_2=\ldots=\xi_n=0$,
i.e. one space is curved and others are Ricci-flat and
$1 \notin I_s$,
$s  \in S$,
i.e. all ``brane'' submanifolds  do not  contain $M_1$.

The potential $(\ref{3.2.17})$ is modified for $\xi_1 \ne0$
as follows (see \rf{4.1.19})
\beq{4.3.6}
V_Q=\frac12  \sum_{s\in S}  \eps_s Q_s^2 \exp[2U^s(\sigma)]
+ \frac12 w\xi_1 d_1  \exp[2U^1(\sigma)],
\eeq
where $U^1(\sigma)$ is defined in  (\ref{4.1.10}) ($d_1 > 1$).

For the scalar products we get from (\ref{4.1.14}) and (\ref{4.1.17})
\bear{4.3.7}
(U^1,U^1)=\frac1{d_1}-1<0, \qquad (U^1,U^{s})=0
\ear
for all $s\in S$.

The solution in the case under consideration
may be obtained   by a little modification
of the solution from the previous section
(using (\ref{4.3.7}), relations
$U^{1i}= - \delta_1^i/d_1$,  $U^{1\alpha}=0$
and Appendix 4).
We get  \cite{IK}
\bear{4.3.19}
g= \biggl(\prod_{s \in S} [f_s(u)]^{2 d(I_s) h_s/(D-2)} \biggr)
\biggl\{[f_1(u)]^{2d_1/(1-d_1)}\exp(2c^1u + 2 \bar c^1)\\ \nn
\times[w du \otimes du+ f_1^2(u) \hat{g}^1] +
\sum_{i = 2}^{n} \Bigl(\prod_{s\in S}
[f_s(u)]^{- 2 h_s  \delta_{i I_s} }
\Bigr)\exp(2c^i u+ 2 \bar c^i) \hat{g}^i \biggr\}.
\\ \label{4.3.19a}
\exp(\varphi^\alpha) =
\left( \prod_{s\in S} f_s^{h_s \chi_s \lambda_{a_s}^\alpha} \right)
\exp(c^\alpha u + \bar c^\alpha),
\ear
and
$F^a= \sum_{s \in S} \delta^a_{a_s} {\cal F}^{s}$
with forms
\bear{4.3.19c}
{\cal F}^s= Q_s
\left( \prod_{s' \in S}  f_{s'}^{- A_{s s'}} \right) du \wedge\tau(I_s),
\qquad s\in S_e,  \\
\label{4.3.19d}
{\cal F}^s= Q_s \tau(\bar I_s), \qquad s \in S_m
\ear
$Q_s \neq 0$, $s \in S$.
Here  $f_s = f_s(u) = \exp(- q^s(u))$ where
$q^s(u)$ obeys  Toda-like equations (\ref{3.2.35})
and
\bear{4.3.10}
f_1(u) =R \sh(\sqrt{C_1}(u-u_1)), \ C_1>0, \ \xi_1 w>0;
\\ \label{4.3.11}
R \sin(\sqrt{|C_1|}(u-u_1)), \ C_1<0, \  \xi_1 w>0;   \\ \label{4.3.12}
R \ch(\sqrt{C_1}(u-u_1)),  \ C_1>0, \ \xi_1w <0; \\ \label{4.3.13}
\left|\xi_1(d_1-1)\right|^{1/2}        , \ C_1=0,  \ \xi_1w>0,
\ear
$u_1, C_1$ are constants and $R =  |\xi_1(d_1-1)/C_1|^{1/2}$.

Vectors $c=(c^A)$ and $\bar c=(\bar c^A)$ satisfy the linear constraints
\bear{4.3.15}
U^r(c)= U^r(\bar c)= 0, \qquad r = s,1,
\ear
(for $r =s$ see (\ref{3.2.47}))
and the zero-energy constraint
\beq{4.3.17}
C_1\frac{d_1}{d_1-1}= 2 E_{TL} +
h_{\alpha\beta}c^\alpha c^\beta+ \sum_{i=2}^nd_i(c^i)^2+
\frac1{d_1-1}\left(\sum_{i=2}^nd_ic^i\right)^2.
\eeq

{\bf Restriction 1} ( see Subsect. 2.3.1) forbids certain intersections of
two $p$-branes with the same color index for  $n_1 \geq 2$.
{\bf Restriction 2} is  satisfied identically in this case.

This solution in a special case
of ${\bf A_m}$ Toda chain was obtained earlier  in \cite{GM1}
(see also \cite{GM2}).
Some special configurations were considered earlier in
\cite{LMPX,LPX,LMMP}.

\subsubsection{Block-orthogonal solutions}

Let  us consider block-orthogonal case:
(\ref{3.1.5}), (\ref{3.1.6}). In this case
due to Appendix 4 we get $f_s = \bar{f}_s^{b_s}$
where $b_s = 2 \sum_{s' \in S} A^{s s'}$,
$(A^{ss'})= (A_{ss'})^{-1}$  and
\bear{4.3.20n}
\bar{f}_s(u)=
R_s \sh(\sqrt{C_s}(u-u_s)), \;
C_s>0, \; \eta_s\eps_s<0; \\ \label{4.3.22n}
R_s \sin(\sqrt{|C_s|}(u-u_s)), \;
C_s<0, \; \eta_s\eps_s<0; \\ \label{4.3.23n}
R_s \ch(\sqrt{C_s}(u-u_s)), \;
C_s>0, \; \eta_s\eps_s>0; \\ \label{4.3.24n}
\frac{|Q^s|}{|\nu_s|}(u-u_s), \; C_s=0, \; \eta_s\eps_s<0,
\ear
where $R_s = |Q_s|/(|\nu_s||C_s|^{1/2})$,
$\eta_s \nu_s^2 = b_s h_s$, $\eta_s= \pm 1$,
$C_s$, $u_s$ are constants, $s \in S$.
The constants $C_s$, $u_s$ are coinciding inside blocks:
$u_s = u_{s'}$, $C_s = C_{s'}$,
$s,s' \in S_i$, $i = 1, \ldots, k$
(see Appendix 4).
The ratios $\eps_s Q_s^2/(b_s h_s)$ are
also coinciding inside blocks, or, equivalently,
\bear{4.3.25n}
\frac{\eps_s Q_s^2}{b_s h_s} = \frac{\eps_{s'} Q_{s'}^2}{b_{s'} h_{s'}},
\ear
$s,s' \in S_i$, $i = 1, \ldots, k$.

Here
\beq{aa}
E_{TL} = \sum_{s \in S} C_s \eta_s \nu_s^2.
\eeq

The solution (\ref{4.3.19})-(\ref{4.3.19d}) with block-orthogonal
set of vectors was obtained in \cite{IMJ1,IMJ2} (for non-composite case see
also \cite{Br1}). In the special  orthogonal case when: $|S_1| = \ldots =
|S_k| = 1$, the solution  was obtained in \cite{IMJ}.
In non-composite case "orthogonal" solutions were considered
in \cite{GrIM,BGIM} (electric case) and \cite{BIM}
(electro-magnetic case). For $n = 1$ see also \cite{LMPX,LPX}.

\subsection{Classical solutions with $\Lambda \neq 0$
on product of Einstein spaces}

Here we describe an important class of
classical solutions appearing when all scale factors are constant
(Freund-Rubin-type solutions).
The solutions with anti-de-Sitter spaces appear in the ``near-horizon''
limit of extremely charged $p$-brane configurations from Sect. 3.

We note, that
recently an interest to  Freund-Rubin-type solutions \cite{FR,DNP}
in multidimensional models with $p$-branes living
on product of Einstein spaces appeared (see, for example,
\cite{torin,DLP}). This interest was
inspired by papers devoted to duality between certain limit of some
superconformal theory in $d$-dimensional space and
string or M-theory compactified on the space $AdS_{d+1} \times
W$, where $W$ is a compact manifold (e.g. sphere) \cite{M}
(see also \cite{GKP,Wi} etc.)
It was shown in \cite{GT} that the solutions in $D = 10,11$ supergravities
representing $D3, M2, M5$ branes interpolate between flat-space vacuum and
compactifications to AdS space.

Here we consider a rather general class of solutions
with spontaneous compactification for the model (\ref{2.1})
defined  on the manifold $M_{0}  \times M_{1} \times \ldots \times M_{n}$,
with the metric
\beq{4.3.f1.2}
g= \hat{g}^0  + \hat{g}^1 + \ldots + \hat{g}^n ,
\eeq
where and  $g^i$ is an Einstein metric on $M_{i}$  satisfying the equation
(\ref{2.13}), $i=0,\ldots,n$.

We note that in the pure gravitational model with cosmological
constant $\Lambda$ the equations of motion
${\rm Ric}[g]= \frac{2 \Lambda}{D-2}g$  have a rather  simple
solution
\beq{4.3.f1.4}
\xi_{i}  = \frac{2 \Lambda}{D-2},
\eeq
$i=0,\ldots,n$, \cite{Zh,BIMZ,BZ1,BZ2} that was also generalized to
some other matter fields, e.g scalar one (see \cite{GZ}
and reference therein).

The solution is given by the relation (\ref{4.3.f1.2})
and (\ref{4.3.f2.9})-(\ref{4.3.f2.12}) (see below).
In the non-composite case
the "cosmological derivation" of the solution was obtained in
\cite{Cosm}. The Freund-Rubin solutions \cite{FR}  in $D =11$
supergravity: $AdS_{4} \times S^7$ and  $AdS_{7} \times S^4$,
correspond to M2-brane "living" on  $AdS_{4}$ and $S^4$ respectively.
The "popular" $AdS_{5} \times S^5$ solution in
$II B$ ($D =10$) supergravity model \cite{DLP}
corresponds to composite self-dual configuration
with two "branes" living on $AdS_{5}$ and $S^5$
and corresponding to 5-form.

We consider the model governed by the action (\ref{2.1}).
The equations of motion were presented in (\ref{2.4})-(\ref{2.6}).
The Hilbert-Einstein equations (\ref{2.4}) may be written in the
equivalent form (\ref{2.2.31}).
Here we keep the multi-index notations with
$\Omega = \Omega_0 $  being a set of all non-empty
subsets of $\{ 0, \ldots,n \}$.

The solution reads as following: the manifold  and the metric
are defined by the relations  (\ref{2.10}), (\ref{4.3.f1.2}),
(\ref{2.13}) (with $i=0,\ldots,n$), the fields of
forms and  scalar fields are the following
\bear{4.3.f2.9}
F^a = \sum_{I \in \Omega_{a}} Q_{aI} \tau(I),
\\ \label{4.3.f2.10}
\varphi^{\alpha} = {\rm const}.
\ear
Here $\Omega_{a} \subset \Omega$ are non-empty subsets
satisfying the {\bf Restriction 3} presented below and
$Q_{aI}$ are constants, $I \in \Omega_{a}$, $a \in \Delta$.
The  parameters of  solution obey the relations \cite{I2}
\bear{4.3.f2.11}
\sum_{a \in \Delta} \theta_a \lambda^{\alpha}_a
e^{2 \lambda_{a}(\varphi)}
\sum_{I \in \Omega_a} (Q_{aI})^2 \eps(I) = 0, \\
\label{4.3.f2.12}
\xi_i =   \frac{2 \Lambda}{D-2}  +
\sum_{a \in \Delta} \theta_a e^{2 \lambda_{a}(\varphi)}
\sum_{I \in \Omega_a} (Q_{aI})^{2} \eps(I) \left[ \delta^i_I
- \frac{n_a -1}{D - 2} \right],
\ear
$i = 0, \ldots, n$.

The solution is valid if the following restriction
on the sets $\Omega_{a}$, $a \in \tri$, (similar to
electric part of {\bf Restriction 1}) is satisfied.

{\bf Restriction 3.}
For any $a \in  \Delta$ and  $I,J \in \Omega_{a}$
\beq{4.3.f2.r}
d(I \cap J) \leq n_a -2
\eeq

Due to relations from the Appendix 2
this restriction guarantees the block-diagonal structure
of the  energy-momentum tensor.
We note that due to (\ref{4.3.f2.9}) the self-consistency condition should
be satisfied
$d(I) = n_a$,
for all $I \in \Omega_a$, $a \in \Delta$.

A simple example: for $D = 11$ supergravity with $n_a =4$
the $p$-brane intersection $d(I_1 \cap I_2) =3$ is ``forbidden'' due to
{\bf Restriction 3.}

The solution mentioned above may be obtained by a straightforward
substitution of the fields (\ref{4.3.f1.2}), (\ref{4.3.f2.9}),
(\ref{4.3.f2.10}) into equations of motion
(\ref{2.4})-(\ref{2.6})  while formulas from the Appendix 2
are keeping in mind. Eq. (\ref{2.4}) should be written
in the form
\beq{2.2.31}
R_{MN}  =   Z_{MN} + \frac{2\Lambda}{D - 2} g_{MN},
\eeq
where
$Z_{MN} \equiv   T_{MN} + \frac{T}{2 -D} g_{MN}$,
and $T = {T_M}^M$.

{\bf Electro-magnetic representation.}
Due to relation
\beq{4.3.f2.24}
* \tau(I) = \eps(I) \delta(I) \tau(\bar{I}),
\eeq
where $*=*[g]$ is the Hodge operator on $(M,g)$,
$\bar I = \{0, \ldots,n \} \setminus I$ is ``dual''
set and $\delta(I) = \pm 1$ is defined by relation
$\tau(I) \wedge \tau(\bar I)=  \delta(I) \tau(\{0, \ldots,n \})$,
the ``electric brane" living on $M_I$ (see (\ref{2.18}))
may be interpreted also as a ``magnetic brane'' living
on $M_{\bar I}$.  The relation (\ref{4.3.f2.9})
may be rewritten in the "electromagnetic" form
\beq{4.3.f2.25}
F^a = \sum_{I \in \Omega_{ae}} Q_{aIe} \tau(I)
+ \sum_{J \in \Omega_{am}} Q_{aJm} * \tau(J),
\eeq
where   $\Omega_{a} = \Omega_{ae} \cup  \Omega_{am}^{*}$,
$ \Omega_{ae} \cap  \Omega_{am}^{*} = \emptyset$,
$\Omega^{*} \equiv \{J| J = \bar I, I \in \Omega  \}$,
and $Q_{aIe} = Q_{aI}$ for $I \in \Omega_{ae}$   and
$Q_{aJm} = Q_{a \bar J}$ for $J \in \Omega_{am}$.

Here we consider some examples of the obtained solutions
when $\eps_0 = -1$ and all $\eps_i = 1$,
$i = 1, \ldots, n$,
i.e. ``our space'' $(M_0,g^0)$ is pseudo-Euclidean space and the
``internal spaces'' $(M_i,g^i)$ are Euclidean ones. We also put
$\theta_a =  1$ and $n_a < D -1$ for all $a \in \Delta$.

{\bf Solution with one $p$-brane.}
Let $\Omega_a = \{ I \}$,
$\lambda_{a} = 0$ for some  $a \in \Delta$ and
$\Omega_b$ are empty for all  $b \neq a$, $b \in \Delta$.

Equations (\ref{4.3.f2.11}) are satisfied identically in this case
and (\ref{4.3.f2.12}) read
\beq{4.3.f2.26}
\xi_i =    \frac{2 \Lambda}{D-2}  +
\eps(I)Q^{2} \left[ \delta^i_I - \frac{n_a -1}{D - 2} \right],
\eeq
$i = 0, \ldots, n$, where $Q = Q_{aI}$.

{\bf $p$-brane does not ``live'' in $M_0$.}
For $I = \{1, \ldots, k \}$,
$1 \leq k \leq n$, we get  $\eps(I) = 1$ and
$\xi_0 = \xi_{k+1} = \ldots =
\xi_n =  \lambda  - Q^{2} r_a$,
$\xi_1 =  \ldots =  \xi_{k} = \lambda + Q^{2} (1 - r_a)$,
where $\lambda  = 2 \Lambda/(D-2)$,
$r_a = (n_a -1)/(D - 2)$.

For $\Lambda = 0$, $Q \neq 0$ we get
$\xi_0 = \xi_{k+1} = \ldots = \xi_n < 0$ and
$\xi_1 =  \ldots =  \xi_{k} > 0$. These solutions
contain the solutions with the manifold
\beq{4.3.f2.29}
M = AdS_{d_0} \times S^{d_1} \times \ldots \times
S^{d_k} \times H^{d_{k+1}} \times \ldots \times M_n.
\eeq
Here $H^d$ is $d$-dimensional Lobachevsky space;
$M_n = H^{d_{n}}$ for $k < n$ and $M_n = S^{d_{n}}$ for $k = n$.

For $2\Lambda = Q^2 (n_a - 1)$ we get a solution with a flat
"our" space: $M = \R^{d_0} \times S^{d_1} \times \ldots \times
S^{d_k} \times \R^{d_{k+1}} \times \ldots$. One may consider
the fine-tuning of the cosmological constant, when
$\Lambda$ and $Q^2$ are of the Planck order but $\xi_0$ is small enough
in agreement with observational data.

{\bf $p$-brane  ``lives'' in $M_0$.}
For $I = \{0, \ldots, k \}$,
$0 \leq k \leq n$, we get  $\eps(I) = - 1$ and
$\xi_{k+1} = \ldots = \xi_n =  \lambda + Q^{2} r_a$,
$\xi_0 =  \ldots =  \xi_{k} = \lambda - Q^{2} (1 - r_a)$.

For $\Lambda = 0$, $Q \neq 0$, we get
$\xi_{k+1} = \ldots = \xi_n > 0$ and
$\xi_0 =  \ldots =  \xi_{k} < 0$. The solutions
contain the solutions with the manifold
\beq{4.3.f2.32}
M = AdS_{d_0} \times H^{d_1} \times \ldots \times
H^{d_k} \times S^{d_{k+1}} \times \ldots \times M_n.
\eeq
Here $M_n = S^{d_{n}}$ for $k < n$ and $M_n = H^{d_{n}}$ for $k = n$.

For $2 \Lambda = Q^2 (D -n_a - 1)$ we get a solution with a flat
our space: $M  = \R^{d_0} \times S^{d_1} \times \ldots \times
S^{d_k} \times \R^{d_{k+1}} \times \ldots$. We may also consider
the fine-tuning mechanism here.

{\bf Solution with two $p$-branes.}
Let $n = 1$, $d_0 = d_1 = n_a = d$,
$\Omega_a = \{ I_0 = \{ 0 \}, I_1 = \{ 1 \} \}$,
for some $a$ and other $\Omega_b$ are empty.
Denote $Q_0 = Q_{a I_0}$ and $Q_1 = Q_{a I_1}$.
For the field of form we get from (\ref{4.3.f2.9})
\beq{4.3.f2.33}
F^a = Q_0 \hat{\tau}_0 + Q_1 \hat{\tau}_1.
\eeq
When $\lambda_{a} \neq 0$ the equations (\ref{4.3.f2.11}) are satisfied
if and only if   $Q_0^2 = Q_1^2 = Q^2$.
Relations (\ref{4.3.f2.12}) read
$\xi_0 = \lambda  - Q^2 e^{2 \lambda_{a}(\varphi)}$,
$\xi_1 = \lambda + Q^2 e^{2 \lambda_{a}(\varphi)}$.

For $\Lambda = 0$ and  $Q \neq 0$ we get
the solution defined on the manifold $M = AdS_{d} \times S^{d}$.
For odd $d$ the form (\ref{4.3.f2.33}) is self-dual (see subsection 3.1).
The solution describes a composite $p$-brane configuration containing
$ AdS_{5} \times S^{5}$  solution in $II B$ supergravity
as a special case.

{\bf Near-horizon limit for $A_2$-dyon in $D = 11$
supergravity.}
Here we consider the dyon solution
(\ref{3.1.2.16}), (\ref{3.1.2.17}) in $D = 11$-supergravity.

Let $g^0 = dR \otimes dR + R^2 \hat{g}[S^2]$, $H = C
+ \frac{M}{R}$, where $C$ and $M$ are constants and
$g[S^2]$ is the metric on $S^2$. For $C = 1$ the
4-dimensional section of the metric describes extremely
charged Reissner-Nordstr\"om black hole of mass $M$ in the
region out of the horizon: $R > 0$.
Now we put $C = 0$ and $M = 1$ (i.e. the so-called ``near-horizon''
limit is considered). We get the solution
\bear{4.3.f2.38}
g= \hat{g}[AdS_2] + \hat{g}[S^2] + \hat{g}^2+ \hat{g}^3, \\
\label{4.3.f2.39}
F^a= \nu_1 \hat{\tau}[AdS_2] \wedge \hat{\tau}_2
+ \nu_2 \hat{\tau}[S^2] \wedge \hat{\tau}_2
\ear
defined on the manifold
\beq{4.3.f2.40}
M = AdS_{2}  \times S^2 \times M_{2}  \times M_{3}.
\eeq
Here
$g[AdS_2] = R^{-2} [dR \otimes dR  - dt \otimes dt]$
is the metric on $AdS_2$,
$(M_i,g^i)$  are Ricci-flat,
$i = 2,3$;  $\eps_2=+1$, $d_2=2$, $d_3=5$   and
$\nu_1^2 = \nu_2^2 =1$.

The solutions (\ref{4.3.f2.38})-(\ref{4.3.f2.40})
may be generalized to $B_D$-models in dimension $D \ge 12$ \cite{IMJ}
(see  (\ref{3.1.2.09})) with  ${\rm rank} F^a \in \{4,\dots, D-7 \}$,
$d_2=a-2$, $d_3=D-2-a$, and all scalar fields are zero.
For $M_2 = \R^2$ and  $M_3 = \R^2 \times M_4$,
the metric (\ref{4.3.f2.38}) may be obtained
also for the solution with two M2 branes and two M5 branes
\cite{DLP,BPS}.

\subsection{Quantum solutions.}

\subsubsection{ Wheeler--De Witt equation.}

Here we fix the gauge as follows
\beq{4.2.1}
\gamma_0-\gamma=f(X),  \quad  {\cal N} = e^f,
\eeq
where $f$: ${\cal M}\to{\bf R}$ is a smooth function. Then we obtain the
Lagrange system with the Lagrangian
\beq{4.2.3}
L_f=\frac\mu2\e^f{\cal G}_{\hat A\hat B}(X)
\dot X^{\hat A}\dot X^{\hat B}-\mu\e^{-f}V_w
\eeq
and the energy constraint
\beq{4.2.4}
E_f=\frac\mu2\e^f{\cal G}_{\hat A\hat B}(X)
\dot X^{\hat A}\dot X^{\hat B}+\mu\e^{-f}V_w=0.
\eeq

Using the standard prescriptions of (covariant and conformally
covariant)  quantization (see, for example,
\cite{IMZ,Mis,Hal}) we are led to the Wheeler-DeWitt (WDW) equation
\beq{4.2.5}
\hat{H}^f \Psi^f \equiv
\left(-\frac{1}{2\mu}\Delta\left[e^f{\cal G}\right]+
\frac{a}{\mu}R\left[e^f{\cal G}\right]
+e^{-f}\mu V_w \right)\Psi^f=0,
\eeq
where
\beq{4.2.6}
a=a_N= (N-2)/8(N-1),
\eeq
$N = n+l +|S|$.

Here $\Psi^f=\Psi^f(X)$ is the so-called ``wave function of the universe''
corresponding to the $f$-gauge (\ref{4.2.1}) and satisfying the relation
\beq{4.2.7}
\Psi^f= e^{bf} \Psi^{f=0}, \quad b = (2-N)/2,
\eeq
($\Delta[{\cal G}_1]$ and
$R[{\cal G}_1]$ denote the Laplace-Beltrami operator and the scalar
curvature corresponding to ${\cal G}_1$, respectively).

{\bf Harmonic-time gauge} The WDW equation (\ref{4.2.5}) for $f=0$
\beq{4.2.10}
\hat H\Psi\equiv\left(-\frac{1}{2\mu}\Delta[{\cal G}]+
\frac{a}{\mu}R[{\cal G}]+\mu V_w \right)\Psi=0,
\eeq
where
\beq{4.2.8}
R[{\cal G}]=-\sum_{s\in S}(U^s,U^s)-
\sum_{s,s'\in S}(U^s,U^{s'}).
\eeq
and
\beq{4.2.9}
\tri[{\cal G}]
=\e^{U(\sigma)}\frac\partial{\partial \sigma^A}\left(\hat G^{AB}
\e^{-U(\sigma)}\frac\partial{\partial \sigma^B}\right)
+\sum_{s\in S}\eps_s\e^{2U^s(\sigma)}
\left(\frac\partial{\partial\Phi^s}\right)^2,
\eeq
with  $U(\sigma)=\sum_{s\in S}U^{s}(\sigma)$.

\subsubsection{Quantum solutions with one curved factor space and
orthogonal $U^s$}

Here as in subsect. 4.2.2  we put $\Lambda=0$,
$\xi_1 \ne0, \quad \xi_2=\ldots=\xi_n=0$,
and $1 \notin I_s$,   $s  \in S$, i.e.
the space $M_1$  is curved and others are Ricci-flat and
all ``brane'' submanifolds  do not  contain $M_1$.
We also put orthogonality restriction on the vectors $U^s$:
$(U^s,U^{s'})= 0$ for $s \neq s'$ and $K_s = (U^s,U^s) \neq 0$
for all $s \in S$.
In this case the  potential (\ref{4.1.9}) reads
$V_w= \frac12w \xi_1d_1\e^{2U^1(\sigma)}$.
The truncated minisuperspace metric
(\ref{4.1.8}) may be diagonalized
by the linear transformation
$z^A=S^A{}_B \sigma^B$, $(z^A)=(z^1,z^a,z^s)$
as follows
\beq{4.2.24}
\hat G=-dz^1\otimes dz^1+
\sum_{s\in S}\eta_sdz^s\otimes dz^s+dz^a\otimes dz^b\eta_{ab},
\eeq
where $a,b=2,\dots,n +l-|S|$; $\eta_{ab} =\eta_{aa} \delta_{ab};
\eta_{aa}= \pm 1$, $\eta_s = {\rm sign}(U^s,U^s)$ and
$q_1 z^1=U^1(\sigma)$,
$q_1\equiv\sqrt{|(U^1,U^1)|} = \sqrt{1 - d_1^{-1}}$,
$q_sz^s=U^s(\sigma)$,
$q_s=\nu_s^{-1} \equiv \sqrt{|(U^s,U^s)|}$
$s=(a_s,v_s,I_s) \in S$.

We are seeking the solution to WDW equation (\ref{4.2.10}) by the method
of the separation of variables, i.e. we put
\beq{4.2.30}
\Psi_*(z)=\Psi_1 (z^1)\left(\prod_{s\in S}\Psi_s(z^s)\right)
\e^{\im P_s\Phi^s}\e^{\im p_az^a}.
\eeq
It follows from  the relation for the Laplace operator
\bear{4.2.27}
 \tri[{\cal G}]=-\left(\frac\partial{\partial z^1}\right)^2+
\eta^{ab}\frac\partial{\partial z^a}\frac\partial{\partial z^b}+
\sum_{s\in S}\eta_s\e^{q_sz^s}\frac\partial{\partial z^s}
\left(\e^{-q_sz^s}\frac\partial{\partial z^s}\right) \\ \nn
+\sum_{s\in  S} \eps_s
\e^{2q_sz^s}\left(\frac\partial{\partial\Phi^s}\right)^2.
\ear
that $\Psi_*(z)$ satisfies WDW equation
(\ref{4.2.10}) if
\bear{4.2.31}
2\hat H_1\Psi_1 \equiv
\left\{\left(\frac\partial{\partial z^1}\right)^2
+\mu^2 w\xi_1d_1\e^{2q_1z^1}\right\}\Psi_1=2{\cal E}_1\Psi_1; \\
\label{4.2.32}
2\hat H_s\Psi_s\equiv\left\{-\eta_s\e^{q_sz^s}\frac\partial{\partial z^s}
\left(\e^{-q_sz^s}\frac\partial{\partial z^s}\right)+
\eps_sP_s^2\e^{2q_sz^s}\right\}\Psi_s=2{\cal E}_s\Psi_s,
\ear
$s\in S$, and
\beq{4.2.33}
2{\cal E}_1+\eta^{ab}p_ap_b+2\sum_{s\in S}{\cal E}_s+
2aR[{\cal G}]=0,
\eeq
where $a$ is from (\ref{4.2.6}) and
$R[{\cal G}]=-2\sum_{s\in S}(U^s,U^s)=
-2\sum_{s\in S}\eta_sq_s^2$.

Using the relations from Appendix 5 we obtain linearly independent
solutions to (\ref{4.2.31}) and (\ref{4.2.32}) respectively
\bear{4.2.24n}
\Psi_1(z^1)=B_{\omega_1}^1
\left(   \sqrt{-w \mu^2 \xi_1d_1}\frac{\e^{q_1z^1}}{q_1}\right), \\
\label{4.2.25n}
\Psi_s(z^s)=\e^{q_sz^s/2}B_{\omega_s}^s
\left(\sqrt{\eta_s\eps_sP_s^2}\frac{\e^{q_sz^s}}{q_s}\right),
\ear
where
$\omega_1=\sqrt{2{\cal E}_1}/q_1$,
$\omega_s=\sqrt{\frac14- 2\eta_s{\cal E}_s\nu_s^2}$,
$s\in S$, and $B_\omega^1,B_\omega^s=I_\omega,K_\omega$
are the modified Bessel function.

The general solution of the WDW equation (\ref{4.2.10}) is a superposition
of the "separated" solutions (\ref{4.2.30}):
\beq{4.2.28a}
\Psi(z)=\sum_B\int dpdPd{\cal E} C(p,P,{\cal E},B)
\Psi_*(z|p,P,{\cal E},B),
\eeq
where $p=(p_a)$, $P=(P_s)$, ${\cal E}=({\cal E}_s, {\cal E}_1)$,
$B=(B^1,B^s)$,  $B^1,B^s=I,K$; and
$\Psi_*=\Psi_*(z|p,P,{\cal E},B)$ is given by relation
(\ref{4.2.30}), (\ref{4.2.24n})--(\ref{4.2.25n}) with ${\cal E}_1$ from
(\ref{4.2.33}). Here $C(p,P,{\cal E},B)$ are smooth enough functions.
In non-composite electric case these solutions were considered
in \cite{GrIM}.

\subsubsection{ WDW equation with fixed charges}

We may consider also another scheme based on zero-energy
constraint relation (\ref{3.2.18}). The corresponding WDW
equation in the harmonic gauge reads
\beq{4.2.34}
\hat{H}_Q \Psi \equiv
\left(-\frac{1}{2\mu}\ {\hat G}^{AB}
\frac{\partial}{\partial x^A} \frac{\partial}{\partial x^B}
+\mu V_Q \right) \Psi=0,
\eeq
where potential $V_Q$ is defined in (\ref{4.1.19}). This equation
describes quantum cosmology with classical fields of forms
and quantum scale factors and dilatonic fields. Such approach
is equivalent to the  scheme of quantization of multidimensional
perfect fluid considered in \cite{IM5}.

Eq. (\ref{4.2.34}) is readily solved in the orthogonal case
(see Appendix 5).  The solutions are given by the following
modifications in eqs. (\ref{4.2.33}), (\ref{4.2.25n}),
respectively,
\bear{4.2.33m}
2{\cal E}_1+\eta^{ab}p_ap_b+2\sum_{s\in S}{\cal E}_s=0,
\\ \label{4.2.25m}
\Psi_s(z^s)  = B_{\bar \omega_s}^s
\left(\sqrt{\eta_s\eps_s Q_s^2}\frac{\e^{q_sz^s}}{q_s}\right),
\ear
where  $\bar \omega_s = \sqrt{- 2\eta_s{\cal E}_s\nu_s^2}$.

This solution for the special case
with one internal space ($n = 1$) and non-composite
$p$-branes was considered in \cite{LMMP}.

\newpage

\section{Black hole solutions}

\subsection{Solutions with a horizon}

Here we consider the spherically symmetric case
of the metric (\ref{4.3.19}), i.e. we put
$w = 1, \quad M_1 = S^{d_1}$,  $g^1 = d \Omega^2_{d_1}$,
where $d \Omega^2_{d_1}$ is the canonical metric on a unit sphere
$S^{d_1}$, $d_1 \geq 2$. In this case $\xi^1 = d_1 -1$.
We put  $M_2 = \R$, $g^2 = - dt \otimes dt$,
i.e.  $M_2$ is a time manifold. We also assume that
$(U^s,U^s) \neq 0$, $s \in S$, and
\beq{5.2.2a}
{\rm det}((U^s,U^{s'})) \neq 0.
\eeq

We put $C_1 \geq 0$.
In this case relations (\ref{4.3.10})-(\ref{4.3.13}) read
$f_1(u) = \bar{d} C_1^{-1/2} \sh(C_1^{1/2} u)$, for $C_1>0$,
and $f_1(u) =\bar du$,  for $C_1=0$.
Here and in what follows $\bar{d} = d_1-1$.

Let us consider the null-geodesic equations
for the light ``moving'' in the radial direction
(following from $ds^2 =0$):
\beq{5.2.4}
\pm \frac{dt}{du} = \Phi
\equiv  f_1^{d_1/(1-d_1)} e^{(c^1 - c^2) u +  \bar c^1 - \bar c^2}
\prod_{s\in S} f_s^{- h_s  \delta_{2 I_s}}.
\eeq
We consider   solutions
defined on some interval $[u_0, +\infty)$ with a  horizon
at $u = + \infty$ satisfying
\beq{5.2.6}
\int_{u_0}^{ + \infty} d  u  \Phi( u) = + \infty.
\eeq

When the matrix $(h_{\alpha\beta})$ is positive
definite  and
\beq{5.2.18}
2 \in I_s, \quad \forall s \in S,
\eeq
i. e. all p-branes have a common time direction $t$,
the horizon condition (\ref{5.2.6}) singles out the unique
solution with $C_1 > 0$ and linear asymptotics at infinity
\beq{5.2.7}
q^s = - \beta^s u + \bar \beta^s  + o(1),
\eeq
$u \to +\infty$, where $\beta^s, \bar \beta^s$ are
constants, $s \in S$, \cite{IMp2,IMp3}.

In this case
\bear{5.2.19}
c^A/\bar{\mu}  = - \delta^{A}_{2} + h_1 U^{1 A}  +
\sum_{s\in S}  h_s b_s U^{s A},  \\ \label{5.2.20}
\beta^s/\bar{\mu} = 2 \sum_{s' \in S} A^{s s'} \equiv b_s,
\ear
where $s \in S$, $A = (i, \alpha)$,
$\bar{\mu} = \sqrt{C_1}$,
the matrix $(A^{s s'})$ is inverse to the
quasi-Cartan matrix  $(A_{s s'})$
and  $h_1 = (U^1, U^1)^{-1} = d_1/(1 - d_1)$.
According to Proposition 1 from \cite{IMp2,IMp3}
the condition (\ref{5.2.18}) is also a necessary condition
for the existence of the horizon at $u \to +\infty$ under
the assumptions assumed: if there exists at least one
brane not containing the time submanifold $\{ t \}$, then
the horizon with reaspect to time $t$ at $u \to +\infty$ is absent.

{\bf Remark 9.} {\em  According to  Lemma 2 from \cite{Br2}
black hole solutions can only exist for
$C_{1} \geq 0$ and the horizon is then at
$u = \infty$. For the extremal case
$C_{1} = 0$ see Subsect. 5.5.
below.}

Let us introduce a new radial variable $R = R(u)$ by relations
\bear{5.2.28}
\exp( - 2\bar{\mu} u) = 1 - \frac{2\mu}{R^{\bar{d}}},
\qquad
\mu = \bar{\mu}/ \bar{d} >0,
\ear
where $u > 0$, $R^{\bar d} > 2\mu$ ($\bar d = d_1 -1$).
We put  $\bar{c}^A = 0$ and
\bear{5.2.27f}
q^s(0) = 0,
\ear
$A = (i, \alpha)$, $s \in S$.
These relations guarantee the asymptotical flatness
(for $R \to +\infty$) of the $(2+d_1)$-dimensional section of the metric.

Let us denote
\beq{5.2.28a}
H_s = f_s e^{- \beta^s u },
\eeq
$s \in S$.
Then,  solutions (\ref{4.3.19})-(\ref{4.3.19d}) may be written
as follows \cite{IMp1,IMp2,IMp3}
\bear{5.2.30}
g= \Bigl(\prod_{s \in S} H_s^{2 h_s d(I_s)/(D-2)} \Bigr)
\biggl\{ \left(1 - \frac{2\mu}{R^{\bar{d}}}\right)^{-1} dR \otimes dR
+ R^2  d \Omega^2_{d_1}  \\ \nn
-  \Bigl(\prod_{s \in S} H_s^{-2 h_s} \Bigr)
\left(1 - \frac{2\mu}{R^{\bar{d}}}\right) dt \otimes dt
+ \sum_{i = 3}^{n} \Bigl(\prod_{s \in S}
  H_s^{-2 h_s \delta_{iI_s}} \Bigr) \hat{g}^i  \biggr\},
\\  \label{5.2.31}
\exp(\varphi^\alpha)=
\prod_{s\in S} H_s^{h_s \chi_s \lambda_{a_s}^\alpha},
\ear
where $F^a= \sum_{s \in S} \delta^a_{a_s} {\cal F}^{s}$,
and
\beq{5.2.32}
{\cal F}^s= - \frac{Q_s}{R^{d_1}}
\left( \prod_{s' \in S}  H_{s'}^{- A_{s s'}} \right) dR \wedge\tau(I_s),
\eeq
$s\in S_e$,
\beq{5.2.33}
{\cal F}^s= Q_s \tau(\bar I_s),
\eeq
$s\in S_m$.
Here $Q_s \neq 0$, $h_s =K_s^{-1}$, $s \in S$, and the
quasi-Cartan matrix $(A_{s s'})$ is non-degenerate.

Functions $H_s > 0$ obey the equations
\beq{5.2.34}
R^{d_1} \frac{d}{dR} \left[ \left(1 - \frac{2\mu}{R^{\bar{d}}}\right)
\frac{ R^{d_1} }{H_s}
\frac{d H_s}{dR} \right] = B_s
\prod_{s' \in S}  H_{s'}^{- A_{s s'}},
\eeq
$s \in S$, where $B_s = \eps_s K_s Q_s^2 \neq 0$.
These equations follow from Toda-type equations (\ref{3.2.35}) and
the definitions   (\ref{5.2.28}) and   (\ref{5.2.28a}).

It follows from (\ref{5.2.7}), (\ref{5.2.28})
and  (\ref{5.2.28a}) that there exist finite limits
\beq{5.2.35a}
H_s  \to H_{s0} \neq 0,
\eeq
for $R^{\bar d} \to 2\mu$, $s \in S$.  In this case the metric (\ref{5.2.30})
has a regular horizon at  $R^{\bar{d}} =   2 \mu$.
From (\ref{5.2.27f})  we get
\beq{5.2.35}
H_s (R = +\infty) = 1,
\eeq
$s \in S$.

The Hawking "temperature" corresponding to
the solution is (see also \cite{Oh,BIM} for orthogonal case)
found to be
 \beq{5.2.36}
T_H=   \frac{\bar{d}}{4 \pi (2 \mu)^{1/\bar{d}}}
\prod_{s \in S} H_{s0}^{- h_s},
\eeq
where $H_{s0}$ are defined in (\ref{5.2.35a})

The boundary conditions (\ref{5.2.35a}) and  (\ref{5.2.35})
play a crucial role here, since they
single out, generally speaking, only few solutions
to eqs. (\ref{5.2.34}).
Moreover, for some values of parameters $\mu = \bar \mu / \bar d$, $\eps_s$
and $Q_s^2$ the solutions to eqs. (\ref{5.2.34})-(\ref{5.2.35})
do not exist \cite{IMp2}.

Thus, we obtained a general family of black hole solutions
defined up to solutions of radial equations (\ref{5.2.34})
with the boundary conditions (\ref{5.2.35a}) and  (\ref{5.2.35}).
In the next sections we consider several exact solutions
to eqs. (\ref{5.2.34})-(\ref{5.2.35}).

{\bf Remark 10.} {\em Let $M_i = \R$ and $g^i = -d \bar t \otimes d\bar t$
for some $i \geq 3$. Then the metric (\ref{5.2.30}) has no a horizon
with respect to the ``second time''  $\bar t$ for $R^{\bar{d}} \to 2\mu$.
Thus, we are led to a ``single-time'' theorem from \cite{Br2}.
Relation (\ref{5.2.18})  coincides with
the ``no-hair'' theorem from \cite{Br2}. }

\subsection{Polynomial structure of $H_s$ for  Lie algebras}

\subsubsection{Conjecture on polynomial structure}

Now we deal  with solutions to second order non-linear
differential equations  (\ref{5.2.34}) that may be rewritten
as follows
\beq{5.3.1}
 \frac{d}{dz} \left( \frac{(1 - 2\mu z)}{H_s}
 \frac{d}{dz} H_s \right) = \bar B_s
\prod_{s' \in S}  H_{s'}^{- A_{s s'}},
\eeq
where $H_s(z) > 0$, $\mu > 0$,
$z = R^{-\bar d} \in (0, (2\mu)^{-1})$ and
$\bar B_s =  B_s/ \bar d^2 \neq 0$.
Eqs. (\ref{5.2.35a}) and  (\ref{5.2.35}) read
\bear{5.3.2a}
H_{s}((2\mu)^{-1} -0) = H_{s0} \in (0, + \infty), \\
\label{5.3.2b}
H_{s}(+ 0) = 1,
\ear
$s \in S$.

It seems rather difficult to find the solutions to a set
of eqs. (\ref{5.3.1})-(\ref{5.3.2b}) for arbitrary
values of parameters $\mu$, $\bar B_s$, $s \in S$, and
quasi-Cartan matrices $A =(A_{s s'})$. But we may
expect a drastical simplification of the problem
under consideration for certain class of parameters and/or
$A$-matrices.

In general we may try to seek solutions of (\ref{5.3.1})  in a class
of functions analytical in a disc $|z| < L$ and continuous
in semi-interval $0 < z \leq (2\mu)^{-1}$. For $|z| < L$
we get
\beq{5.3.3}
H_{s}(z) = 1 + \sum_{k = 1}^{\infty} P_s^{(k)} z^k,
\eeq
where $P_s^{(k)}$ are constants, $s \in S$. Substitution
of (\ref{5.3.3})  into (\ref{5.3.1}) gives us an infinite
chain of relations on parameters $P_s^{(k)}$  and
$\bar B_s$.  In general case it seems to be impossible
to solve this chain of equations.

Meanwhile there exist solutions to eqs. (\ref{5.3.1})-(\ref{5.3.2b})
of polynomial type. The simplest example occurs in orthogonal
case \cite{CT,AIV,Oh,IMJ,BIM}:
$(U^s,U^{s'})= 0$, for  $s \neq s'$, $s, s' \in S$. In this case
$(A_{s s'}) = {\rm diag}(2,\ldots,2)$ is a Cartan matrix
for semisimple Lie algebra
${\bf A_1} \oplus  \ldots  \oplus  {\bf A_1}$
and
\beq{5.3.5}
H_{s}(z) = 1 + P_s z,
\eeq
with $P_s \neq 0$, satisfying
\beq{5.3.5a}
P_s(P_s + 2\mu) = -\bar B_s,
\eeq
$s \in S$.

In \cite{Br1,IMJ2,CIM} this solution was generalized to a
block-orthogonal case  (\ref{3.1.5}), (\ref{3.1.6}). In this case
(\ref{5.3.5}) is modified as follows \beq{5.3.8} H_{s}(z) = (1 +
P_s z)^{b_s}, \eeq where $b_s$ are defined in  (\ref{5.2.20}) and
parameters $P_s$  and are coinciding inside blocks, i.e. $P_s =
P_{s'}$ for $s, s' \in S_i$, $i =1,\dots,k$. Parameters $P_s \neq
0 $ satisfy the relations

 $$P_s(P_s + 2\mu) = - \bar B_s/b_s, $$

 $b_s \neq 0$, and parameters $\bar B_s/b_s$  are also  coinciding
inside blocks, i.e. $\bar B_s/b_s = \bar B_{s'}/b_{s'}$ for $s, s'
\in S_i$, $i =1,\dots,k$. In this case $H_s$ are analytical in the
disc $|z| < L$, where $L = {\rm min} (| P_ s|^{-1}, s \in S$).

Let $(A_{s s'})$ be  a Cartan matrix  for a  finite-dimensional
semisimple Lie  algebra $\cal G$. In this case all powers in
(\ref{5.2.20})  are  natural numbers  coinciding with the components
of twice the  dual Weyl vector in the basis of simple coroots
\cite{FS} (see Appendix 3)
and  hence, all functions $H_s$ are polynomials, $s \in S$.

{\bf Conjecture 1.} {\em Let $(A_{s s'})$ be  a Cartan matrix
for a  semisimple finite-dimensional Lie algebra $\cal G$.
Then  the solution to eqs. (\ref{5.3.1})-(\ref{5.3.2b})
(if exists) is a polynomial
\beq{5.3.12}
H_{s}(z) = 1 + \sum_{k = 1}^{n_s} P_s^{(k)} z^k,
\eeq
where $P_s^{(k)}$ are constants,
$k = 1,\ldots, n_s$; $n_s = b_s = 2 \sum_{s' \in S} A^{s s'} \in \N$
and $P_s^{(n_s)} \neq 0$,  $s \in S$}.

In extremal case ($\mu = + 0$) an a analogue of this conjecture
was suggested previously in \cite{LMMP}.
Conjecture 1 was verified for ${\bf A_m}$ and ${\bf C_{m+1}}$ series of Lie
algebras in \cite{IMp2,IMp3}.

\subsection{Examples}

\subsubsection{Solution for $A_2$}

Here we consider solutions related
to the Lie algebra ${\bf A_2} = sl(3)$. According to the
results of previous section we  seek the solutions
to eqs. (\ref{5.3.1})-(\ref{5.3.2b}) in the following
form (see  (\ref{5.3.12}); here $n_1 = n_2 =2$):
\beq{5.4.1}
H_{s} = 1 + P_s z + P_s^{(2)} z^{2},
\eeq
where $P_s= P_s^{(1)}$ and $P_s^{(2)} \neq 0$ are constants,
$s = 1,2$.

From (\ref{5.3.1}) we get for $P_1 +P_2 + 4\mu \neq 0$ \cite{IMp1}
\bear{5.4.5}
P_s^{(2)} = \frac{ P_s P_{s +1} (P_s + 2 \mu )}{2 (P_1 +P_2 + 4\mu)},
\qquad
\bar B_s = - \frac{ P_s (P_s + 2 \mu )(P_s + 4 \mu )}{P_1 +P_2 + 4\mu},
\ear
$s = 1,2$.
Here we denote  $s+ 1 = 2, 1$ for $s = 1,2$, respectively.
For $P_1 +P_2 + 4\mu = 0$ we get a special (exceptional) solution with
$P_1= P_2 = -2 \mu$, $2 P_{s}^{(2)} = \bar B_s >0$ and
$\bar B_1 + \bar B_2 = 4 \mu^2$.

Thus, in the ${\bf A_2}$-case the non-exceptional solution is
described by relations  (\ref{5.2.30})-(\ref{5.2.33}) with $S =
\{s_1,s_2\}$ (identified with $\{1,2 \})$, intersection rules
\ber{5.4.7}
d(I_{s_1}\cap I_{s_2})=\Delta(s_1,s_2) - K,
\eer
where  symbol $\Delta(s_1,s_2)$ is defined in (\ref{3.1.30}) and
$K = K_{s_i} = (U^{s_i},U^{s_i})\neq 0$; functions $H_{s_i} = H_i$
are defined by relations   (\ref{5.4.1}) and (\ref{5.4.5})
with $z = R^{-\bar d}$, $i =1,2$.

\subsubsection{$A_2$-dyon in $D = 11$ supergravity}

Consider the  "truncated"  bosonic sector of
$D= 11$ supergravity  with the action  (\ref{3.1.31}).
Let us consider a dyonic black-hole solutions
with  electric $2$-brane and magnetic  $5$-brane
defined on the manifold
\beq{5.4.8}
M =  (2\mu, +\infty )  \times
(M_1 = S^{2})  \times (M_2 = \R) \times M_{3} \times M_{4},
\eeq
where ${\dim } M_3 =  2$ and ${\dim } M_4 =  5$. The solution reads,
\bear{5.4.9}
g=  H_1^{1/3} H_2^{2/3} \biggl\{ \frac{dR \otimes dR}{1 - 2\mu / R} +
R^2  d \Omega^2_{2} \\ \nn
 -  H_1^{-1} H_2^{-1} \left(1 - \frac{2\mu}{R} \right) dt\otimes dt
+ H_1^{-1} \hat{g}^3 + H_2^{-1} \hat{g}^4 \biggr\}, \mm
\label{5.4.10}
F =  - \frac{Q_1}{R^2} H_1^{-2} H_2  dR \wedge dt \wedge \hat{\tau}_3+
Q_2 \hat{\tau}_1 \wedge \hat{\tau}_3,
\ear
where metrics $g^2$ and  $g^3$ are
Ricci-flat metrics of Euclidean signature,
and $H_s$  are defined
by (\ref{5.4.1}),
where $z = R^{- 1}$ and parameters
$P_s$,  $P_s^{(2)}$, $\bar B_s = B_s = - 2 Q_s^2$,
$s =1,2$, satisfy  (\ref{5.4.5}).

The  solution describes ${\bf A_2}$-dyon consisting
of electric  $2$-brane with worldvolume isomorphic
to $(M_2 = \R) \times M_{3}$ and magnetic  $5$-brane
with worldvolume isomorphic to $(M_2 = \R) \times M_{4}$.
The ``branes'' are intersecting on the time manifold $M_2 = \R$.
Here  $K_s = (U^s,U^s)=2$, $\eps_s = -1$ for all $s \in S$.
The ${\bf A_2}$ intersection rule reads: $3 \cap 6= 1$.

The field configurations  (\ref{5.4.9}), (\ref{5.4.10})
also satisfies to equations of motion for  $D =11$ supergravity
(see  (\ref{3.1.40}),  (\ref{3.1.41}); in this case $F \wedge F = 0$ ).

This solution in a special case $H_1 = H_2 = H^2$
($P_1 = P_2$, $Q_1^2 =  Q_2^2$) was considered in \cite{CIM}.
The 4-dimensional section of the metric (\ref{5.4.9})
in this special case coincides   with the Reissner-Nordstr\"om  metric.
For the extremal case, $\mu \to + 0$,  and multi-black-hole generalization
see also  \cite{IMBl}.

\subsection{Extremal case}

\subsubsection{"One-pole" solution}

Here we consider the extremal case: $\mu \to +0$.
The relation  for the metric (\ref{5.2.30}) reads in this
case as follows \cite{IMp2}
\bear{5.6.1}
g= \Bigl(\prod_{s \in S} H_s^{2 h_s d(I_s)/(D-2)} \Bigr)
\biggl\{ dR \otimes dR  + R^2  d \Omega^2_{d_1}  \\ \nn
-  \Bigl(\prod_{s \in S} H_s^{-2 h_s} \Bigr) dt \otimes dt
+ \sum_{i = 3}^{n} \Bigl(\prod_{s\in S}
  H_s^{-2 h_s \delta_{iI_s}} \Bigr) \hat{g}^i  \biggr\},
\ear
and the relations for scalar fields and  fields
of forms (\ref{5.2.31})-(\ref{5.2.33}) are unchanged.
Here $H_s =  H_s(z) > 0$, $z = R^{-\bar d} \in (0,+\infty)$ and
the following relations  are satisfied
\bear{5.6.2}
 \frac{d^2}{dz^2} \ln H_s =
 \bar B_s \prod_{s' \in S}  H_{s'}^{- A_{s s'}},
 \qquad H_{s}(+ 0) = 1, \\
\label{5.6.3}
E_{TL} = \frac{\bar{d}^2}{4}  \sum_{s,s' \in S} h_s
A_{s s'} (\frac{d}{dz}{\ln H_s}) \frac{d}{dz}{\ln H_{s'}}
+  \sum_{s \in S}   A_s \prod_{s' \in S}  H_{s'}^{- A_{s s'}} =0,
\ear
where $\bar B_s =  B_s/ \bar d^2 \neq 0$, and
$B_s = 2K_sA_s$, $A_s=  \frac12  \eps_s Q_s^2$,
$s \in S$.
These solution may be  obtained as a special case of
solutions from Subsect. 4.2.2 with
\beq{5.6.4}
C_1 =  E_{TL} = c^A = 0,
\eeq
$A = (i,\alpha)$ and $u = z/ \bar d$ ($H_s = f_s$, see (\ref{5.2.28a})).

{\bf Conjecture 2.} {\em Let $(A_{s s'})$ be  a Cartan matrix  for a
simple finite-dimensional Lie algebra. Then
there exists a  polynomial solution
\beq{5.6.5}
H_{s}(z) = 1 + \sum_{k = 1}^{n_s} P_s^{(k)} z^k,
\eeq
to eqs. (\ref{5.6.2})-(\ref{5.6.3}) for $\bar B_s < 0$, $s \in S$,
where $P_s^{(k)}$ are constants,
$k = 1,\ldots, n_s$; numbers $n_s = b_s$ are
defined in (\ref{5.2.20}) and $P_s^{(n_s)} \neq 0$, $s \in S$.}

This conjecture  was considered (in fact) previously in
\cite{LMMP}.  The coefficients $P_s^{(n_s)} = C_s > 0 $ may be calculated
by substitution of asymptotical relations
\beq{5.6.6}
H_{s}(z) \sim  C_s z^{b_s},
\quad
z \to + \infty,
\eeq
into eqs.  (\ref{5.6.2}), (\ref{5.6.3}), $s \in S$. This results in
the relations
\beq{5.6.7}
C_s = \prod_{s' \in S} ( - b_0^{s'} \bar B_{s'})^{A^{s s'}},
\eeq
$s \in S$.
We note that the asymptotical relations (\ref{5.6.6})
are satisfied in a more general case, when
$\bar{B}_s b_s < 0$, $s \in S$.

Let us consider  the metric (\ref{5.6.1}) with $H_s$
obeying asymptotical relations (\ref{5.6.6}).
We have a horizon for $R  \to +0$, if
\bear{5.6.8}
\xi = \sum_{s\in S} h_s b_s - \frac1{d_0-2}\ge0,
\ear
where $d_0 = d_1 +1$.
This relation follows from the requirement of infinite time
propagation of light to $R \to +0$.

For flat internal spaces  $M_i=\R^{d_i}$, $i=3,\dots,n$,
we get for the Riemann tensor squared (Kretschmann scalar)
from Appendix 1
\beq{5.6.9}
{\cal K}[g]= [C + o(1)] R^{4(d_0-2)\eta}
\eeq
for $R \to +0$, where
\ber{6.10}
\eta = \sum_{s \in S} h_s b_s \frac{d(I_s)}{D-2}-
\frac{1}{d_0-2},
\eer
and $C \geq 0$ ($C = {\rm const}$).
Due to (\ref{5.6.9}) the metric (\ref{5.6.1})
with flat internal spaces has no
curvature singularity when $R \to + 0$, if
\bear{5.6.11}
\eta \ge 0.
\ear

For $h_s b_s > 0$, $d(I_s)<D-2$, $s \in S $, we get
$\eta < \xi$ and relation (\ref{5.6.11}) single out extremal
charged  black  $p$-branes and with flat internal spaces.

\subsubsection{Multi-black-hole extension}

The solutions under consideration have a Majumdar-Papapetrou-type
extension defined on the manifold
\beq{5.6.12a}
M =    M_0  \times (M_2 = \R) \times  \ldots \times M_{n},
\eeq
The solution reads
\bear{5.6.13}
g= \Bigl(\prod_{s \in S} H_s^{2 h_s d(I_s)/(D-2)} \Bigr)
\biggl\{ \hat{g}^0 -
\Bigl(\prod_{s \in S} H_s^{-2 h_s} \Bigr)  dt \otimes dt
+ \sum_{i = 3}^{n} \Bigl(\prod_{s\in S}
  H_s^{-2 h_s \delta_{iI_s}} \Bigr) \hat{g}^i  \biggr\},
\\  \label{5.6.14}
\exp(\varphi^\alpha)=
\prod_{s\in S} H_s^{h_s \chi_s \lambda_{a_s}^\alpha},
\\  \label{5.6.15}
F^a= \sum_{s \in S} \delta^a_{a_s} {\cal F}^{s},
\ear
where
\bear{5.6.16}
{\cal F}^s= Q_s
( \prod_{s' \in S}  H_{s'}^{- A_{s s'}}) dH \wedge\tau(I_s),
\quad s\in S_e,
\\ \label{5.6.17}
{\cal F}^s= Q_s (*_0 d H) \wedge \tau(\hat I_s),
\quad s\in S_m.
\ear
Here $\hat I \equiv \{2,\ldots,n\}\setminus I$,
$g^0=g_{\mu\nu}^0(x)dx^\mu\otimes dx^\nu$
is a Ricci-flat metric on $M_0$ and $*_0 = *[g^0]$ is the
Hodge operator on $(M_0,g^0)$ and
\beq{5.6.18}
H_s =  H_s(H(x)),
\eeq
where functions $H_s =  H_s(z) > 0$, $z \in (0, +\infty)$, $s\in S$,
satisfy the relations (\ref{5.6.2}) and  (\ref{5.6.3}) and
$H = H(x) > 0$ is a harmonic function on $(M_0,g^0)$, i.e. $\tri[g^0]H=0$.
This solution is a special case of the solutions
from Subsect. 3.2.2
corresponding to restrictions (\ref{5.6.4}).

Let us consider as an example a flat space:
$M_0=\R^{d_0} \setminus X$, $d_0>2$, and
$g^0=\delta_{\mu\nu}dx^\mu\otimes dx^\nu$  and
\beq{5.6.20}
H(x) = \sum_{b \in X}\frac{q_{b}}{|x- b|^{d_0-2}},
\eeq
where $X$ is finite non-empty subset $X \subset M_0$ and
all $q_{b}>0$ for $b \in X$. For flat internal spaces  $M_i=\R^{d_i}$,
$i=3,\dots,n$, and non-negative indices $\eta$ and $\xi$ (see (\ref{5.6.8})
and (\ref{5.6.11}))  the solution describes
a set of $|X|$ extremal $p$-brane black holes.
Here relations $H(x) \to 0$ for $|x| \to + \infty$
and $H_{s}(+ 0) = 1$ ($s \in S$) imply
the asymptotical flatness of the $(1+d_0)$-dimensional section of the
metric. A black hole corresponding to a ``point'' (horizon) $b \in X$
carries brane charges $Q_s q_b$, $s \in S$.  Since the solution is
invariant under the replacement  of parameters:  $Q_s \mapsto \alpha Q_s $,
$q_b \mapsto  Q_s/\alpha $, $\alpha >0$, $b \in X$, $s \in S$,
we may normalize parameters $q_b$ by the restriction
$\sum_{b \in X} q_{b} = 1$.

\newpage

\section{Conclusions and discussions}

Here we reviewed several rather general families of exact solutions
in multidimensional gravity with a set of scalar fields and fields
of forms. These solutions describe composite non-localized
electromagnetic $p$-branes defined on products of {\em Ricci-flat}
(or sometimes Einstein) spaces of {\em arbitrary signatures}.
The metrics are block-diagonal and
all scale factors, scalar fields and fields of forms depend
on points of some (mainly Ricci-flat) manifold $M_0$.
The solutions include those depending upon harmonic functions,
cosmological and spherically-symmetric solutions
(e.g. black-brane ones) and static configurations
on product of Einstein spaces (i.e. generalizations of Freund-Rubin
solution). Our scheme is based on the {\em sigma-model representation}
obtained in \cite{IMC} (for special configurations see
\cite{IM11,IM12, IMR}) under the rather general assumption on
intersections of composite $p$-branes: namely, when
stress-energy tensor $T^M_N$ has a diagonal structure.
Here we presented three types of {\em restrictions}:
 electric-electric,  magnetic-magnetic
and  electric-magnetic, which give a sufficient
condition for the diagonality of $T^M_N$. In \cite{IMC}
we obtained general and sufficient condition in terms
of a set of constraints on all moduli scalar fields.

In this review (as in other authors publications) we
were dealing with the most {\em general intersection
rules}. Here we showed that in general intersection rules
correspond to the {\em quasi-Cartan} matrix related to
certain "brane" vectors $U^s$ belonging to some vector space
(dual to truncated minisuperspace). A major part of well-known
$p$-brane solutions have intersection rules equivalent to
the orthogonality condition: $(U^s,U^{s'}) =0$, $s \neq s'$,
where $(.,.)$ is the bilinear form dual to truncated minisupermetric
(first it was understood in \cite{IM11}). The
{\em quasi-Cartan} matrix in this case is nothing more
than  the Cartan matrix for the semisimple Lie algebra
${\bf A_1} \oplus   \dots \oplus {\bf A_1}$ ($r$ terms).
For Majumdar-Papapetrou type
solutions we got $r$ independent harmonic functions on $M_0$.
Solutions of such class  contain  a large variety of
supersymmetric (BPS saturated) solutions in
supergravitational models. Here we also considered a more general
class of "block-orthonal" solutions governed by harmonic
functions and also considered a generalization in a one-block
case governed by several functions (of one harmonic function)
obeying Toda-type equations  (e.g. Euclidean Toda lattices).
The $p$-brane solutions may be considered as a nice tool
for utilization of Euclidean and hyperbolic Toda lattices
related to finite dimensional Lie algebras and
hyperbolic Kac-Moody (KM) algebras.
Here
we considered also the Wheeler-DeWitt (WDW) equation
for $p$-brane cosmology in d'Alembertian (covariant)
and conformally covariant form
and integrated it for orthogonal $U$-vectors.
The open problem is to find the solutions of WDW equation for
more general $p$-brane configurations.

We also considered general classes of "cosmological"
and spherically symmetric solutions governed by Toda-type
equations, e.g. black brane configurations. An interesting
point here is the appearance of polynomials for black hole
solutions for brane intersections governed by Cartan
matrices of finite-dimensional simple Lie algebras
\cite{IMp1,IMp2,IMp3}.
An open problem here is to find all polynomials
related to classical series of Lie algebras  (at present
only ${\bf A_1}, {\bf A_2}, {\bf A_3}$ and ${\bf C_2}$ solutions are
known \cite{GrIvMel}). In extremal case $\mu \to +0$ several examples
were presented earlier in \cite{LMMP}. Another open problem
is to verify the Verlinde-Cardy formula (or modifications)
for the entropy of black branes related to Lie algebras.

Due to dimensional restrictions, only a small number of
 solutions related to simple Lie algebras of different ranks
appear in $D \leq 11$ supergravities (among them
${\bf A_2}$-dyon solutions). We considered
not obviously supergravitational theories, but also
a chain of $B_D$-models in dimensions $D \geq 12$
(we remind the reader
that $B_{12}$ model appear in the field limit of $F$-theory
\cite{KKLP}).

Another topic of interest  was the investigation
of Toda-type $p$-brane solutions related to hyperbolic Kac-Moody
(KM) algebras.  This looks  promising not only because of existence of
$p$-brane solutions related  to hyperbolic KM algebras (see subsect. 3.1.2)
but also due to the explanation of never ending oscillating
behaviour in supergravitational cosmologies and pure gravitational in terms
of hyperbolic algebras \cite{DamH3,DamHJN} (for general construction of
$p$-brane billiards see \cite{IMb1,IM}).

It should be noted that here we were also trying
to present the most general classes of classical composite non-localized
$p$-branes solutions up to solutions of Laplace equations
and/or Toda-like equations. Thus, any future progress in
investigations of Toda-like differential equations may be
immediately "recorded" in terms of new $p$-brane solutions.
In other words, all "kynematics" (intersections, power coefficients, etc)
may be solved and presented here, the only progress may be expected
in "dynamics."

\newpage

\section{Appendix}

\subsection{Appendix 1}

\subsubsection{Ricci-tensor components}

The nonzero Ricci tensor components
for the metric (\ref{2.11}) are the following \cite{IM0}
\bear{A1.1}
R_{\mu \nu}[g]  =   R_{\mu \nu}[g^0 ] +
          g^0 _{\mu \nu} \Bigl[- \Delta_0 \gamma
          +(2-d_0)  (\p \gamma)^2
- \p \gamma \sum_{j=1}^{n} d_j \p \phi^j ]
\\ \nn
+ (2 - d_0) (\gamma_{;\mu \nu} - \gamma_{,\mu} \gamma_{,\nu})
 - \sum_{i=1}^{n} d_i ( \phi^i_{;\mu \nu} - \phi^i_{,\mu} \gamma_{,\nu}
 - \phi^i_{,\nu} \gamma_{,\mu} + \phi^i_{,\mu} \phi^i_{,\nu}),
\\
\label{A1.2}
R_{m_{i} n_{i}}[g]  = {R_{m_{i} n_{i}}}[g^i ]
     - e^{2 \phi^{i} - 2 \gamma} g^i _{m_{i} n_{i}}
      \biggl\{ \Delta_0 \phi^{i}
+ (\p \phi^{i}) [ (d_0 - 2) \p \gamma  +
          \sum_{j=1}^{n} d_j \p \phi^j ] \biggr\},
\ear
Here
$\p \beta \,\p \gamma \equiv g^{0\  \mu \nu} \beta_{, \mu} \gamma_{, \nu}$
and  $\Delta_0 = \Delta[g^0]$ is the Laplace-Beltrami operator corresponding
to  $g^0 $ and all covariant derivatives correspond to $g^0$.
The scalar curvature for (\ref{2.11}) is \cite{IM0}
\bear{A1.3}
  R[g] =  \sum_{i =1}^{n} e^{-2 \phi^i} {R}[g^i ]
          + e^{-2 \gamma} \biggl\{ {R}[g^0 ]
          - \sum_{i =1}^{n} d_i (\p \phi^i)^2
\\ \nn
  -  (d_0 {-} 2) (\p \gamma)^2
     -  (\p f)^2 - 2 \Delta_0 (f +  \gamma) \biggr\},
\ear
where $f$ is defined in  (\ref{2.2.40}).
Relations for the Ricci tensor may be obtained using
the relations for the Riemann tensor from the next subsection
and the relations for the conformal transformations from the last subsection.

\subsubsection{Riemann tensor.}

Let us denote $\bar{g}^0 = e^{2 \gamma} g^0$.
The non-zero components of the Riemann tensor corresponding
to metric  The set $S$ consists of elements $s=(a_s,v_s,I_s)$,
where $a_s \in \tri$, $v_s = e, m$ and $I_s \in \Omega_{a_s,v_s}$.
 have the following form
\bear{A1.4}
&&R_{\mu \nu \rho \sigma }[g] =
R_{\mu \nu \rho \sigma }[\bar{g}^0], \\   \label{A1.5}
&&R_{\mu m_i \nu n_i }[g]    = - R_{m_i \mu \nu n_i }[g] =
- R_{\mu m_i n_i \nu }[g]    =  \nonumber \\
&& R_{m_i \mu n_i \nu }[g] =  - \exp(2 \phi^i) g^{i }_{ m_i n_i}
[\btd_{\mu}[\bar{g}^0](\p_{\nu} \phi^i)  +
(\p_{\mu} \phi^i) (\p_{\nu} \phi^i)],  \\     \label{A1.6}
&& R_{m_i n_j p_k q_l }[g]  = \exp(2 \phi^i)
\delta_{ij} \delta_{kl} \delta_{ik} R_{m_i n_i p_i q_i }[g^{i }] +
\nonumber \\
&&\exp(2 \phi^i + 2 \phi^j)
\bar{g}^{0 \mu \nu}(\p_{\mu} \phi^i) (\p_{\nu} \phi^j)
[ \delta_{il} \delta_{jk}   g^{i }_{ m_i q_i} g^{j}_{ n_j p_j}
-  \delta_{ik} \delta_{jl}
g^{i }_{ m_i p_i} g^{j}_{ n_j q_j} ],
\ear
where indices $\mu, \nu, \rho, \sigma$ correspond to $M_0$ and
$m_i, n_i, p_i, q_i$  to $M_i$; $i,j,k,l= 1, \ldots, n$;
${\btd}[g^0]$ is covariant derivative with respect to $g^0$.
Here we consider the chart $C_0 \times \ldots \times C_n$
on the manifold $M_{0}  \times M_{1} \times \ldots \times M_{n}$,
where $C_{\nu}$ is a chart on $M_{\nu}$, $\nu= 0,\ldots,n$.

The relations (\ref{A1.4})-(\ref{A1.6}) may be obtained from the following
relations for the non-zero components of the
Christophel-Schwarz symbols
\bear{A1.7}
&&\Gamma^{\mu}_{ \nu \rho }[g] =
\Gamma^{\mu}_{ \nu \rho }[\bar{g}^0], \\   \label{A.6}
&&\Gamma^{m_i}_{n_i \nu }[g] = \Gamma^{m_i}_{\nu n_i }[g]
= \delta ^{m_i}_{n_i} \p_{\nu} \phi^i,   \\  \label{A.7}
&&\Gamma^{\mu}_{ m_i n_i }[g] =
- \bar{g}^{0 \mu \nu} (\p_{\nu} \phi^i)
\exp(2 \phi^i)  g^{i }_{ m_i n_i},  \\   \label{A.8}
&&\Gamma^{m_i}_{ n_i p_i }[g] =
\Gamma^{m_i}_{ n_i p_i }[g^{i }],
\ear
$i= 1,\ldots,n$.

\subsubsection{Riemann tensor squared (Kretchmann scalar).}

It follows from the relations (\ref{A1.4})-(\ref{A1.7})
that the Riemann tensor squared  for the metric
(\ref{2.11}) with $\gamma = 0$ has the following form  \cite{IM8}
\bear{A.10}
&&{\cal K}[g] = I [\bar{g}^0] +
\sum_{i=1}^{n} \{ e^{-4 \phi^{i}} I[g^{i}]
- 4 e^{-2 \phi^i} {U}[\bar{g}^0, \phi^i] {R}[g^{i }]
\nonumber \\
&&- 2 d_i {U^2}[\bar{g}^0, \phi^i]  +
4 d_i {V}[\bar{g^0}, \phi^i] \} +
\sum_{i,j =1}^{n} 2 d_i d_j [\bar{g}^{0 \mu \nu} (\p_{\mu} \phi^i)
\p_{\nu} \phi^j]^2 ,
\ear
where
${R}[g^{i }]$ is  scalar curvature
of $g^{i }$  and  $d_i = {\rm dim} M_i$  is
dimension of $M_i$, $i = 1, \ldots, n$.  In (\ref{A.10})
\bear{A.11}
&&{U}[g,\phi] \equiv g^{MN} (\p_{M} \phi) \p_{N} \phi, \\  \label{A.12}
&&{V}[g,\phi] \equiv
g^{M_{1}N_{1}} g^{M_{2}N_{2}}
[\btd_{M_1}(\p_{M_2} \phi)  +  (\p_{M_1} \phi) \p_{M_2} \phi] \times
\nonumber \\
&&[\btd_{N_1}(\p_{N_2} \phi)  +  (\p_{N_1} \phi) \p_{N_2} \phi],
\ear
where $\btd = {\btd}[g]$ is covariant derivative with respect to $g$.

\subsubsection{The cosmological case.}

Now we consider the special case of the metric (\ref{2.11}) with
$M_0 = (t_1, t_2)$, $t_1 < t_2$.  Thus, we consider the metric
\beq{A.13}
g_c = - {B}(t) dt \otimes dt +  \sum_{i=1}^{n} {A_{i}}(t) \hat{g}^{i},
\eeq
defined on the manifold
\beq{A.14}
M = (t_1, t_2) \times M_{1} \times \ldots \times M_{n}.
\eeq

Here $g^{i }$ is a metric on  $M_{i}$ and
${B}(t), {A_{i}}(t) \neq 0$ are smooth functions, $i = 1, \ldots, n$.

From  (\ref{A.11}) we obtain the Riemann tensor squared for the metric
(\ref{A.14}) \cite{IM8}
\bear{A.15}
{\cal K}[g_c] =
&& \sum_{i=1}^{n} \{ A_{i}^{-2} {\cal K}[g^{i }] + A_{i}^{-3} B^{-1}
\dot{A}_{i}^{2} {R}[g^{i }]  - \frac{1}{8}d_{i} B^{-2} A_{i}^{-4}
\dot{A}_{i}^{4}
 \nonumber \\
&&+\frac{1}{4} d_{i} B^{-2}(2 A_{i}^{-1}
\ddot{A}_{i} - B^{-1} \dot{B} A_{i}^{-1} \dot{A}_{i} - A_{i}^{-2}
 \dot{A}_{i}^{2})^{2} \}
 \nonumber \\
&&+ \frac{1}{8}  B^{-2} [\sum_{i=1}^{n} d_{i}
(A_{i}^{-1} \dot{A}_{i})^{2}]^{2}.
\ear

\subsubsection{Parameter $C = C(b)$.}

Here we also present the relation for the parameter
$C = C(b)$, $b \in X$, from (\ref{3.1.3.29})
\ber{A.15.1}
C = C_0 + C_1 + C_2, \mm
\label{A.15.2}
C_0 = 2 (d_0-1)(d_0 -2) \alpha^2 (\alpha -2)^2, \mm
\label{A.15.3}
C_1 = 4 [(d_0 - 1) \alpha^2  + (\alpha - 1)^2]
\sum_{i=1}^{n} d_i \alpha_i^2, \mm
\label{A.15.4}
C_2 = 2 (\sum_{i=1}^{n} d_i \alpha_i^2)^2 - 2
\sum_{i=1}^{n} d_i \alpha_i^4,
\eer
where
\ber{A.15.5}
\alpha = \alpha(b) \equiv
(d_0-2) \sum_{s\in S(b)}(-\eps_s)\nu_s^2 \frac{d(I_s)}{D-2}
= (d_0 -2) \eta(b) + 1, \mm
\label{A.15.6}
\alpha_i = \alpha_i(b) \equiv
(d_0-2) \sum_{s\in S(b)}(-\eps_s)\nu_s^2
\left[\delta_{iI_s} - \frac{d(I_s)}{D-2} \right],
\eer
$i=1, \dots,n$.

It follows from definitions  (\ref{A.15.1})-(\ref{A.15.4})
that $C \geq 0$ and
\ber{A.15.7}
C =0 \Leftrightarrow (\alpha = 0, 2, \quad \alpha_i =0, \ i=1, \dots,n).
\eer

Parameter $C$ appears in the Kretschmann scalar
(\ref{3.1.2.31e}) for the  metric
\ber{A.15.8}
g_*= r^{-2 \alpha}[dr \otimes dr + r^2  d \Omega^2_{d_0 -1} ]
+ \sum_{i = 1}^{n} r^{2\alpha_i} \hat{g}^i,
\eer
with $R[g^i]= {\cal K}[g^i]=0$, $i=1,\dots,n$.
Using formula the \ref{A.15},
we obtain
\ber{A.15.9}
{\cal K}[g_*] = C r^{-4 +4 \alpha}.
\eer

\subsubsection{Conformal transformation}

Here  we also present for a convenience the well-known
relations \cite{KSMH}
\bear{A.16}
&& e^{-2 \gamma} R_{\mu \nu \rho \sigma }[e^{2 \gamma} g^0]
  =    R_{\mu \nu \rho \sigma }[g^0]  + \nonumber \\
&&Y_{\nu \rho} g^0_{\mu \sigma } -
Y_{\mu \rho} g^0_{\nu \sigma } -
Y_{\nu \sigma}  g^0_{\mu \rho } +
Y_{\mu \sigma}  g^0_{\nu \rho },  \\   \label{A.17}
&&R_{\mu \nu}[e^{2 \gamma} g^0]
= R_{\mu \nu}[ g^0] + (2 - d_0) Y_{\mu \nu}
- g^0_{\mu \nu} (g^{0 \rho \tau}  Y_{\rho \tau}), \\ \label{A.18}
&& \btu[e^{2 \gamma} g^0] =
e^{- 2 \gamma} \{ \btu[g^0]  +
(d_0 - 2) g^{0 \mu \nu} (\p_{\mu} \gamma) \p_{\nu} \}
\ear
where  the metric $g^0$  is defined
on $M_0$, ${\rm dim}M_0 = d_0$, $\btu[g^0]$ is
Laplace-Beltrami operator on $M_0$  and
\beq{A.19}
Y_{\mu \nu} = \gamma_{; \mu \nu} -  \gamma_{\mu } \gamma_{\nu }
+ \frac{1}{2} g^0_{\mu \nu }  \gamma_{\rho } \gamma^{\rho}.
\eeq

\subsection{Appendix 2. Product of forms}

Let $F_1$ and $F_2$ be forms of rank $r$ on $(M,g)$ ($M$ is a manifold and
$g$ is a metric on it). We define
\bear{A2.1}
(F_1\cdot F_2)_{MN}\equiv
{(F_1)_{MM_2\dots M_r}(F_2)_N}^{M_2\cdots M_r};
\\ \label{A2.2}
F_1F_2 \equiv {(F_1\cdot F_2)_M}^M=
(F_1)_{M_1\dots M_r}(F_2)^{M_1\dots M_r}.
\ear
It is clearly, that
\beq{A2.3}
(F_1\cdot F_2)_{MN}=(F_2\cdot F_1)_{NM},\quad F_1F_2=F_2F_1.
\eeq

For the volume forms (\ref{2.16}) we get
\bear{A2.4}
\tau(I) \tau(I)=   d(I)! \eps(I), \\
\label{A2.5}
( \tau(I) \cdot \tau(I))_{m_i n_i} = (d(I) -1)! \eps(I) \delta_{iI},
\ear
where indices $m_i, n_i$ correspond to the manifold $M_i$,
$i = 1, \ldots, n$. The symbols $\eps(I)$ and $\delta_{iI}$
are defined in (\ref{2.17})  and (\ref{2.2.12}) respectively.

For the form $F^{(a,e,I)}$ from (\ref{2.1.2}) and metric $g$ from (\ref{2.11})
we obtain from (\ref{A2.4})-(\ref{A2.5})
\bear{A2.6}
\frac1{n_a!}(F^{(a,e,I)}\cdot F^{(a,e,I)})_{\mu\nu}=
\frac{A(I)}{n_a}
\partial_\mu\Phi^{(a,e,I)}\partial_\nu\Phi^{(a,e,I)} \exp(2\gamma);
\\ \label{A2.7}
\frac1{n_a!}(F^{(a,e,I)}\cdot F^{(a,e,I)})_{m_in_i}=
\delta_{iI} g^i_{m_i n_i} \frac{A(I)}{n_a}(\partial\Phi^{(a,e,I)})^2
\exp(2\phi^i),
\ear
where indices $m_i, n_i$ correspond to the manifold $M_i$, $i = 1, \ldots, n$, and
\beq{A2.8}
A(I)= \eps(I)\exp\Bigl(-2\gamma-2\sum_{i\in I}d_i\phi^i \Bigr),
\eeq
$I\in\Omega_{a,e}$. All other components of $(F^{(a,e,I)}\cdot
F^{(a,e,I)})_{MN}$ are zero. For the scalar invariant we have
\beq{A2.9}
\frac1{n_a!}({\cal F}^{(a,e,I)})^2\equiv\frac1{n_a!}{\cal F}^{(a,e,I)}{\cal F}^{(a,e,I)}=
A(I)(\partial\Phi^{(a,e,I)})^2,
\eeq
$I\in\Omega_{a,e}$. Here, as above, we use the notations:
$\partial\Phi_1\partial\Phi_2=
g^{0\mu\nu}\partial_\mu\Phi_1\partial_\nu\Phi_2$ and
$(\partial\Phi_1)^2=\partial\Phi_1\partial\Phi_1$ for functions
$\Phi_1=\Phi_1(x)$ and $\Phi_2=\Phi_2(x)$ on $M_0$.

Analogous relations for magnetic case may be
obtained using the formulas
\bear{A2.10}
\frac1{k_*!}(*F_1)(*F_2)=\frac{\eps[g]}{k!}F_1F_2, \\
\label{A2.11}
\frac1{(k_*-1)!}[(*F_1)\cdot(*F_2)]_{MN}=
\frac{\eps[g]}{k!}\{g_{MN}(F_1F_2)-k(F_2\cdot F_1)_{MN}\},
\ear
where $k=\rank F_i$ and $k_*= \rank(*F_i) = D - k$, $i =1,2$.

Let $I,J \in \Omega$, $I \neq J$ and $d(I) = d(J)$. Then
\beq{A2.12}
\tau(I) \tau(J) = 0.
\eeq
Due to {\bf Restriction 1}
from Section 2 (or {\bf Restriction 3} from Sect. 4.2)
\beq{A2.13}
(\tau(I) \cdot \tau(J))_{MN} = 0.
\eeq

It follows from (\ref{A2.10}) and  (\ref{A2.12}) that for
 $I\ne J$
\beq{A2.14}
{\cal F}^{(a,v,I)} {\cal F}^{(a,v,J)}=0,
\eeq
$I,J \in \Omega_{a,v}$, $v = e,m$.
For  composite field
\ber{A2.15}
F^{a,v}=\sum_{I\in\Omega_{a,v}} {\cal F}^{(a,v,I)},
\eer
$a \in \Delta$, $v = e,m$ we  get (see (\ref{A2.14}))
\bear{A2.16}
(F^{a,v})^2= \sum_{I\in\Omega_{a,v}}({\cal F}^{(a,v,I)})^2,
\\  \label{A2.17}
(F^{a,v}\cdot F^{a,v})_{MN}=
\sum_{I \in \Omega_{a,v}}({\cal F}^{(a,v,I)}\cdot
{\cal F}^{(a,v,I)})_{MN}+
\sum_{I,J \in \Omega_{a,v}\atop I\ne J}
({\cal F}^{(a,v,I)}\cdot {\cal F}^{(a,v,J)})_{MN}.
\ear
The last term in (\ref{A2.17}) gives rise to off-block-diagonal
components of stress-energy tensor.

For $a \in \Delta$ ($d_0\neq 2$) we obtain
\beq{A2.18}
{\cal F}^{(a,e,I)} {\cal F}^{(a,m,J)}=
{\cal F}^{(a,m,J)} {\cal F}^{(a,e,I)}=0,
\eeq
$I\in\Omega_{a,e}$, $J\in\Omega_{a,m}$, and hence
\beq{A2.19}
F^{a,e}F^{a,m}=F^{a,m}F^{a,e}=0, \qquad
(F^a)^2=(F^{a,e})^2+(F^{a,m})^2,
\eeq
where $F^a =F^{a,e} + F^{a,m}$. We also get
\bear{A2.21}
(F^a\cdot F^a)_{MN}=(F^{a,e}\cdot F^{a,e})_{MN}+
(F^{a,m}\cdot F^{a,m})_{MN} \nn \\
+(F^{a,e}\cdot F^{a,m})_{MN}+(F^{a,m}\cdot F^{a,e})_{MN}
\ear
for $a \in \Delta$. The last two terms in (\ref{A2.21}) give rise to
off-block-diagonal components of stress-energy tensor.

\subsection{Appendix 3. Simple finite dimensional Lie algebras}

In summary \cite{FS}, there are four infinite series of simple Lie algebras,
which are denoted by
$${\bf A_r} \ (r\ge1), \quad {\bf  B_r} \ (r\ge3), \quad
{\bf  C_r} \ (r\ge2), \quad {\bf D_r} \ (r\ge4),$$
and in addition five isolated cases, which are called
$${\bf E_6}, \quad {\bf E_7}, \quad {\bf E_8}, \quad {\bf G_2}, \quad {\bf
F_4}.$$
In all cases the subscript denotes the rank of the algebra. The algebras
in the infinite series of simple Lie algebras are called the classical
(Lie) algebras. They are isomorphic to the matrix algebras
$${\bf A_r } \cong \SL(r+1), \quad {\bf  B_r} \cong \SO(2r+1), \quad
{\bf  C_r} \cong \SP(r), \quad {\bf D_r} \cong \SO(2r).$$
The five isolated cases are referred to as the exceptional Lie algebras.

{\bf $A_r$ series.} Let $A$ be $r\times r$ Cartan matrix for the Lie
algebra ${\bf A_r}= \SL(r+1)$, $r\ge 1$.
This matrix is described graphically by
the Dynkin diagram pictured on Fig. A.1.
\begin{center}
\bigskip
\begin{picture}(66,11)
\put(5,10){\circle*{2}}
\put(15,10){\circle*{2}}
\put(25,10){\circle*{2}}
\put(55,10){\circle*{2}}
\put(65,10){\circle*{2}}
\put(5,10){\line(1,0){30}}
\put(45,10){\line(1,0){10}}
\put(5,5){\makebox(0,0)[cc]{1}}
\put(15,5){\makebox(0,0)[cc]{2}}
\put(25,5){\makebox(0,0)[cc]{3}}
\put(40,10){\makebox(0,0)[cc]{$\dots$}}
\put(55,5){\makebox(0,0)[cc]{$r-1$}}
\put(65,5){\makebox(0,0)[cc]{$r$}}
\put(55,10){\line(1,0){10}}
\end{picture} \\[5pt]
\small

Fig. A.1. \it Dynkin diagram for $A_r$ Lie algebra
\end{center}

Using the relation for the inverse matrix $A^{-1}=(A^{ss'})$
(see Sect.7.5 in \cite{FS})
\beq{A3.12}
A^{ss'}=\frac1{r+1}\min(s,s')[r+1-\max(s,s')]
\eeq
we get for $n_s \equiv 2\sum_{s'=1}^{r} A^{ss'}$:
\beq{A3.13}
n_s = s(r+1 -s),
\eeq
$s = 1,\dots,r$.

{\bf ${\bf  B_r}$ and ${\bf  C_r}$ series.} Dynkin diagrams for these
cases are  pictured on Fig. A.2.
\begin{center}
\bigskip
\begin{picture}(66,12)
\put(5,10){\circle*{2}}
\put(15,10){\circle*{2}}
\put(25,10){\circle*{2}}
\put(55,10){\circle*{2}}
\put(65,10){\circle*{2}}
\put(5,10){\line(1,0){30}}
\put(45,10){\line(1,0){10}}
\put(5,5){\makebox(0,0)[cc]{1}}
\put(15,5){\makebox(0,0)[cc]{2}}
\put(25,5){\makebox(0,0)[cc]{3}}
\put(40,10){\makebox(0,0)[cc]{$\dots$}}
\put(55,5){\makebox(0,0)[cc]{$r-1$}}
\put(65,5){\makebox(0,0)[cc]{$r$}}
\put(55,11){\line(1,0){10}}
\put(55,9){\line(1,0){10}}
\put(60,10){\makebox(0,0)[cc]{\large $>$}}
\end{picture} \qquad
\begin{picture}(66,12)
\put(5,10){\circle*{2}}
\put(15,10){\circle*{2}}
\put(25,10){\circle*{2}}
\put(55,10){\circle*{2}}
\put(65,10){\circle*{2}}
\put(5,10){\line(1,0){30}}
\put(45,10){\line(1,0){10}}
\put(5,5){\makebox(0,0)[cc]{1}}
\put(15,5){\makebox(0,0)[cc]{2}}
\put(25,5){\makebox(0,0)[cc]{3}}
\put(40,10){\makebox(0,0)[cc]{$\dots$}}
\put(55,5){\makebox(0,0)[cc]{$r-1$}}
\put(65,5){\makebox(0,0)[cc]{$r$}}
\put(55,11){\line(1,0){10}}
\put(55,9){\line(1,0){10}}
\put(60,10){\makebox(0,0)[cc]{\large $<$}}
\end{picture} \\[5pt]
\small
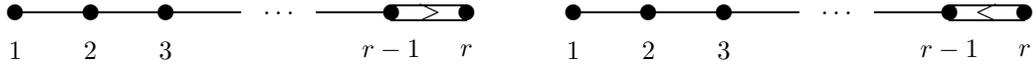
Fig.A.2. \it Dynkin diagrams for ${\bf  B_r}$ and ${\bf  C_r}$ Lie algebras
\end{center}

In these cases we have the following formulas for inverse Cartan matrices
\ber{A3.14}
A^{ss'}=\left\{\begin{array}{ll}
\min(s,s') & \mbox{for }s\ne r, \\[5pt]
\frac12s' & \mbox{for }s=r,
\end{array}\right. \quad
A^{ss'}=\left\{\begin{array}{ll}
\min(s,s') & \mbox{for }s'\ne r, \\[5pt]
\frac12s & \mbox{for }s'=r
\end{array}\right.
\eer
and
\ber{A3.15}
n_s =\left\{\begin{array}{ll}
s(2r+1 -s) & \mbox{for }s\ne r,\\
\frac r2 (r+1) & \mbox{for }s=r;
\end{array}\right.\quad
n_s =s(2r -s),
\eer
for ${\bf  B_r}$ and ${\bf  C_r}$ series respectively, $s =1,\dots,r$.

{\bf ${\bf D_r}$ series.} We have the following Dynkin diagram for this case
(Fig. A.3):
\begin{center}
\bigskip
\begin{picture}(63,31)
\put(5,20){\circle*{2}}
\put(15,20){\circle*{2}}
\put(25,20){\circle*{2}}
\put(55,20){\circle*{2}}
\put(5,20){\line(1,0){30}}
\put(45,20){\line(1,0){10}}
\put(5,15){\makebox(0,0)[cc]{1}}
\put(15,15){\makebox(0,0)[cc]{2}}
\put(25,15){\makebox(0,0)[cc]{3}}
\put(40,20){\makebox(0,0)[cc]{$\dots$}}
\put(55,15){\makebox(0,0)[rc]{$r-2$}}
\put(62,27){\circle*{2}}
\put(62,13){\circle*{2}}
\put(55,20){\line(1,1){7}}
\put(55,20){\line(1,-1){7}}
\put(62,31){\makebox(0,0)[cc]{$r$}}
\put(62,8){\makebox(0,0)[cc]{$r-1$}}
\end{picture} \\[5pt]
\small

Fig.A.3. \it Dynkin diagram for ${\bf D_r}$ Lie algebra

\end{center}

and formula for the inverse matrix \cite{FS}:
\ber{A3.16}
A^{ss'}=\left\{\begin{array}{ll}
\min(s,s') & \mbox{for }s,s'\notin\{r,r-1\}, \\[5pt]
\frac12s & \mbox{for }s\notin\{r,r-1\}, \ s'\in\{r,r-1\}, \\[5pt]
\frac12s' & \mbox{for }s\in\{r,r-1\}, \ s'\notin\{r,r-1\}, \\[5pt]
\frac14r & \mbox{for }s=s'=r \mbox{ or } s=s'=r-1, \\[5pt]
\frac14(r-2) & \mbox{for }s=r, \ s'=r-1 \mbox{ or vice versa.}
\end{array}\right.
\eer
Then
\ber{A3.17}
n_s =\left\{\begin{array}{ll}
s(2r-1 -s) & \mbox{for }s\notin\{r,r-1\},\\
\frac r2(r-1) & \mbox{for }s\in\{r,r-1\},
\end{array}\right.
\eer
$s =1,\dots,r$.

Let us consider the exceptional Lie algebras. Dynkin diagrams of
these algebras are pictured on Fig. 4.A.
\begin{center}
\bigskip
\begin{picture}(46,21)
\put(5,10){\circle*{2}}
\put(15,10){\circle*{2}}
\put(25,10){\circle*{2}}
\put(35,10){\circle*{2}}
\put(45,10){\circle*{2}}
\put(25,20){\circle*{2}}
\put(5,10){\line(1,0){40}}
\put(25,10){\line(0,1){10}}
\put(5,5){\makebox(0,0)[cc]{1}}
\put(15,5){\makebox(0,0)[cc]{2}}
\put(25,5){\makebox(0,0)[cc]{3}}
\put(35,5){\makebox(0,0)[cc]{4}}
\put(45,5){\makebox(0,0)[cc]{5}}
\put(28,20){\makebox(0,0)[lc]{6}}
\end{picture} \qquad
\begin{picture}(56,21)
\put(5,10){\circle*{2}}
\put(15,10){\circle*{2}}
\put(25,10){\circle*{2}}
\put(35,10){\circle*{2}}
\put(45,10){\circle*{2}}
\put(25,20){\circle*{2}}
\put(5,10){\line(1,0){40}}
\put(25,10){\line(0,1){10}}
\put(5,5){\makebox(0,0)[cc]{1}}
\put(15,5){\makebox(0,0)[cc]{2}}
\put(25,5){\makebox(0,0)[cc]{3}}
\put(35,5){\makebox(0,0)[cc]{4}}
\put(45,5){\makebox(0,0)[cc]{5}}
\put(28,20){\makebox(0,0)[lc]{7}}
\put(55,10){\circle*{2}}
\put(45,10){\line(1,0){10}}
\put(55,5){\makebox(0,0)[cc]{6}}
\end{picture} \\[15pt]
\begin{picture}(66,21)
\put(5,10){\circle*{2}}
\put(15,10){\circle*{2}}
\put(25,10){\circle*{2}}
\put(35,10){\circle*{2}}
\put(45,10){\circle*{2}}
\put(45,20){\circle*{2}}
\put(5,10){\line(1,0){40}}
\put(45,10){\line(0,1){10}}
\put(5,5){\makebox(0,0)[cc]{1}}
\put(15,5){\makebox(0,0)[cc]{2}}
\put(25,5){\makebox(0,0)[cc]{3}}
\put(35,5){\makebox(0,0)[cc]{4}}
\put(45,5){\makebox(0,0)[cc]{5}}
\put(48,20){\makebox(0,0)[lc]{8}}
\put(55,10){\circle*{2}}
\put(65,10){\circle*{2}}
\put(45,10){\line(1,0){20}}
\put(55,5){\makebox(0,0)[cc]{6}}
\put(65,5){\makebox(0,0)[cc]{7}}
\end{picture} \qquad
\begin{picture}(36,12)
\put(5,10){\circle*{2}}
\put(15,10){\circle*{2}}
\put(25,10){\circle*{2}}
\put(35,10){\circle*{2}}
\put(5,10){\line(1,0){10}}
\put(15,11){\line(1,0){10}}
\put(15,9){\line(1,0){10}}
\put(25,10){\line(1,0){10}}
\put(20,10){\makebox(0,0)[cc]{\large $>$}}
\put(5,5){\makebox(0,0)[cc]{1}}
\put(15,5){\makebox(0,0)[cc]{2}}
\put(25,5){\makebox(0,0)[cc]{3}}
\put(35,5){\makebox(0,0)[cc]{4}}
\end{picture} \qquad
\begin{picture}(16,12)
\put(5,10){\circle*{2}}
\put(15,10){\circle*{2}}
\put(5,11){\line(1,0){10}}
\put(5,10){\line(1,0){10}}
\put(5,9){\line(1,0){10}}
\put(10,10){\makebox(0,0)[cc]{\large $>$}}
\put(5,5){\makebox(0,0)[cc]{1}}
\put(15,5){\makebox(0,0)[cc]{2}}
\end{picture} \\[5pt]
\small

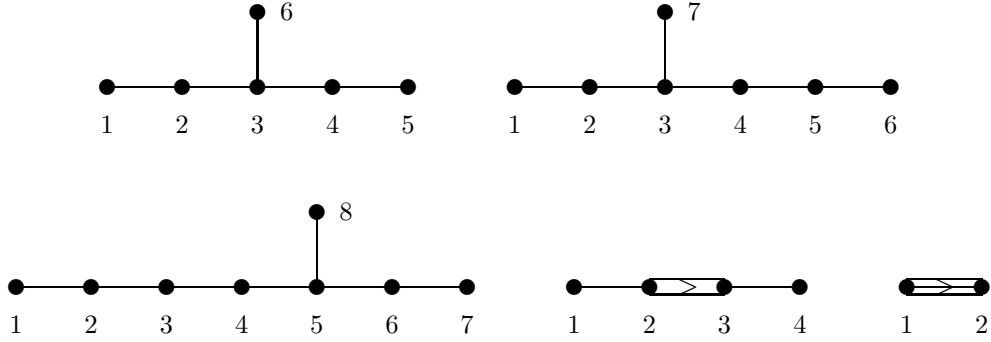
Fig.4.A. \it Dynkin diagrams for ${\bf E_6}$, ${\bf E_7}$, ${\bf E_8}$, ${\bf F_4}$ and ${\bf G_2}$
Lie algebras, respectively
\end{center}
Using relations for inverse Cartan matrices from \cite{FS} we get
\ber{A3.18}
 n_s/2  =\left\{\begin{array}{ll}
8,15,21,15,8,11 & \mbox{for }{\bf E_6}, \ s=1,\dots,6; \\[5pt]
17,33,48,\ds\frac{75}2,26,\ds\frac{27}2,\ds\frac{49}2 &
\mbox{for }{\bf E_7}, \ s=1,\dots,7; \\[7pt]
29,57,84,110,135,91,46,68 & \mbox{for }{\bf E_8}, \ s=1,\dots,8; \\[5pt]
11,21,15,8 & \mbox{for }{\bf F_4}, \ s=1,\dots,4; \\[5pt]
5,3 & \mbox{for }{\bf G_2}, \ s=1,2.
\end{array}\right.
\eer

\subsection{Appendix 4: Solutions for Toda-like system }

\subsubsection{General solutions}

Let
\bear{A4.1}
L=\frac12<\dot x,\dot x>- \sum_{s \in S} A_s\exp(2<u_s,x>)
\ear
be a Lagrangian, defined on $V\times V$, where $V$ is
$n$-dimensional vector space over $\R$, $A_s\ne0$, $s \in S$;
$S \ne\emptyset$, and $<\cdot,\cdot>$ is
non-degenerate real-valued quadratic form on $V$.
Let $K_s = <u_s,u_{s}> \neq 0$,
for all $s \in S$.

Then, the Euler-Lagrange equations for the Lagrangian (\ref{A4.1})
\beq{A4.4}
\ddot{x} + \sum_{s \in S} 2 A_s u_s \exp(2<u_s,x>) =0,
\eeq
have the following  solutions
\beq{A4.5}
x(t)=  \sum_{s \in S} \frac{q^s(t)  u_s}{<u_s,u_{s}>} +
\alpha t + \beta,
\eeq
where $\alpha,\beta\in V$,
\beq{A4.6}
<\alpha,u_s>=<\beta,u_s>=0,
\eeq
$s \in S$, and functions $q^s(u)$
satisfy the Toda-like  equations
\beq{A4.7}
\ddot{q^s} = -  2 A_s  K_s \exp( \sum_{s' \in S} A_{s s'} q^{s'} ),
\eeq
with
\beq{A4.8}
 A_{s s'} = \frac{2 <u_s,u_{s'}>}{<u_{s'},u_{s'}>},
\eeq
$s, s'  \in S$. Let the matrix  $(A_{s s'})$ be a non-degenerate one.
In this case vectors $u_s$, $s \in S$, are linearly independent.
Then eqs. (\ref{A4.7})  are field equations corresponding to the  Lagrangian
\beq{A4.9}
 L_{TL} = \frac{1}{4}  \sum_{s,s' \in S} K^{-1}_s
   A_{s s'} \dot{q^s} \dot{q^{s'}}
  -  \sum_{s \in S} A_s  \exp( \sum_{s' \in S} A_{s s'} q^{s'} ).
\eeq

For the energy corresponding to the solution (\ref{A4.5}) we get
\beq{A4.10}
E=\frac12<\dot x,\dot x> + \sum_{s \in S} \exp(2<u_s,x>)
= E_{TL} + \frac12 <\alpha,\alpha>,
\eeq
where
\beq{A4.11}
 E_{TL} = \frac{1}{4}  \sum_{s,s' \in S} K^{-1}_s
  A_{s s'} \dot{q^s} \dot{q^{s'}}
  + \sum_{s \in S} A_s  \exp( \sum_{s' \in S} A_{s s'} q^{s'} ),
\eeq
is the energy function  corresponding to the Lagrangian (\ref{A4.10}).

For dual vectors $u^s\in V^*$ defined as
$u^s(x)=<u_s,x>$, $\forall x \in V$, we have $<u^s,u^l>_*=<u_s,u_l>$,
where $< \cdot, \cdot>_*$ is dual form on  $V^*$.  The orthogonality
conditions (\ref{A4.6}) read
\beq{A4.18}
u^s(\alpha)=u^s(\beta)=0 ,
\eeq
$s \in S$.

\subsubsection{Solutions with block-orthogonal set of vectors }

Let us consider the Lagrangian (\ref{A4.1}) with the
set
$S=S_1\sqcup\dots\sqcup S_k$,
all $S_i\ne\emptyset$, and
\beq{A4.1.2b}
<u_s,u_{s'}>=0,
\eeq
for all $s \in S_i$, $s' \in S_j$,
$i\ne j$; $i,j=1,\dots,k$.

Let $h_s = K_s^{-1}$,  $(A^{ss'})= (A_{ss'})^{-1}$,
\beq{A4.1.2}
b_s = 2 \sum_{s' \in S} A^{s s'},
\eeq
for all $s \in S$, and
\beq{A4.1.3}
A_s/(b_s h_s) = A_{s'}/(b_{s'} h_{s'}),
\eeq
$s,s' \in S_i$, $i = 1, \ldots, k$,
(the ratio $A_s/(b_s h_s)$ is constant inside  $S_i$).

Then,  there exists a special solution to eqs. (\ref{A4.7})
\bear{A4.1.4}
q^s(t) = -b_s \ln[y_s(t) |2A_s/(b_s h_s)|]
\ear
where functions $y_s(t) \neq 0$ satisfy to equations
\beq{A4.1.7}
\frac{d}{dt}\left(y_s^{-1} \frac{d y_s}{dt}\right) = - \xi_s y_s^{-2},
\eeq
with
\beq{A4.1.8}
\xi_s = {\rm sign} \left(\frac{A_s}{b_s h_s}\right),
\eeq
$s \in S$, and coincide inside blocks:
\bear{A4.1.9}
y_s(t) = y_{s'}(t),
\ear
$s,s' \in S_i$, $i = 1, \ldots, k$.
More explicitly
\bear{A4.1.10}
y_s(t) = s(t - t_s, \xi_s, C_s),
\ear
where constants $t_s, C_s \in \R$ coincide inside blocks
\bear{A4.1.11}
t_s = t_{s'}, \qquad C_s = C_{s'},
\ear
$s,s' \in S_i$, $i = 1, \ldots, k$,
and
\bear{A4.1.12}
s(t, \xi, C) \equiv
\frac{1}{\sqrt{C}} \sh(t \sqrt{C}), \ \xi = +1, \quad C>0; \\
\label{A4.1.13}
\frac{1}{\sqrt{-C}} \sin(t \sqrt{-C}), \ \xi = +1, \quad C <0; \\
\label{A4.1.14}
t, \ \xi = +1, \quad C = 0; \\
\label{A4.1.15}
\frac{1}{\sqrt{C}} \ch(t \sqrt{C}), \ \xi = -1, \quad C>0.
\ear

For "Toda" part of energy  we get
\beq{A4.1.17}
E_{TL}= \frac12 \sum_{s \in S} C_s b_s h_s .
\eeq

\subsection{Appendix 5: Solutions with  Bessel functions}

Let us consider two differential operators
\bear{A5.1}
2\hat H_0 = - \frac{\partial^2}{\partial z^2} + 2 A \e^{2qz}, \\
\label{A5.2}
2\hat H_1 =
- \e^{qz}\frac\partial{\partial z}
\left(\e^{-qz}\frac\partial{\partial z}\right)+ 2 A \e^{2qz}.
\ear
Equation
\beq{A5.3}
H_k \Psi_k ={\cal E} \Psi_k
\eeq
has the following linearly independent solutions for $q \neq 0$
\beq{A5.4}
\Psi_k(z)=  \e^{kqz/2}  B_{\omega_k({\cal E})}
\left(   \sqrt{2 A} \frac{\e^{qz}}{q}\right), \qquad
\omega_k({\cal E})=\sqrt{\frac{k}{4}- \frac{2{\cal E}}{q^2}},
\eeq
where $k = 0,1$ and $B_\omega,B_\omega=I_\omega,K_\omega$
are modified Bessel function.


\begin{center}
{\bf Acknowledgments}
\end{center}

This work was supported in part
by the DFG grant  436 RUS 113/236/O(R),
by the Russian Ministry of
Science and Technology and  Russian Foundation for Basic Research
grant 98-02-16414, Project SEE and CONACYT, Mexico.


\end{document}